# 2022 Roadmap on Neuromorphic Computing and Engineering

Dennis V. Christensen[1], Regina Dittmann[2], Bernabé Linares-Barranco[3], Abu Sebastian[4], Manuel Le Gallo[4], Andrea Redaelli[5], Stefan Slesazeck[6], Thomas Mikolajick[6,7], Sabina Spiga[8], Stephan Menzel[9], Ilia Valov[9], Gianluca Milano[10], Carlo Ricciardi[11], Shi-Jun Liang[12], Feng Miao[12], Mario Lanza[13], Tyler J. Quill[14], Scott T. Keene[15], Alberto Salleo[14], Julie Grollier[16], Danijela Marković[16], Alice Mizrahi[16], Peng Yao[17], J. Joshua Yang[17], Giacomo Indiveri[18], John Paul Strachan[19], Suman Datta[20], Elisa Vianello[21], Alexandre Valentian[22], Johannes Feldmann[23], Xuan Li[23], Wolfram H.P. Pernice[24,25], Harish Bhaskaran[23], Steve Furber[26], Emre Neftci[27], Franz Scherr[28], Wolfgang Maass[28], Srikanth Ramaswamy[29], Jonathan Tapson[30], Priyadarshini Panda[31], Youngeun Kim[31], Gouhei Tanaka[32], Simon Thorpe[33], Chiara Bartolozzi[34], Thomas A. Cleland[35], Christoph Posch[36], Shih-Chii Liu[18], Gabriella Panuccio[37], Mufti Mahmud[38], Arnab Neelim Mazumder[39], Morteza Hosseini[39], Tinoosh Mohsenin[39], Elisa Donati[18], Silvia Tolu[40], Roberto Galeazzi[40], Martin Ejsing Christensen[41], Sune Holm[42], Daniele Ielmini[43], and N. Pryds[1,44].

---

[1] Department of Energy Conversion and Storage, Technical University of Denmark, DK-2800 Kgs. Lyngby, Denmark
[2] Peter Gruenberg Institute 7, Forschungszentrum Juelich GmbH, 52425 Juelich, Germany and JARA-FIT, RWTH Aachen University, 52056 Aachen, Germany
[3] Instituto de Microelectrónica de Sevilla (IMSE-CNM), CSIC and Universidad de Sevilla, 41092 Seville, Spain
[4] IBM Research – Zurich, Switzerland
[5] STMicroelectronics, Agrate, Italy
[6] NaMLab gGmbH, 01187 Dresden, Germany
[7] Institute of Semiconductors and Microsystems, TU Dresden; Dresden, Germany
[8] CNR-IMM, Unit of Agrate Brianza, via C. Olivetti 2, Agrate Brianza (MB), Italy
[9] FZ Juelich (PGI-7), Juelich , Germany
[10] Advanced Materials Metrology and Life Science Division, INRiM (Istituto Nazionale di Ricerca Metrologica), Torino, Italy
[11] Department of Applied Science and Technology, Politecnico di Torino, Torino, Italy
[12] National Laboratory of Solid State Microstructures, School of Physics, Collaborative Innovation Center of Advanced Microstructures, Nanjing University, Nanjing, China
[13] Physical Sciences and Engineering Division, King Abdullah University of Science and Technology (KAUST), 23955-6900 Thuwal, Saudi Arabia
[14] Department of Materials Science and Engineering, Stanford University, Stanford, CA 94305, United States of America
[15] Department of Engineering, University of Cambridge, Cambridge CB2 1PZ, United Kingdom.
[16] Unité Mixte de Physique, CNRS, Thales, Université Paris-Saclay, 91767 Palaiseau, France
[17] Electrical and Computer Engineering Department, University of Southern California, Los Angeles, CA, USA
[18] Institute of Neuroinformatics, University of Zurich and ETH Zurich, Switzerland
[19] Hewlett Packard Laboratories, Hewlett Packard Enterprise, San Jose, CA, USA
[20] Department of Electrical Engineering, University of Notre Dame, Notre Dame, IN, USA
[21] CEA, LETI, Université Grenoble Alpes, Grenoble, France
[22] CEA, LIST, Université Grenoble Alpes, Grenoble, France
[23] Department of Materials, University of Oxford, Parks Road, OX1 3PH Oxford, UK
[24] Institute of Physics, University of Münster, Heisenbergstr. 11, 48149 Münster, Germany Heisenbergstr.
[25] Center for Soft Nanoscience, University of Münster, 48149 Münster, Germany
[26] The University of Manchester, UK
[27] Department of Cognitive Sciences, University of California, Irvine, Irvine, CA, USA
[28] Institute of Theoretical Computer Science, Graz University of Technology, Graz, Austria
[29] École Polytechnique Fédérale de Lausanne, Geneva, Switzerland
[30] School of Electrical and Data Engineering, University of Technology, Sydney, Australia
[31] Department of Electrical Engineering, New Haven, Yale University, USA
[32] International Research Center for Neurointelligence (IRCN), The University of Tokyo, 7-3-1 Hongo Bunkyo-ku, Tokyo 113-0033, Japan
[33] CerCo, Université Toulouse 3, CNRS, CHU Purpan, Pavillon Baudot, 31059 Toulouse, France
[34] Event Driven Perception for Robotics, Italian Institute of Technology, iCub Facility, Genoa, Italy
[35] Dept. of Psychology, Cornell University, Ithaca, NY, USA
[36] Prophesee, Paris, France
[37] Enhanced Regenerative Medicine, Istituto Italiano di Tecnologia, Italy
[38] Department of Computer Science and Medical Technologies Innovation Facility, Nottingham Trent University, UK
[39] University of Maryland, Baltimore County, Catonsville, USA
[40] Technical University of Denmark, Denmark
[41] The Danish Council on Ethics, Denmark
[42] Department of Food and Resource Economics, University of Copenhagen, Denmark
[43] Politecnico di Milano, 20133 Milano, Italy
[44] Email: Nini Pryds, nipr@dtu.dk

**Abstract:**

Modern computation based on the von Neumann architecture is today a mature cutting-edge science. In the Von Neumann architecture, processing and memory units are implemented as separate blocks interchanging data intensively and continuously. This data transfer is responsible for a large part of the power consumption. The next generation computer technology is expected to solve problems at the exascale with $10^{18}$ calculations each second. Even though these future computers will be incredibly powerful, if they are based on von Neumann type architectures, they will consume between 20 and 30 megawatts of power and will not have intrinsic physically built-in capabilities to learn or deal with complex data as our brain does. These needs can be addressed by neuromorphic computing systems which are inspired by the biological concepts of the human brain. This new generation of computers has the potential to be used for the storage and processing of large amounts of digital information with much lower power consumption than conventional processors. Among their potential future applications, an important niche is moving the control from data centers to edge devices.

The aim of this Roadmap is to present a snapshot of the present state of neuromorphic technology and provide an opinion on the challenges and opportunities that the future holds in the major areas of neuromorphic technology, namely materials, devices, neuromorphic circuits, neuromorphic algorithms, applications, and ethics. The Roadmap is a collection of perspectives where leading researchers in the neuromorphic community provide their own view about the current state and the future challenges for each research area. We hope that this Roadmap will be a useful resource by providing a concise yet comprehensive introduction to readers outside this field, for those who are just entering the field, as well as providing future perspectives for those who are well established in the neuromorphic computing community.

# Contents:

**Introduction**

# Introduction

Computers have become essential to all aspects of modern life and are omnipresent all over the globe. Today, the recent data-intensive applications have placed a high demand on hardware performance, in terms of short access latency, high capacity, large bandwidth, low cost, and ability to execute artificial intelligence (AI) tasks. However, the ever-growing pressure for big data creates additional challenges: on the one hand, *energy consumption* has become a remarkable challenge, due to the rapid development of sophisticated algorithms and architectures. Currently, about 5–15% of the world's energy is spent in some form of data manipulation, such as transmission or processing[1], and this fraction is expected to rapidly increase due to the exponential increase of data generated by ubiquitous sensors in the era of internet of things. On the other hand, data processing is increasingly limited by the *memory bandwidth* due to the Von-Neumann's architecture with physical separation between processing and memory units. While the Von Neumann computer architecture has made an incredible contribution to the world of science and technology for decades, its performance is largely inefficient due to the relatively slow and energy demanding data movement.

Conventional Von-Neumann computers based on complementary metal oxide semiconductor (CMOS) technology do not possess the intrinsic capabilities to learn or deal with complex data as the human brain does. To address the limits of digital computers, there are significant research efforts worldwide in developing profoundly different approaches inspired by biological principles. One of these approaches is the development of *neuromorphic systems*, namely computing systems mimicking the type of information processing in the human brain.

The term "neuromorphic" was originally coined in the 1990s by Carver Mead to refer to mixed signal analog/digital very large scale integration computing systems that take inspiration from the neuro-biological architectures of the brain[2]. "Neuromorphic engineering" emerged as an interdisciplinary research field that focused on building electronic neural processing systems to directly "emulate" the bio-physics of real neurons and synapses[3]. More recently, the definition of the term neuromorphic has been extended in two additional directions[4]. Firstly, the term neuromorphic was used to describe spike-based processing systems engineered to explore large-scale computational neuroscience models. Secondly, neuromorphic computing comprises dedicated electronic neural architectures that implement neuron and synapse circuits. Note that this concept is distinct from AI machine learning approaches which are based on pure software algorithms developed to minimize the recognition error in pattern recognition tasks[5]. However, a precise definition of neuromorphic computing is somewhat debated. It can range from very strict high-fidelity mimicking of neuroscience principles where very detailed synaptic chemical dynamics are mandatory, to very vague high-level loosely brain-inspired principles, such as the simple vector (input) times matrix (synapses) multiplication. In general, as of today, there is a wide consensus that neuromorphic computing should at least encompass some time-, event-, or data-driven computation. In this sense, systems like spiking neural networks (SNN), sometimes referred to as the third generation of neural networks[6], are strongly representative. However, there is an important cross-fertilization between the technologies required to develop efficient SNNs and those for more traditional non-spiking neural networks, referred to as artificial neural networks (ANN), which are typically more time-step-driven. While the former definition of neuromorphic computing is more plausible, in this Roadmap we aim at broadening the scope to emphasize the cross-fertilization between ANN and SNN.

Nature is a vital inspiration for the advancement to a more sustainable computing scenario, where neuromorphic systems display much lower power consumption than conventional processors, due to the integration of non-volatile memory and analog/digital processing circuits as well as the dynamic learning capabilities in the context of complex data. Building ANNs that mimic a biological counterpart is one of the remaining challenges in computing. If the fundamental technical issues are solved in the next few years, the neuromorphic computing market is projected to rise from $0.2 billion in 2025 to $22 billion in 2035[7] as neuromorphic computers with ultra-low power consumption and high speed advance and drive demands for neuromorphic devices.

In line with these increasingly pressing issues, the general aim of the ***Roadmap on Neuromorphic Computing and Engineering*** is to provide an overview of the different fields of research and development that contribute to the advancement of the field, to assess the potential applications of neuromorphic technology in cutting edge technologies and to highlight the necessary advances required to reach these. The Roadmap addresses:

- Neuromorphic materials and devices
- Neuromorphic circuits
- Neuromorphic algorithms
- Applications
- Ethics

*Neuromorphic materials and devices:*
To advance the field of neuromorphic computing and engineering, the exploration of novel materials and devices will be of key relevance in order to improve the power efficiency and scalability of state-of-the-art CMOS solutions in a disruptive manner[4,8]. Memristive devices, which can change their conductance in response to electrical pulses[9–11], are promising candidates to act as energy- and space-efficient hardware representation for synapses and neurons in neuromorphic circuits. Memristive devices have originally been proposed as binary non-volatile random-access memory and research in this field has been mainly driven by the search for higher performance in solid-state drive technologies (e.g., Flash replacement) or storage class memory[12]. However, thanks to their analog tunability and complex switching dynamics, memristive devices also enable novel computing functions such as analog computing or the realisation of brain-inspired learning rules. A large variety of different physical phenomena has been reported to exhibit memristive behaviour, including electronic effects, ionic effects as well as structural or ferroic ordering effects. The material classes range from magnetic alloys, metal oxides, chalcogenides to 2D van de Waals materials or organic materials. Within this Roadmap, we cover a broad range of materials and phenomena with different maturity levels with respect to their use in neuromorphic circuits. We consider emerging memory devices that are already commercially available as binary non-volatile memory such as phase-change memory, magnetic random-access memory, ferroelectric memory as well as redox-based resistive random-access memory and review their prospects for neuromorphic computing and engineering. We complement it with nanowire networks, 2D materials, and organic materials that are less mature but may offer extended functionalities and new opportunities for flexible electronics or 3D integration.

*Neuromorphic circuits:*
Neuromorphic devices can be integrated with conventional CMOS transistors to develop fully functional neuromorphic circuits. A key element in neuromorphic circuits is their non-von Neumann architecture, for instance consisting of multiple cores each implementing distributed computing and memory. Both SSNs, adopting spikes to represent, exchange and compute data in analogy to action potentials in the brain, as well as circuits that are only loosely inspired by the brain, such as ANNs, are generally included in the roster of neuromorphic circuits, thus will be covered in this Roadmap. Regardless of the specific learning and processing algorithm, a key processing element in neuromorphic circuits is the neural network, including several synapses and neurons. Given the central role of the neural network, a significant research effort is currently aimed at technological solutions to realize dense, fast, and energy-efficient neural networks by in-memory computing[13]. For instance, a memory array can accelerate the matrix-vector multiplication (MVM)[14]. This is a common feature of many neuromorphic circuits, including spiking and non-spiking networks, and takes advantage of Ohm's and Kirchhoff's laws to implement multiplication and summation in the network. The MVM crosspoint circuit allows for the straightforward hardware implementation of synaptic layers with high density, high real-time processing speed, and high energy efficiency, although the accuracy is challenged by stochastic variations in memristive devices in particular, and analog computing in general. An additional circuit challenge is the mixed analog-digital computation, which results in the need for large and energy-hungry analog-digital converter circuits at the interface between the analog crosspoint array and the digital system. Finally, neuromorphic circuits seem to take the most benefit from hybrid integration, combining front-end CMOS technology with novel memory devices that can implement MVM and neuro-biological functions, such as spike integration, short-term memory, and synaptic plasticity[15]. Hybrid integration may also need to extend, in the long term, to alternative nanotechnology concepts, such as bottom-up nanowire networks[16], and alternative computing concepts, such as photonic[17] and even quantum computing[18], within a single system or even a single chip with 3D integration. In this scenario, a Roadmap for the development and assessment of each of these individual innovative concepts is essential.

*Neuromorphic algorithms:*
A fundamental challenge in neuromorphic engineering for real application systems is to train them directly in the spiking domain in order to be more energy-efficient, more precise, and also be able to continuously learn and update the knowledge on the portable devices themselves without relying on heavy cloud computing servers. Spiking data tend to be sparse with some stochasticity and embedded noise, interacting with non-ideal non-linear synapses and neurons. Biology knows how to use all this to its advantage to efficiently acquire knowledge from the surrounding environment. In this sense, computational neuroscience can be a key ingredient to inspire neuromorphic engineering, and learn from this discipline how brains perform computations at a variety of scales, from small neurons ensembles, mesoscale aggregations, up to full tissues, brain regions and the complete brain interacting with peripheral sensors and motor actuators. On the other hand, fundamental questions arise on how information is encoded in the brain using nervous spikes. Obviously, to maximize energy efficiency for both processing and communication, the brain maximizes information per unit spike[19]. This means unravelling the information encoding and processing by exploiting spatio-temporal signal processing to maximize information while minimizing energy, speed, and resources.

*Applications:*
The realm of applications for neuromorphic computing and engineering continues to grow at a steady rate, although remaining within the boundaries of research and development. While it is becoming clear that many applications are well suited to neuromorphic computing and engineering, it is also important to identify new potential applications to further understand how neuromorphic materials and hardware can address them. The Roadmap includes some of these emerging applications as examples of biologically-inspired computing approaches for implementation in robots, autonomous transport capability or in perception engineering where the applications are based on integration with sensory modalities of humans.

*Ethics:*
While the future development and application of neuromorphic systems offer possibilities beyond the state of the art, the progress should also be addressed from an ethical point of view where, e.g., lack of transparency in complex neuromorphic systems and autonomous decision making can be a concern. The Roadmap thus ends with a final section addressing some of the key ethical questions that may arise in the wake of advancements in neuromorphic computation.

We hope that this Roadmap represents an overview and updated picture of the current state-of-the-art as well as being the future projection in these exciting research areas. Each contribution, written by leading researchers in their topic, provides the current state of the field, the open challenges, and a future perspective. This should guide the expected transition towards efficient neuromorphic computations and highlight the opportunities for societal impact in multiple fields.

**Acknowledgements**

D.V.C and N.P. acknowledge the funding from Villum Fonden, for the NEED project (00027993), Danish Council for Independent Research Technology and Production Sciences for the DFF Research Project 3 (Grant No. 00069B), the European Union's Horizon 2020, Future and Emerging Technologies (FET) programme (Grant No. 801267) and Danish Council for Independent Research Technology and Production Sciences for the DFF- Research Project 2 (Grant No. 48293). R.D. acknowledges funding from the German Science foundation within the SFB 917 "Nanoswitches", by the Helmholtz Association Initiative and Networking Fund under project number SO-092 (Advanced Computing Architectures, ACA) and the Federal Ministry of Education and Research (project NEUROTEC grant no. 16ES1133K). B.L.B. acknowledges funding from the European Union's Horizon 2020 (grants 824164, 871371, 871501, and 899559). D.I. acknowledges funding from the European Union's Horizon 2020 (grants 824164, 899559 and 101007321).

Nini Pryds, Professor, Technical University of Denmark
Dennis Valbjørn Christensen, Scientist, Technical University of Denmark
Bernabe Linares-Barranco, Professor, Sevilla Microelectronics Institute, CSIC
Daniele Ielmini, Professor, Politecnico di Milano
Regina Dittmann, Professor, Forschungszentrum Jülich

## 1.1 Phase-change memory devices

Abu Sebastian, IBM Research – Zurich, Switzerland
Manuel Le Gallo, IBM Research – Zurich, Switzerland
Andrea Redaelli, ST Microelectronics, Agrate, Italy

**Status**

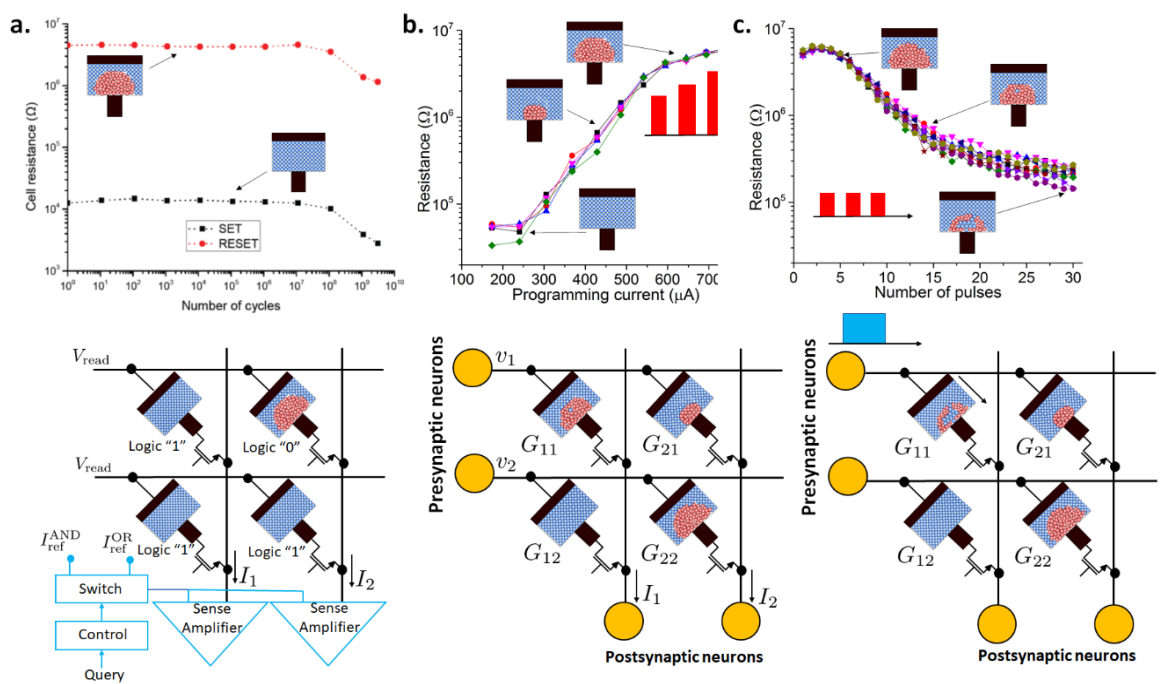

**Figure 1. Key physical attributes that enable neuromorphic computing. a**. Non-volatile binary storage facilitates in-memory logical operations relevant for applications such as hyper-dimensional computing. **b**. Analog storage enables efficient matrix-vector multiply operations that are key to applications such as deep neural network inference. **c**. The accumulative behaviour facilitates applications such as deep neural network training and emulation of neuronal and synaptic dynamics in spiking neural networks.

Phase-change memory (PCM) exploits the behaviour of certain phase-change materials, typically compounds of Ge, Sb and Te, that can be switched reversibly between amorphous and crystalline phases of different electrical resistivity [1]. A PCM device consists of a certain nanometric volume of such phase change material sandwiched between two electrodes.

In recent years, PCM devices are being explored for brain-inspired or neuromorphic computing mostly by exploiting the physical attributes of these devices to perform certain associated computational primitives in-place in the memory itself [2,3]. One of the key properties of PCM that enables such in-memory computing (IMC) is simply the ability to store two levels of resistance/conductance values in a non-volatile manner and to reversibly switch from one level to the other (binary storage capability). This property facilitates in-memory logical operations enabled through the interaction between the voltage and resistance state variables [3]. Applications of in-memory logic include database query [4] and hyper-dimensional computing [5].

Another key property of PCM that enables IMC is its ability to achieve not just two levels but a continuum of resistance values (analogue storage capability) [1]. This is typically achieved by creating intermediate phase configurations through the application of partial RESET pulses. The analogue storage capability facilitates the realization of matrix-vector multiply (MVM) operations in O(1) time complexity by exploiting Kirchhoff's circuit laws. The most prominent application for this is deep neural network (DNN) inference [6]. It is possible to map each synaptic layer of a DNN to a crossbar array of PCM devices. There is a widening industrial interest in this application owing to the promise of significantly improved latency and energy consumption with respect to existing solutions. This in-memory MVM operations also enable non-neuromorphic applications such as linear-solvers and compressed sensing recovery [3].

The third key property that enables IMC is the accumulative property arising from the crystallization kinetics. This property can be utilized to implement DNN training [7,8]. It is also the central property that is exploited for realizing local learning rules like spike-timing-dependent plasticity in spiking neural networks [9,10]. In both cases, the accumulative property is exploited to implement the synaptic weight update in an efficient manner. It has also been exploited to emulate neuronal dynamics [11].

Note that, PCM is at a very high maturity level of development with products already on the market and a well-established roadmap for scaling. This fact, together with the ease of embedding PCM on logic platforms (embedded PCM) [12] make this technology of unique interest for neuromorphic computing and IMC in general.

**Current and Future Challenges**

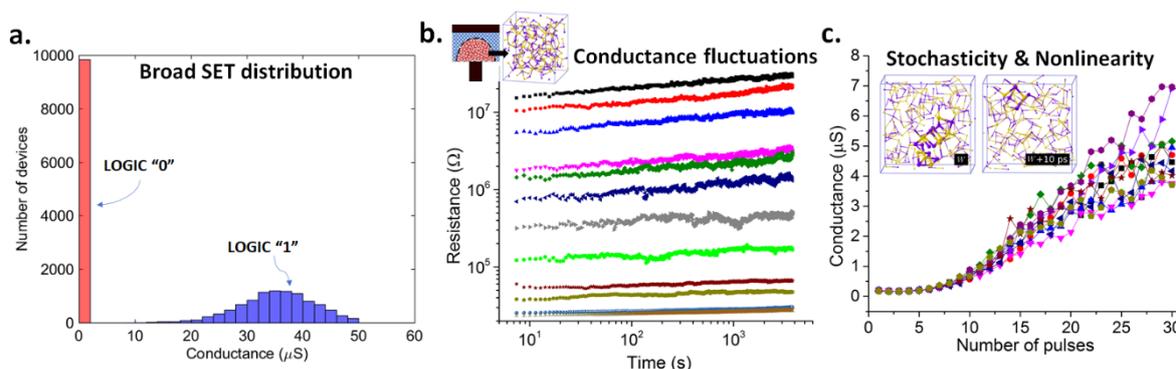

**Figure 2. Key challenges associated with PCM devices a.** The SET/RESET conductance values exhibit broad distributions which is detrimental for applications such as in-memory logic. **b.** The drift and noise associated with analogue conductance values results in imprecise matrix-vector multiply operations. **c.** The nonlinear and stochastic accumulative behaviour result in imprecise synaptic weight updates.

PCM devices have several attractive properties such as the ability to operate them at timescales on the order of tens of nanoseconds. The cycling endurance is orders of magnitude higher for PCM compared to other non-volatile memory devices such as Flash memory. The retention time can also be tuned relatively easily with the appropriate choice of materials, although the retention time associated with the intermediate phase configurations could be substantially lower than that of the full amorphous state.

However, there are also several device-level challenges as shown in Figure 2. One of the key challenges associated with the use of PCM for in-memory logic operations is the wide distribution of the SET states. These distributions could detrimentally impact the evaluation of logical operations. The central challenge associated with in-memory MVM operations is the limited precision arising from the 1/f noise as well as conductance drift. Drift is attributed to the structural relaxation of the melt-quenched amorphous phase [13]. Temperature-induced conductance variations could also pose challenges. One additional challenge is related to the stoichiometric stability during cycling where ion migration effects can occur [14]. Moreover, the accumulative behaviour in PCM is highly nonlinear and stochastic. While one could exploit this intrinsic stochasticity to realize stochastically firing neurons and for stochastic computing, this behaviour is detrimental for applications such as DNN training in which the conductance must be precisely modulated.

PCM-based IMC has the potential for ultra-high compute density since PCM devices can be scaled to nanoscale dimensions. However, it is not straightforward to fabricate such devices in a large array due to fabrication challenges such as etch damage and deposition of materials in high-aspect ratio pores [15]. The integration density is also limited by the access device, which could be a selector in the back-end-of-the-line (BEOL) or front-end bipolar junction transistors (BJT) or Metal-Oxide-Semiconductor Field Effect Transistors (MOSFET). The threshold voltage must be overcome when SET operations are performed, so the access device must be able to manage voltages at least as high as the threshold voltage. While MOSFET selector size is mainly determined by the PCM RESET current, the BJT and BEOL selectors can guarantee a minimum cell size of $4F^2$, leading to very high density [16]. However, BEOL selector-based arrays have some drawbacks in terms of precise current control, while the management of parasitic drops is more complex for BJT-based arrays [17].

**Advances in Science and Technology to Meet Challenges**

A promising solution towards addressing the PCM nonidealities such as 1/f noise and drift is that of projected phase-change memory (Projected PCM) [18, 19]. In these devices, there is a non-insulating projection segment in parallel to the phase-change material segment. By exploiting the highly non-linear I-V characteristics of phase-change materials, one could ensure that during the SET/RESET process, the projection segment has minor impact on the operation of the device. An increase in the reset current is anyway expected and some work should be done on material engineering side to compensate for that. However, during read, the device conductance is mostly determined by the projection segment that appears parallel to the amorphous phase-change segment. Recently, it was shown that it is possible to achieve remarkably high precision in-memory scalar multiplication (equivalent to 8-bit fixed point arithmetic) using projected PCM devices [20]. These projected PCM devices also facilitate array-level temperature compensation schemes. Alternate multi-layered PCM devices have also been proposed that exhibit substantially lower drift [21].

There is a perennial focus on trying to reduce the RESET current via scaling the switchable volume of the PCM device. Either by shrinking the overall dimension of the device in a confined geometry or by scaling the bottom electrode dimensions of a mushroom-type device. The exploration of new material classes such as single elemental Antimony could help with the scaling challenge [22].

The limited endurance and various other non-idealities associated with the accumulative behaviour such as limited dynamic range, nonlinearity and stochasticity can be partially circumvented with multi-PCM synaptic architectures. Recently, a multi-PCM synaptic architecture was proposed that employs an efficient counter-based arbitration scheme [23]. However, to improve the accumulation behaviour at the device level, more research is required on the effect of device geometries as well as the randomness associated with crystal growth.

Besides conventional electrical PCM devices, photonic memory devices based on phase-change materials, which can be written, erased, and accessed optically, are rapidly bridging a gap towards all-photonic chip-scale information processing. By integrating phase-change materials onto an integrated photonics chip, the analogue multiplication of an incoming optical signal by a scalar value encoded in the state of the phase change material was achieved [24]. It was also shown that by exploiting wavelength division multiplexing, it is possible to perform convolution operations in a single time step [25]. This creates opportunities to design phase-change materials that undergo faster phase transitions and have a higher optical contrast between the crystalline and amorphous phases [26].

**Concluding Remarks**

The non-volatile binary storage, analogue storage and accumulative behaviour associated with PCM devices can be exploited to perform in-memory computing. Compared to other non-volatile memory technologies, the key advantages of PCM are the well understood device physics, volumetric switching and easy embeddability in a CMOS platform. However, there are several device and fabrication-level challenges that need be overcome to enable PCM-based IMC and this is an active area of research.

It will also be rather interesting to see how PCM-based neuromorphic computing will eventually be commercialized. Prior to true IMC, a hybrid architecture where PCM memory chips are used to store synaptic weights in a non-volatile manner while the computing is performed in a stacked logic chip is likely to be considered as an option by the industry. Despite the tight interconnect between the stacked chips, data transfer will remain a bottleneck for this approach. A better solution could be PCM directly embedded with the logic itself (BEOL) without any interconnect bottleneck and eventually we could foresee full-fledged non-von Neumann accelerator chips where the embedded PCM is also used for analogue in-memory computing.

**Acknowledgements**

This work was supported in part by the European Research Council through the European Union's Horizon 2020 Research and Innovation Programme under grant no. 682675.

## 1.2 – Ferroelectric Devices

Dr.-Ing. Stefan Slesazeck[1], Prof. Dr.-Ing. Thomas Mikolajick[1,2]
[1] NaMLab gGmbH; [2] Institute of Semiconductors and Microsystems, TU Dresden; Dresden, Germany

**Status**

Ferroelectricity was firstly discovered in 1920 by Valasek in Rochelle salt [1] and describes the ability of a non-centrosymmetric crystalline material to exhibit a permanent and switchable electrical polarization due to the formation of stable electric dipoles. Historically, the term ferroelectricity stems from the analogous behavior with the magnetization hysteresis of ferromagnets when plotting the ferroelectric polarization versus the electrical field. Regions of opposing polarization are called domains. The polarization direction of such domains can be switched typically by 180° but based on the crystal structure also other angles are possible. Since the discovery of the stable ferroelectric barium titanate (BTO) in 1943 ferroelectrics found application in capacitors in electronics industry. Already in the 1950s ferroelectric capacitor (FeCAP) based memories (FeRAM) have been proposed [2], where the information is stored as polarization state of the ferroelectric material. Read and write operation are performed by applying an electric field larger than the coercive field $E_C$. The destructive read operation determines the switching current of the FeCAP upon polarization reversal, thus requiring a write-back operation after readout. Thanks to the development of mature processing techniques for ferroelectric lead zirconium tantalate (PZT) FeRAMs are commercially available since the early 1990s [3]. However, the need for a sufficiently large capacitor together with the limited thin-film manufacturability of the perovskite materials so far restricted their use to niche applications [4].

The ferroelectric field effect transistors (FeFET) that was proposed in 1957 [5] features a ferroelectric capacitor as gate insulator, modulating the transistor's threshold voltage that can be sensed non-destructively by measuring the drain-source current. Perovskite based FeFET memory arrays with up to 64kBit have been demonstrated [6]. But due to difficulties in the technological implementation, limited scalability and data retention issues, no commercial devices became available.

The ferroelectric tunneling junction (FTJ) was proposed by L. Esaki et al. in 1970s as a "polar switch" [7] and was firstly demonstrated in 2009 using a BaTiO3 ferroelectric layer [8]. The FTJ features a ferroelectric layer sandwiched between two electrodes, thus modifying the tunneling electro-resistance. A polarization-dependent current is measured non-destructively when applying electrical fields smaller than $E_C$.

Since the fortuitous discovery of ferroelectricity in hafnium oxide ($HfO_2$) in 2008 and its first publication in 2011 [9] the well-established and CMOS-compatible fluorite-structure material has been extensively studied and recently gained a lot of interest in the field of nonvolatile memories and beyond von-Neumann computing [10] [11].

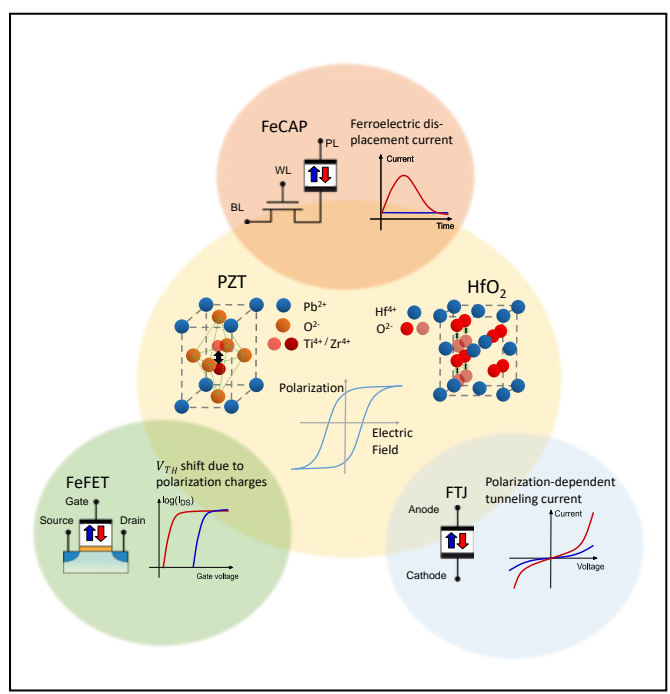

**Figure 1.** The center shows two typical ferroelectric crystals and the corresponding PV-hysteresis curve. The top figure illustrates a FeCAP based FeRAM, the figure on the bottom left shows a FeFET and the bottom right an FTJ.

## Current and Future Challenges

Very encouraging electrical results of fully front-end-of-line (FEOL) integrated FeFET devices featuring switching speeds <50ns at <5V pulse voltage have been reported recently based on >1Mbit memory arrays [12]. The ability of fine-grained co-integration of FeFET memory devices together with CMOS logic transistors paves the way for the realization of brain-inspired architectures to overcome the limitations of the van-Neumann bottleneck, which restricts the data transfer due to limited memory and data bus bandwidth [13]. However, one of the main challenges for the FeFET devices and topic of intense research is the formation of ferroelectric $HfO_2$-based thin films featuring a uniform polarization behavior at nano-scale as an important prerequisite for the realization of small scaled devices with feature sizes <100nm.

Another important challenge for many application cases is the limited cycling endurance of silicon-based FeFETs that is typically in the range of $10^5$ cycles. This value is mainly dictated by the breakdown of the dielectric $SiO_2$ interfacial layer that forms between the Si channel and the ferroelectric gate insulator.

Ferroelectric capacitors have been successfully integrated into the back-end-of-line (BEOL) of modern CMOS technologies and operation of a $HfO_2$-based based FeRAM memory array at 2.5V and 14ns switching pulses was successfully demonstrated [14]. At this point the main challenge is the decrease of the ferroelectric layer thickness well below 10nm to allow scaling of 3D capacitors towards the 10nm node. Moreover, phenomenon such as the so called "wake-up effect" with increasing of $P_r$ for low cycle counts as well as the "fatigue effect" resulting in a reduction of $P_r$ at high cycle counts due to oxygen vacancy redistribution [15] and defect generation have to be tackled. That is especially important for fine-grained circuit implementations where the switching properties of single ferroelectric devices impact the designed operation point of analogue circuits.

One of the most interesting benefits of FTJ devices is the small current density making them very attractive for applications requiring massive parallel operations such as analogue matrix-vector-multiplications in larger cross-bar structures [16]. However, increasing the ratio between the on-current density and the self-capacitance of the FTJ devices turns out as one of the main challenges to increase the reading speed for these devices. The tunneling current densities depend strongly on the thickness of the ferroelectric layer and the composition of the multi-layer stacks. The formation of very thin ferroelectric layers is hindered by unintentional formation of interfacial dead layers towards the electrodes and increasing leakage currents due to defects and grain-boundaries in the poly-crystalline thin films.

**Advances in Science and Technology to Meet Challenges**

Although ferroelectricity in hafnium oxide has been extensively studied for over one decade now, there are still many open questions in understanding the formation of the ferroelectric *Pca2$_1$* phase and regarding the interaction with material layers such as electrodes, dielectric tunneling barriers in multi-layer FTJs or interfacial layers in FeFETs. Moreover, the interplay between charge trapping phenomenon and ferroelectric switching mechanisms [17], the trade-off between switching speed and voltage of the nucleation limited switching and its impact on device reliability or the different behavior of abrupt single domain switching [11] and smooth polarization transitions in negative capacitance devices that were observed in the very similar material stacks are still not completely understood. However, that knowledge will be an important ingredient for proper optimization of material stacks as well as electrical device operation conditions.

On the materials side the stabilization of the ferroelectric orthorhombic *Pca*2$_1$ phase in crystallized HfO$_2$ thin films has to be optimized further. Adding dopants, changing oxygen vacancy densities or inducing stress by suitable material stack and electrode engineering are typical measures. In most cases a poly-crystalline material layer is attained consisting of a mixture of different crystalline ferroelectric and non-ferroelectric phase fractions. Moreover, ferroelectric grains that differ in size or orientation of the polarization axis, electronically active defects as well as grain size dependent surface energy effects give rise to the formation of ferroelectric domains that possess different electrical properties in terms of coercive field $E_C$ (typical values ~1 MV/cm) or remnant polarization $P_r$ (typical values 10 – 40 µC/cm$^2$) with impact on the device-to-device variability and the gradual switching properties that are important especially for analog synaptic devices. Some drawbacks of the poly-crystallinity of ferroelectric HfO$_2$- and ZrO$_2$-based thin films could be tackled by the development of epitaxial growth of monocrystalline ferroelectric layers [18] where domains might extend over a larger area. Especially in the case of FTJs the effect of domain wall motion might allow a more gradual and analogue switching behavior even in small scaled devices. The utilization of an anti-ferroelectric hysteretic switching that was demonstrated in ZrO$_2$ thin films bears the potential to overcome some limitations that are related to the high coercive field of ferroelectric HfO$_2$, such as operation voltages being larger than the typical core voltages in modern CMOS technologies or the limited cycling endurance [19].

Finally, besides the very encouraging results adopting ferroelectric HfO$_2$ in 2019 another promising material was realized. The AlScN is a semiconductor processing compatible and already utilized piezoelectric material that was made ferroelectric [20].

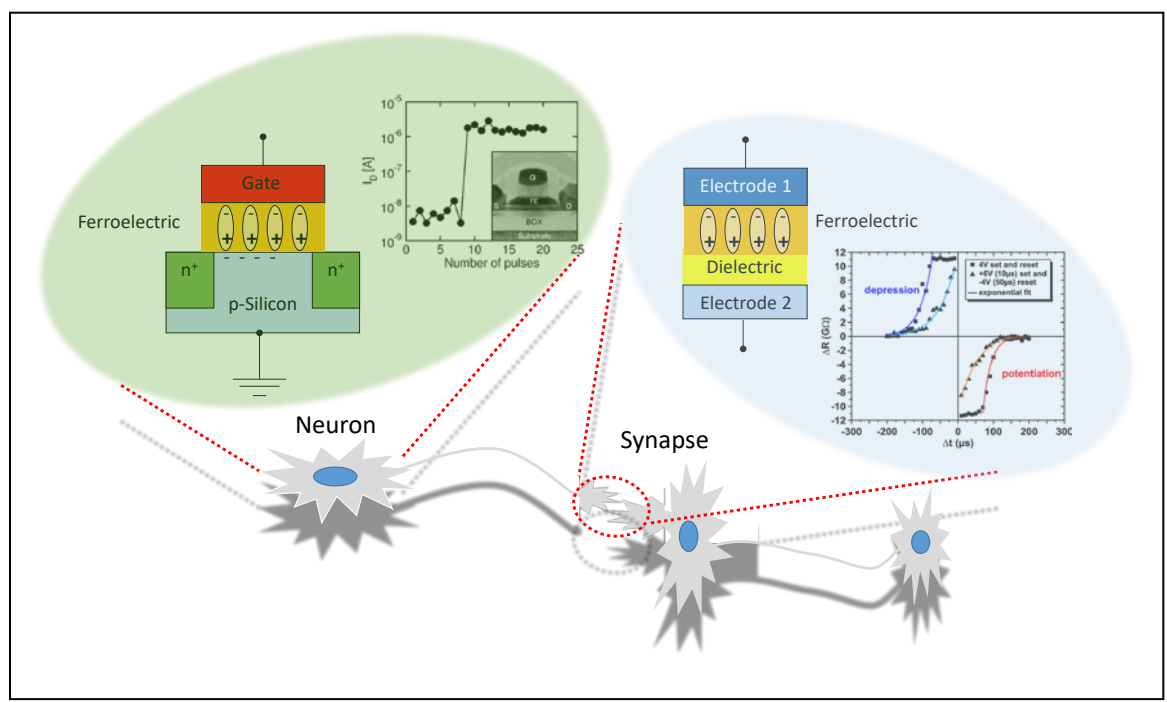

Figure 2. Main elements of a neural network. Neurons can be realized using scaled down FeFETs [11] while synapses can be realized using FTJs [10] or medium to large scale FeFETS. Adapted with permission from [10] Copyright (2020) American Chemical Society and [11] Copyright (2018) The Royal Society of Chemistry.

**Concluding Remarks**

The discovery of ferroelectricity in hafnium oxide has led to a resumption in the research on ferroelectric memory devices, since hafnium oxide is a well-established and fully CMOS compatible material in both front end of line and back end of line processing. Besides the expected prospective realization of densely integrated non-volatile and ultra-low-power ferroelectric memories in near future, this development directly leads to the adoption of the trinity of ferroelectric memory devices – FeCAP, FeFET and FTJ - for beyond von Neumann computing. While in the memory application the important topic of reliability on the array level is yet to be solved, for neuromorphic applications the linear switching to many different states, especially in scaled down devices, is a topic that needs further attention. Moreover, very specific properties of the different ferroelectric device types demand for the development of new circuit architectures that facilitate a proper device operation taking into account the existing non-idealities. A thorough design technology co-optimization will be the key to fully exploit their potential in neuromorphic and edge computing. Finally, large scale demonstrations of ferroelectrics based neuromorphic circuits need to be investigated to identify all possible issues.

**Acknowledgements**

This work was financially supported out of the State budget approved by the delegates of the Saxon State Parliament.

## 1.3 Valence change memory

Sabina Spiga, CNR-IMM, Unit of Agrate Brianza, via C. Olivetti 2, Agrate Brianza (MB), Italy
Stephan Menzel, FZ Juelich (PGI-7), Juelich , Germany

**Status**

Resistive random access memories (RRAMs), also named memristive devices, change their resistance state upon electrical stimuli. They can store and compute information at the same time, thus enabling in-memory and brain-inspired computing [1, 2]. RRAM devices relying on oxygen ion migration effects and subsequent valence changes are named valence change memory (VCM) [3]. They have been proposed to implement electronic synapses in hardware neural networks, due to the ability to adapt their strength (conductance) in an analogue fashion as a function of incoming electrical pulses (synaptic plasticity), leading to long-term (short-term) potentiation and depression. In addition, learning rules such as spike-time or spike-rate dependent plasticity, paired-pulse facilitation or the voltage threshold–based plasticity have been demonstrated; the stochasticity of the switching process has been exploited for stochastic update rules [4-6]. Most of the VCM devices are based on a two-terminal configuration, and the switching geometry involves either confined filamentary, or interfacial regions (Fig.1A). Filamentary VCMs are today the most advanced in terms of integration and scaling. Their switching mechanism relies on the creation and rupture of conductive filaments (CF), formed by a localized concentration of defects, shorting the two electrodes. The modulation/control of the CF diameter and/or CF dissolution can lead to two or multiple stable resistance states [7, 8]. Prototypes of neuromorphic chips have been recently shown, integrating $HfO_x$ and $TaO_x$-based filamentary-VCM as synaptic nodes in combination with CMOS neurons [9-11]. In interfacial VCM devices, the conductance scales with the junction area of the device, and the mechanism is related to a homogenous oxygen ion movement through the oxides, either at the electrode/oxide or oxide/oxide interface. Reference material systems are based on complex oxides, such as bismuth ferrite [12] and praseodymium calcium manganite [13]; or bilayers stacks, e.g. $TiO_2/TaO_2$ [14] and $a-Si/TiO_2$ [15]. Finally, 3-terminal VCM redox transistors have been recently studied (Fig.1A-right), where the switching mechanism is related to the control of the oxygen vacancy concentration in the bulk of the transistor channel [16, 17]. While interfacial and redox-transistor devices are today at low technological readiness, and most of the studies are reported at single device level, they promise future advancement in neuromorphic computing in terms of analogue control, higher resistance values, improved reliability, reduced stochasticity with respect to filamentary devices [18]. To design neuromorphic circuits including VCM devices, compact models are requested. For filamentary devices compact models including variability are available [18, 19], but lacking for interfacial VCM and redox-based transistors.

**Current and Future Challenges**

VCM devices have been developed in the last 15 years mainly for storage applications, but for neuromorphic applications the required properties differ. In general, desirable properties of memories for neural networks include (i) analogue behaviour or controllable multilevel states, (ii) compatibility with learning rules supporting also online learning, (iii) tuneable short-term and long-term stability of the weights to implement various dynamics and timescales in synaptic and neuronal circuits [4-6]. A significant debate still refers to the linear/non-linear and symmetric/asymmetric conductance update of experimental devices, synaptic resolution (number of resistance levels), and how to exploit or mitigate these features (Figs1-B,C).

**Filamentary** devices are the most mature type of VCMs. Nevertheless, many issues are pending: e.g. control of multi-level operation, device variability, intrinsic stochasticity, program and read disturbs, and the still too low resistance level range for neuromorphic circuits [20]. Moreover, the understanding/modelling of their switching mechanism is still under debate. Whereas first models including switching variability and read noise are available [18, 19], retention modelling, and the modelling of volatile effects and device failures are current challenges. First hybrid CMOS-VCM chips have been developed demonstrating inference application, but so far they do not support on-chip learning [9-11].

**Interfacial** VCM devices show in general less variability, less (no) read instability and a very analogue tuning of the conductance states, which can leads to a more deterministic and linear conductance update compared to filamentary devices [13]. Still these properties are not characterized on a high statistical basis. The retention, especially for thin oxide devices, is lower than for filamentary devices, which may be still compatible with some applications. As the conductance scales with area, the achievable high resistance levels promise a low power operation. Typical devices, however, have a large area or thick switching oxides, and scaling them to the nanoscale is an open issue. Moreover, devices showing a large resistance modulation require high switching voltages, not easily compatible with scaled CMOS nodes. The fabrication and characterization of interfacial VCM arrays needs to be further addressed. Simulation models for interfacial VCM are not available yet and need to be developed.

**Redox-based VCM transistors** have been only shown on a single device level [16, 17]. Thus, reliable statistical data on cycle-to-cycle variability, device-to-device variability and stability of the programmed states is not available yet. Moreover, the trade-off between switching speed and voltage has not been studied in detail. Another challenge is the understanding of the switching mechanism and the development of suitable models for circuit design.

The open challenges for all three types of VCM devices are summarized in Table I.

**Advances in Science and Technology to Meet Challenges**

The current challenges for VCM-type devices push the research in various but connected directions, which span from material, to theory, devices and architecture. A better understanding of material properties and microscopic switching mechanisms is definitely required. However, the key step is to demonstrate the device integration in complex circuits and hybrid CMOS-VCM hardware neuromorphic chips. While VCMs are not ideal devices, many issues can be solved or mitigated at circuit level still taking advantage of their properties in term of power, density, and dynamic properties.

In this context, **filamentary** VCM devices are the most mature technology, but their deployment into neuromorphic computing hardware is still at its infancy. A comprehensive compact model, depicting complete dynamics including retention effects, e.g. to accurately simulate online learning, is required for the development of optimized circuits. On the material level, the biggest issues are read noise and switching variability. Due to the inherent Joule heating effect, the transition time of the conductance switching is very short and depends strongly on the device state [21]. This makes it hard to control the conductance update. Future research could explore very fast pulses in the range of the transition time to update the cell conductance, or use thermal engineering of the device stacks to increase the transition time. Finally, to achieve low power operation, resistance state values should be moved to the MΩ regime.

**For interfacial** and **redox-transistor** VCM devices, one of the next important steps is to shift from single device research to large arrays, possibly co-integrated with CMOS. This step enables to collect a large amount of data, which is required for modelling and demonstrating robust neuromorphic functions. It would be highly desirable to identify a reference material system with a robust switching mechanism supported by a comprehensive understanding and modelling from underlying physics to compact and circuits modelling. Indeed, the modelling of these devices are still at its infancy. One open question for both devices is the trade-off between data retention and switching speed. In contrast to the filamentary devices, the velocity of the ions are probably not accelerated by Joule heating. Thus, the voltage needs to be increased more than in filamentary devices, to operate the devices at fast speed [22]. This might limit the application of these device to a certain time domain as the CMOS might not be able to provide the required voltage. By using thinner device layers or material engineering this issue could be addressed.

**Concluding Remarks**
The VCM device technologies can integrate novel functionalities in hardware as key elements of the synaptic nodes in neural networks, i.e. to store the synaptic weight. Moreover, they can enable new learning algorithms that enable bio-plausible functions over multiple timescales. At the moment, it is still not clear which can be the best "final" VCM material system and/or VCM device type, having each of them advantages and disadvantages. The missing "killer" system, with consolidated properties/understanding/easy manufacturing, prevents to concentrate the efforts of the scientific community in single direction to bring VCM device to industrial real applications beyond a niche market. While filamentary VCMs are already been implemented in neuromorphic computing hardware, interfacial VCM or redox transistor can open new perspectives in the long term. To this end, there is an urgent request to further develop VCM devices enhancing new properties through a combined synergetic development based on materials design, physical and electrical characterizations and multiscale modelling to support the microscopic understanding of the link between the device structure and the electrical characteristics. Moreover, the device development targeting brain-inspired computing systems can only go hand-in-hand with theory and architectures design in a holistic view.

**Acknowledgements**
*This work was partially supported by the Horizon 2020 European projects MeM-Scales (Grant No. 871371), MNEMOSENE (Grant No.780215), and NEUROTECH (Grant No. 824103); in part by the Deutsche Forschungsgemeinschaft (SFB 917); in part by the Helmholtz Association Initiative and Networking Fund under project number SO-092 (Advanced Computing Architectures, ACA) and in part by the Federal Ministry of Education and Research (BMBF, Germany) in the project NEUROTEC (project numbers 16ES1134 and 16ES1133K).*

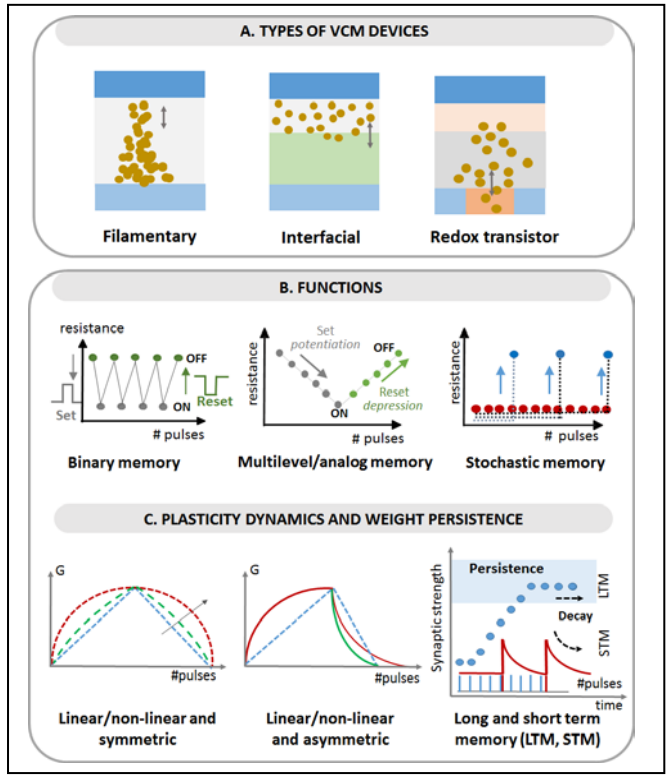

**Figure 1.** A. Sketch of the three types of VCM devices (filamentary, interfacial and redox transistor). B. Possible functionalities that can be implemented by VCM devices, namely binary memory (left), analog/multilevel (centre) and stochastic (right) memory. In the figures, the device resistance evolution is plotted as a function of applied electrical stimuli (pulses). C. Schematic drawing of some of the interesting properties of VCM for neuromorphic applications, i.e. synaptic plasticity dynamics and type of memory with different long or short retention scales (LTM, STM) . Many experimental VCM devices show a non-linear and asymmetric modulation of the conductance (G) update, but plasticity dynamics can be as well modulated by programming strategies or materials engineering.

| | Filamentary | Interfacial | Redox Transistor |
|---|---|---|---|
| Single Device | Various materials systems and scaling down to nm scale, low V operation | Various materials, scaling at nm scales, still overall high V and thick oxide | Recently studied, promising but still at initial development stage |
| Array | Demonstrated as 1T-1R and 1S-1R. Total capacity up to 1-2 Mb in recent works. | Published in few works, up to few Kb, mainly 1R array only | To be addressed |
| Monolithic CMOS Integration | Demonstrated, down to the 2x techn. nodes, and in 3D architectures | Partial demonstration, mainly for oxide bilayer and not for perovskites | To be addressed |
| Compact Models | Available, retention and variability modelling to be optimized | Almost not available | Almost not available |
| Binary Memory | Available, with good endurance (>$10^6$-$10^9$) and retention (> years) | Possible, but lack of statistical data. Endurance and long retention to be optimized | Very new devices, mostly proposed for multilevel applications, lack of statistical data |
| Analog Multilevel Memory | Hard to control, multilevel demonstrated in array using program/verify algorithms | Promising, less variability, high R, but reduced dynamic range and mainly data for single devices | Promising, high R value, deterministic control of G update, but shown for single devices |
| Stochastic Memory | Proposed in some works, to be further validated | - | - |
| Long term memory (LTM) | Yes, retention at high T and for 6-10 years. Depends on R levels. | Possible, lack of statistic data on array, single device retention up to years for some material stacks | Possible. Few studies. Lack of statistical data. |
| Short term memory (STM) | Usually difficult to achieve controlled decay | Possible, to be further addressed and optimized | Possible, to be further addressed. Few studies |

Table I. Summary of status and open challenges of the three types of VCM devices.

## 1.4 Electrochemical metallization cells
Ilia Valov, Research Centre Juelich

**Status**

Electrochemical metallization memories were introduced in nanoelectronics with perspective to be used as memory, optical, programmable resistor/capacitor devices, sensors and as well for crossbar arrays and rudimentary neuromorphic circuits by M. Kozicki[1, 2] under the name programmable metallization cells (PMC). These type devices are termed also conductive bridging random access memories (CBRAM) or atomic switches[3]. The principle of operation of these two electrode devices using thin layers as ion transporting media is schematically shown in Figure 1. As electrochemically active electrodes Ag, Cu, Fe or Ni are mostly used and as counter electrodes Pt, Ru, Pd, TiN or W are preferred. Electrochemical reactions at the electrodes and ionic transport within the device trigged by internal[4] or applied voltage the formation of metallic filament (bridge) short-circuiting the electrodes and defining low resistance state (LRS). Voltage of opposite polarity is used to dissolve the filament, returning the resistance to high ohmic state (HRS). LRS and HRS are used to define Boolean 1 and 0, respectively.

Apart from prospective for a paradigm shift in computing and information technology offered by memrsitive devices in general[5], ECMs provide particular advantages compared to other redox-based resistive memories. They operate at low voltages (~ 0.2 V to ~ 1 V) and currents (from nA to µA range) allowing for low power consumption. Huge spectrum of materials can be used as solid electrolytes, ionic conductors, mixed conductors, semiconductors, macroscopic insulators and even high-k materials such as $SiO_2$, $HfO_2$, $Ta_2O_5$ etc. predominantly in amorphous but also in crystalline state[6]. The spectrum of these materials includes also 1D and 2D materials but also different polymers, bio-inspired / bio-compatible materials, proteins and other organic and composite materials[7, 8]. The metallic filament can vary in thickness and may either completely bridge the device, or be only partially dissolved providing multilevel to analog behaviour. Very thin filaments are extremely unstable and dissolve fast (down to $10^{10}$ sec)[9]. The devices are stable against radiation/cosmic rays, high energy particles and electromagnetic waves and can operate in large temperature range[10, 11]. Due to these properties, ECMs can be implemented to various environments, systems and technologies. The typical applications are as selector devices, volatile, non-volatile digital and analog memories, transparent and flexible devices, sensors, artificial neurons and synapses[12-14]. The devices can combine more functions and are thought as basic units for the fields of autonomous systems, beyond von Neumann computing and artificial intelligence. Further development in the field is essential to realise the full potential of this technology.

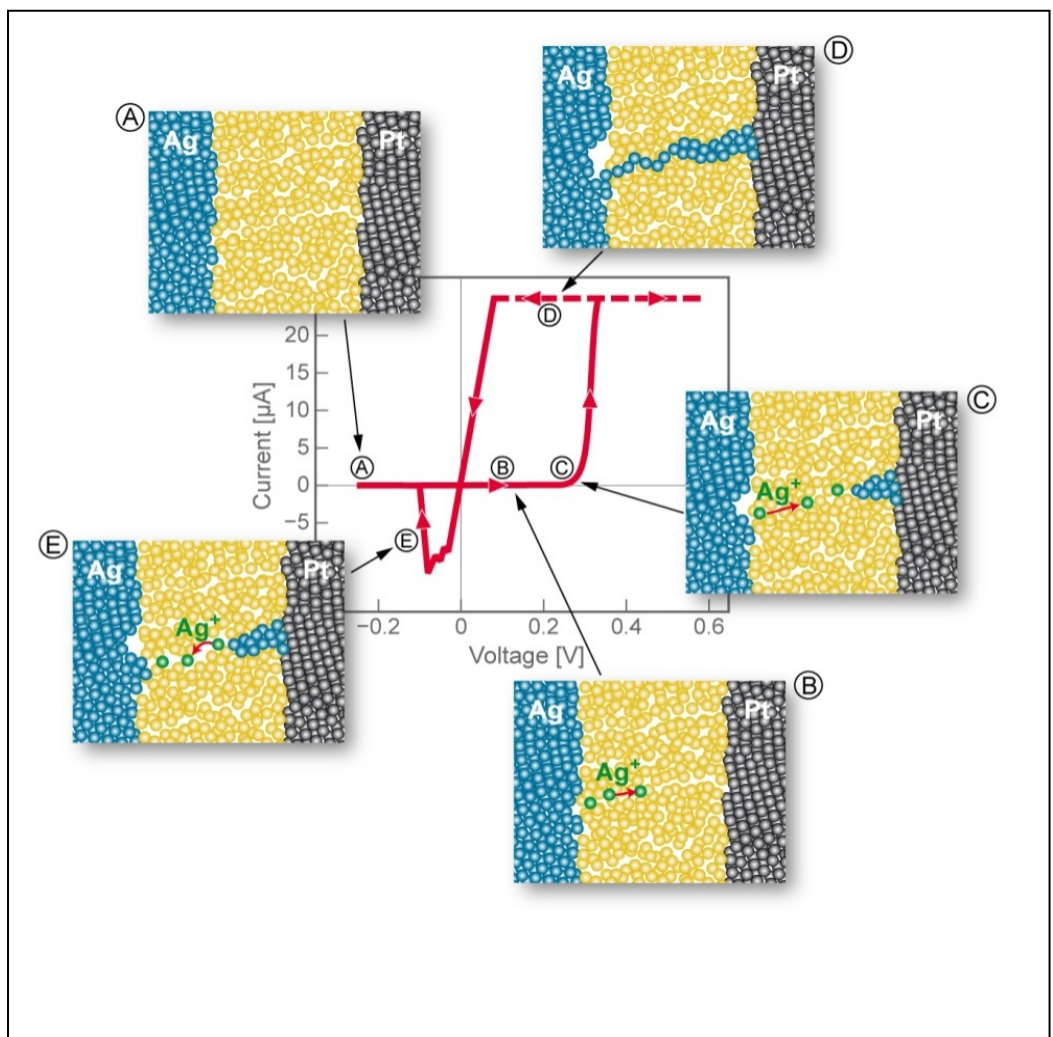

**Figure 1.** Principle operation and current-voltage characteristics of electrochemical metallization devices. The individual physical processes are related to the corresponding part of the I-V dependence. The figure is reproduced from[15]

**Current and Future Challenges**

Despite the apparent simplicity and ease operation ECM cells are complex nanoscale systems, relying on redox reactions and ion transport at extreme conditions[16]. Despite low absolute voltages and currents, the devices are exposed to electric fields of up to $10^8$ V cm$^{-1}$ and current densities of up to ~ $10^{10}$ A cm$^{-2}$. There is no other example in the entire field of electrochemical applications even approaching these conditions. Small device volume, harsh and strongly non-equilibrium conditions is making the understanding of fundamental processes and their control extremely challenging. The latter results in less precise (or missing) control over the functionalities and reliable operation. Indeed, maybe the most serious disadvantage of ECMs is the large variability in switching voltages, currents and resistive states. Additional problems are fluctuations and drift of the resistance states, as well their chemically and/or physically determined instabilities.

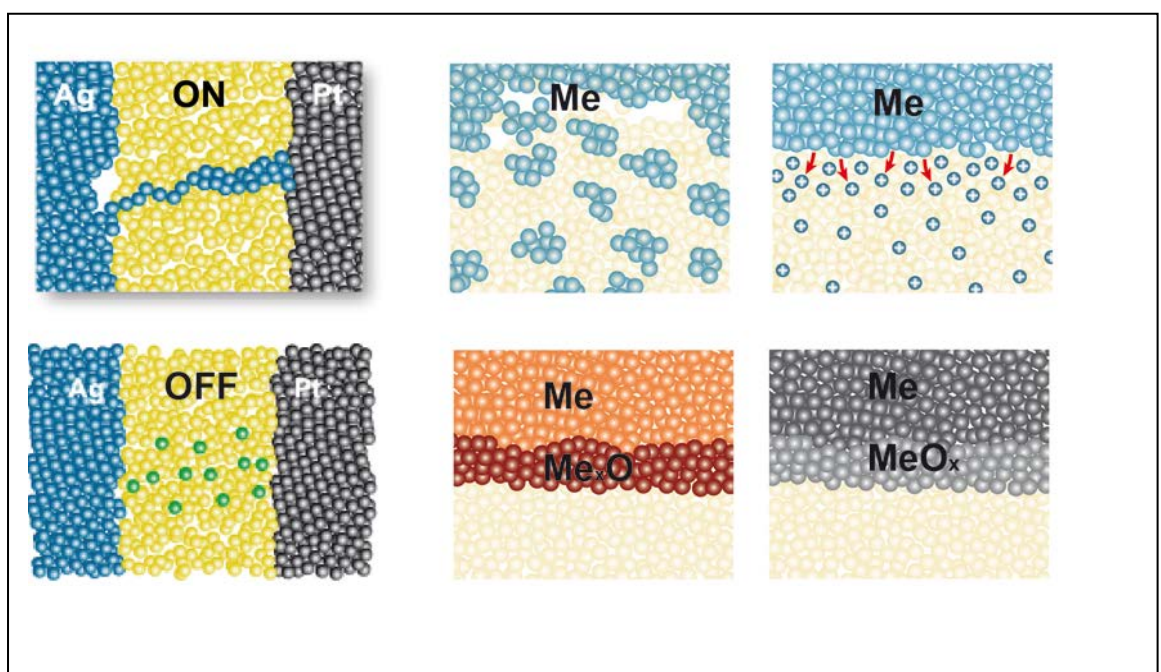

Figure 2. Schematic differences between ideal cells (left) and real cells accounting for interface interactions occurring due to sputtering conditions, chemical interactions or environmental influences. Physical instabilities/dissolution of the electrode, leading to clustering and formation of conductive oxides in ECM devices (middle). Chemical dissolution of the electrode and formation of insulating oxides (right) The figure is modified from[17].

Several notable issues should be taken in consideration: 1) Missing unequivocal experimental value about what part of the applied current is carried by ions and by electrons. Whereas in macroscopic systems these numbers are constant, in nanoscale ECMs it may vary depending on the conditions and charge concentration. 2) The charge/ion concentration may vary with time. Due to the small volume, it is easy to enrich or deplete the film with mobile ions (acting as donors/acceptors) during the operation cycles, resulting in deviation of the switching voltages and currents and finally to failures. 3) Again due to small volume, even low number of foreign atoms/ions (impurities) will cause considerable changes in the electronic properties. Impurities or dopants and as well the matrix significantly alter the characteristics due to effects on the switching layer[18, 19] or on the electrodes[20] . 4) Effects of protons and oxygen. Both can be incorporated either during device preparation (e.g. lithography steps, or deposition technique e.g. ALD etc. ) or from the environment[21], even if capping layer is used. Many devices even cannot operate without presence of protons and many electrode materials such as Cu, Mo, W or TiN etc. can be partially or even are fully oxidized by environmental factors. 5) Interfacial interactions are commonly occurring at the electrode/solid electrolyte interface. The thickness of these interfacial layers can sometime even exceed the thickness of the switching layer and inhibit or support reliable operation[17].

All these effects have their origin in the nanosize of the devices and highly non-equilibrium operating conditions.

**Advances in Science and Technology to Meet Challenges**

Addressing the challenges and issues that still limit the implementation of ECM devices in the praxis, should be considered on different levels. On a fundamental level, an in-depth understanding of the nanoscale processes and rate-limiting steps that determine the resistive switching mechanism is essential. To overcome the current limitations the theory should be further improved to account not only quantitatively but also qualitatively for the fundamental differences in thermodynamics and kinetics on the nanoscale compared to the macroscale. The scientific equipment needs to be improved to address the demand on sufficient mass and charge sensitivity and as well lateral and vertical resolution. Accent should be set on *in situ* and *in operando* techniques at real conditions enhanced by high time and imaging resolutions.

On a materials level, efforts should be made to understand and effectively use the relation between physical and chemical material properties, such as chemical composition, non-stoichiometry, purity, doping, density, thickness and mechanical properties and device performance and functionalities. A more narrow selection from the vast sea of ECM materials should be made on which systematic research should be performed. Final task to be achieved by these selective materials research approach is establishing a universal materials treasure map.

On a device/circuit/technology level, common problems such as sneak path problem still need to be addressed. Limitation of interactions between devices and high-density integration (also within CMOS) needs to be further improved. The control during the deposition of layer materials should be adjusted to avoid layer intermixing, contaminations and incorporation of impurities. In many cases, deposition of thin films of non-oxidized elements or components with higher affinity to oxygen such as W, Mo, TiN or oxygen-free containing chalcogenides is possible only after special pre-care. The technological processes must be adapted and regularly controlled to ensure high quality and defined chemical composition. Additional efforts should also be made to integrate devices utilizing different functionalities and allowing for higher degree of complexity. The internal electromotive force should be further explored and utilized in respect autonomous systems and as well applications in space technologies and medicine should be further developed.

These issues are in fact highly interrelated and closely depend on each other. Most important on the current stage of development of ECM devices is to understand and control the relation between material properties, physical processes and device performance and functionalities. This knowledge will result in improved reliability of the devices and advanced technology.

**Concluding Remarks**

ECM devices have been intensively developed in the last 20 years however, still not reaching their full potential. Opportunities for various applications in the fields of nanoelectronics, nanoionics, magnetics, optics, sensorics etc. and prospective for implementation as basic units in neuromorphic computing, big data processing, autonomous systems and artificial intelligence are impeded by insufficient control of the nanoscale processes and incomplete knowledge on the relation between material properties, fundamental processes and devices characteristics and functionalities. To achieve these tasks, not only existing theory but also the scientific equipment and characterization techniques should be further improved allowing a direct insight in the complex nanoscale phenomena. Interacting and complementing fundamental and applied research is the key to address these issues in order to deploy the advantages and opportunities offered by the electrochemical metallization cells into modern information and communications technologies.

## 1.5 Nanowire Networks

Gianluca Milano[1], Carlo Ricciardi[2]

[1]Advanced Materials Metrology and Life Sciences Division, INRiM (Istituto Nazionale di Ricerca Metrologica), Torino, Italy.

[2]Department of Applied Science and Technology, Politecnico di Torino, Torino, Italy.

**Status**

The human brain is a complex network of about $10^{11}$ neurons connected by $10^{14}$ synapses, anatomically organized over multiple scales of space, and functionally interacting over multiple scales of time [1]. Synaptic plasticity, i.e. the ability of synaptic connections to strengthen or weaken over time depending on external stimulation, is at the root of information processing and memory capabilities of neuronal circuits. As building blocks for the realization of artificial neurons and synapses, memristive devices organized in large crossbar arrays with a top-down approach have been recently proposed [2]. Despite the state-of-art of this rapidly growing technology demonstrated hardware implementation of supervised and unsupervised learning paradigms in artificial neural networks (ANN), the rigid top-down and grid-like architecture of crossbar arrays fails in emulating the topology, connectivity and adaptability of biological neural networks, where the principle of self-organization governs both structure and functions [1]. Inspired by biological systems (Figure 1a), more biologically plausible nanoarchitectures based on self-organized memristive nanowire (NW) networks have been proposed [3]–[8] (Figure 1b and c). Here, the main goal is to focus on the emergent behaviour of the system arising from complexity rather than on learning schemes that require addressing of single ANN elements. Indeed, in this case main players are not individual nano objects but their interactions [9]. In this framework, the cross-talk in between individual devices, that represents an unwanted source of sneak currents in conventional crossbar architectures, here represents an essential component for the network emerging behaviour needed for the implementation of unconventional computing paradigms. NW networks can be fabricated by randomly dispersing NWs with a metallic core and an insulating shell layer on a substrate by a low-cost drop casting technique that does not require nanolithography or cleanroom facilities. The obtained NW network topology shows small-world architecture similarly to biological systems [10]. Both single NW junctions and single NWs show memristive behaviour due to the formation/rupture of a metallic filament across the insulating shell layer and to breakdown events followed by electromigration effects in the formed nanogap, respectively (Figure 1e-h) [7]. Emerging network-wide memristive dynamics were observed to arise from the mutual electrochemical interaction in between NWs, where the information is encoded in "winner-takes-all" conductivity pathways that depend on the spatial location and temporal sequence of stimulation [11]–[13]. By exploiting these dynamics, NW networks in multiterminal configuration can exhibit homosynaptic, heterosynaptic and structural plasticity with spatiotemporal processing of input signals [7]. Also, nanonetworks have been reported to exhibit fingerprints of self-organized criticality similarly to our brain [3], [14], [15], a feature that is considered responsible for optimization of information transfer and processing in biological circuits. Because of both topological structure and functionalities, NW networks are considered as very promising platforms for hardware realization of biologically plausible intelligent systems.

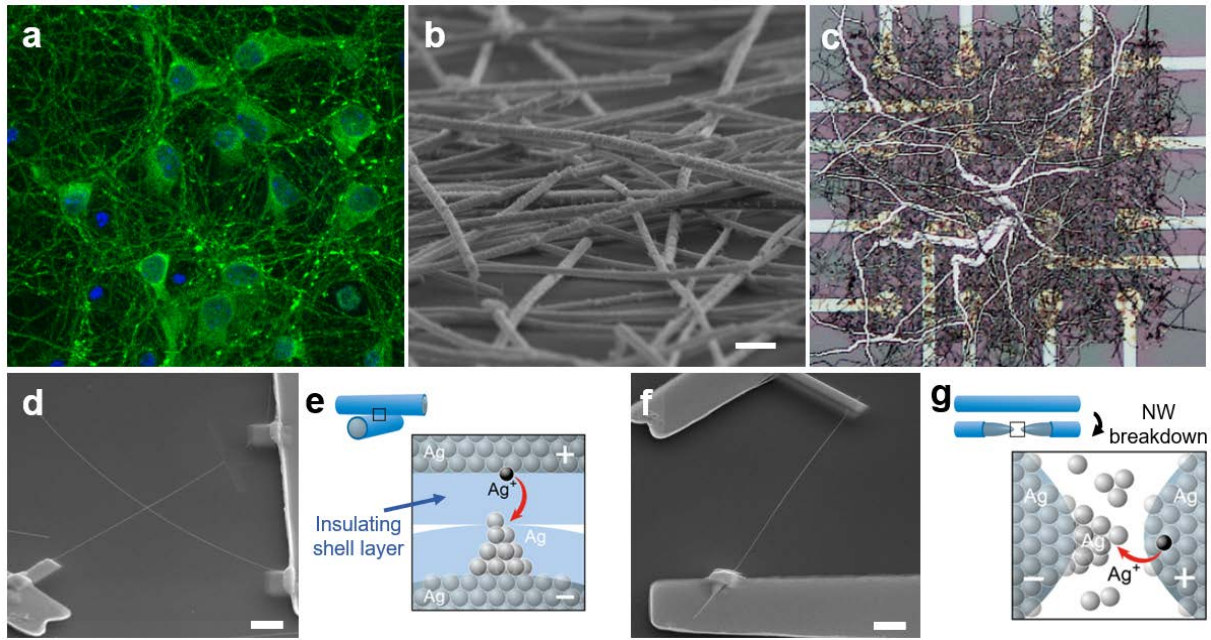

**Figure 1.** Bio-inspired memristive NW networks. (a) Biological neural networks where synaptic connections between neurons are represented by bright fluorescent boutons (image of primary mouse hippocampal neurons); (b) self-organizing memristive Ag NW networks realized by drop-casting (scale bar, 500 nm). Adapted from [7] under the terms of Creative Commons Attribution 4.0 License, Copyright 2020, Wiley-VCH. (c) Atomic switch network of Ag wires. Adapted from [8], Copyright 2013, IOP Publishing. (d-e) Single NW junction device where the memristive mechanism rely on the formation/rupture of a metallic conductive filament in between metallic cores of intersecting NWs under the action of an applied electric field and (f-g) single NW device where the switching mechanism, after the formation of a nanogap along the NW due to an electrical breakdown, is related to the electromigration of metal ions across this gap. Adapted from [7] under the terms of Creative Commons Attribution 4.0 License, Copyright 2020, Wiley-VCH.

**Current and Future Challenges**

Current and future challenges for hardware implementation of neuromorphic computing in the bottom-up NW network will need integrated theoretical and experimental multidisciplinary approaches involving material physics, electronics engineering, neuroscience and network science (an overview of the roadmap is shown in Figure 2). In NW networks, unconventional computing paradigms that emphasize the network as a whole rather than the role of single elements need to be developed. In this framework, great attention has recently been devoted to the reservoir computing (RC) paradigm where a complex network of nonlinear elements is exploited to map input signals into a higher dimensional feature space that is then analysed by means of a readout function. In this framework, nano-networks have been proposed [16] and experimentally exploited as 'physical' reservoirs for *in materia* implementation of the RC paradigm [17,18,19]. However, fundamental research is needed to address remaining challenges. The design and fabrication of multiterminal memristive NW networks able to process multiple spatio-temporal inputs with nonlinear dynamics, fading memory (short-term memory) and echo-state properties minimizing energy dissipation are needed. Importantly, these NW networks have to operate at low voltages and currents to be implemented with conventional electronics. These represent challenges from the material science point of view, since to achieve this goal NWs have to be optimized in terms of core-shell structures for tailoring ionic dynamics underlying resistive switching mechanism. Also, a fully-hardware RC system requires hardware implementation of the readout function for processing outputs of the NW network physical reservoir. Despite the neural network readout can be implemented by means of crossbar arrays of ReRAM devices to realize a fully-memristive architecture as demonstrated in ref. [17], the software/hardware for interfacing the NW network with the ReRAM readout represents a challenge from the electronic engineering point of view. To fully investigate the computing capabilities of these self-organized systems, modelling of the emergent behaviour is required for understanding the interplay in between network topology and functionalities. This relationship can be explored with a complex network approach by means of graph theory metrics. Current challenges in understanding and modelling the emergent behaviour of NW networks rely on the experimental investigation of resistive switching mechanism in single network elements, including a statistical analysis of inherent stochastic switching features of individual memristive elements. Also, combined experiment and modelling are essential to investigate hallmarks of criticality including short and long-range correlations among network elements, power-law distributions of events and avalanche effects by means of an information theory approach. Despite scale-free networks operating near the critical point similarly to the cortical tissue are expected to enhance information processing, understanding how critical phenomena affect computational capabilities of self-organized NW networks still remain an open challenge.

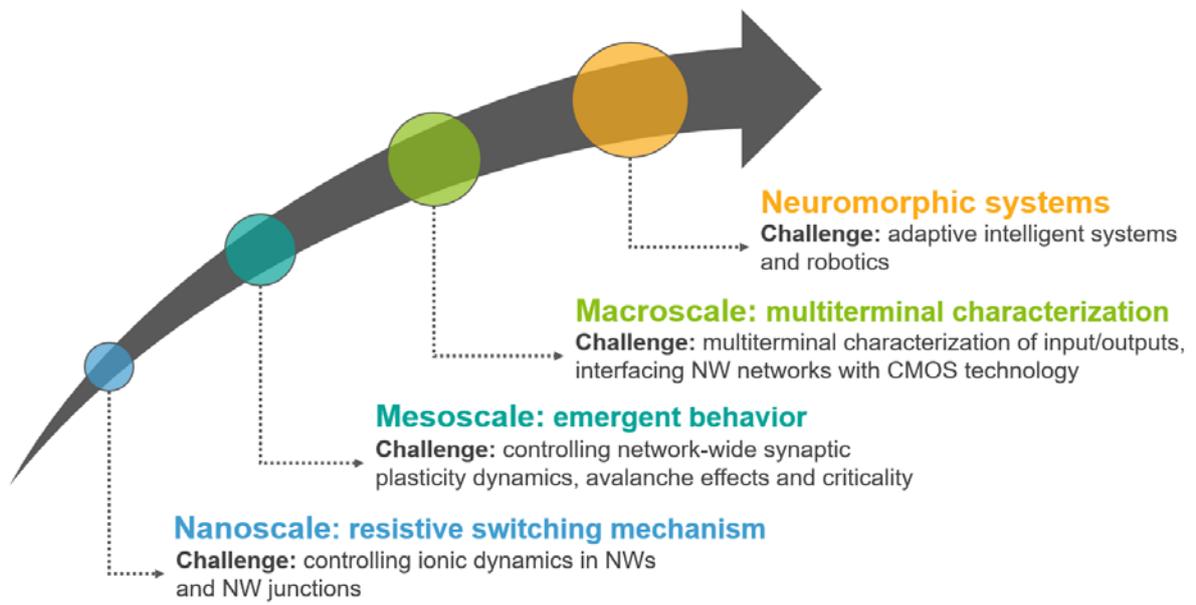

**Figure 2.** Roadmap for the development of neuromorphic systems based on NW networks.

**Advances in Science and Technology to Meet Challenges**

Understanding dynamics from the nanoscale, at the single NW/NW junction level, to the macroscale where a collective behaviour emerges is a key requirement for implementing neuromorphic-type of data processing in NW networks. At the nanoscale, scanning probe microscopy (SPM) techniques can be employed to assess local network dynamics. In particular, Conductive Atomic Force Microscopy (C-AFM), that provides information on the local NW network conductivity, can be exploited not only as a tool to investigate changes of conductivity after switching events, but also for locally manipulating the electrical connectivity at the single NW/NW junction level [20]. Scanning Thermal Microscopy (SThM) can be employed to locally measure the network temperature with spatial resolution < 10 nm, well below the resolution of the conventional Lock-in Thermography (LIT) [12], providing information about nanoscale current pathways across the sample. At the macroscale, advances in electrical characterization techniques are required for analysing the spatial distribution of electrical properties across the network and their evolution over time upon stimulation. In this framework, one-probe electrical mapping can be adopted for spatially visualizing voltage equipotential lines across the network [21], even if this scanning technique does not allow an analysis of the network evolution over time. In contrast, non-scanning electrical resistance tomography (ERT) have been recently demonstrated as a versatile tool for mapping the network conductivity over time at the macroscale (~ cm$^2$) [22]. Thus, ERT can allow *in-situ* direct visualization of the formation and spontaneous relaxation of conductive pathways, providing quantitative information on the conductivity and morphology of conductive pathways in relation with the spatio-temporal location of stimulation. Advancements in the synthesis of core-shell NWs are required for engineering the insulating shell layer surrounding the metallic inner core that acts as a solid electrolyte. Taking into advantage of the possibility of producing conformal thin films with control of thickness and composition at the atomic level, Atomic Layer Deposition (ALD) represents one of the most promising techniques for the realization of metal-oxide shell layers. Also, alternative bottom-up nanopatterning techniques such as Direct Self-Assembly (DSA) of Block Copolymers (BCPs) can be explored for the fabrication of self-organizing NW networks with the possibility of controlling correlation lengths and degree of order [23]. This approach can allow a statistical control of network topology. Customized characterization techniques, from the nanoscale to the macroscale, coupled with a proper engineering of NW structure/materials and network topology, will ultimately enable the control of network dynamics needed for efficient computing implementations.

**Concluding Remarks**

Self-organized NW networks can provide a new paradigm for the realization of neuromorphic hardware. The concept of nanoarchitecture, where the mutual interaction among a huge number of nano parts causes new functionalities to emerge, resembles our brain, where an emergent behaviour arises from the synaptic interactions among a huge number of neurons. Besides reservoir computing that represents one of the most promising computing paradigms to be implemented on these nanoarchitectures, unconventional computing frameworks able to process sensor inputs from the environment can be explored for online adapting of robot behavior. In perspective, more complex network dynamics can be explored by realizing computing nanoarchitectures composed of multiple interconnected networks or by stimulating networks with heterogeneous stimuli. In this scenario, NW networks that can learn and adapt when externally stimulated - thus mimicking the processes of experience-dependent synaptic plasticity that shapes connectivity of our nervous system - would not only represent a breakthrough platform for neuro-inspired computing but could also facilitate the understanding of information processing in our brain, where structure and functionalities are intrinsically related.

**Acknowledgements**

This work was supported by the European project MEMQuD, code 20FUN06. This project (EMPIR 20FUN06 MEMQuD) received funding from the EMPIR programme co-financed by the participating states and from the European Union's Horizon 2020 research and innovation programme.

## 1.6 - 2D materials

Shi-Jun Liang[1], Feng Miao[1] and Mario Lanza[2]

[1]National Laboratory of Solid State Microstructures, School of Physics, Collaborative Innovation Center of Advanced Microstructures, Nanjing University, Nanjing, China.

[2]Physical Sciences and Engineering Division, King Abdullah University of Science and Technology (KAUST), 23955-6900 Thuwal, Saudi Arabia.

**Status**

With more and more deployed edge devices, the huge volumes of data are being generated each day and waiting for real-time analysis. To process these raw data, these data have to be collected and stored, which are accomplished in sensors, memory unit and computing unit, respectively. This usually gives rise to large delay and high energy consumption, which becomes severe with an explosive growth in data generation. Computing in sensory or memory devices allows for reducing latency and power consumption associated with data transfer [1] and is promising for real-time analysis. Functional diversity and performances of these two distinct computing paradigms are largely determined by the type of functional materials. Two-dimensional (2D) materials represent a novel class of materials and show many promising properties, such as atomically thin geometry, excellent electronic properties, electrostatic doping, gate-tuneable photoresponse, superior thermal stability, exceptional mechanical flexibility and strength, etc. Stacking distinct 2D materials on top of each other enables creation of diverse van der Waals (vdW) heterostructures with different combinations and stacking orders, not only retaining the properties of dividual 2D components but also exhibiting additional intriguing properties beyond those of individual 2D materials.

2D materials and vdW heterostructures has recently shown great potential on achieving in-sensor computing and in-memory computing, as shown in Fig. 1. There has intense interest in exploring unique properties of 2D materials and their vdW heterostructures for designing computational sensing devices. For example, photovoltaic properties of gate-tuneable p-n homojunction based on ambipolar material $WSe_2$ were exploited for ultrafast vision sensor capable of processing images within 50 ns [2]. Employing gate-tuneable optoelectronic response of $WSe_2$/h-BN vdW heterostructure can emulate the hierarchical architecture and biological functionalities of human retina to design reconfigurable retinomorphic sensor array [3].

2D materials and their associated vdW heterostructures were also introduced for in-memory computing devices and circuits to improve the switching characteristics and offering additional functionalities. Several switching mechanisms such as conductive filament [4], charging-discharging [5-7], grain boundary migration [8], ionic intercalation [9, 10], lattice phase transition [11], etc., have been reported in 2D materials-based planar and vertical devices. With strict limitation in available space and the number of references, only a few representative works are mentioned in this roadmap. Interested readers are encouraged to refer to previous review article [12]. Based on superior thermal stability and atomically-sharp interface of graphene/$MoS_{2-x}O_x$/graphene vdW heterostructure, a robust memristive device was reported to exhibit endurance of $10^7$ at room temperature and stable switching performance in a record-high operating temperature of 340 $^0$C [13]. Different from oxide-based memristive devices, metal/2D material/metal vertical devices with layered-structure feature of switching medium were used to mimic high-performance electronic synapses with good energy efficiency [14], which holds promise for modelling artificial neural network in a high-density memristive crossbar array [15]. Reducing the thickness of switching medium down to monolayer allows for fabrication of thinnest resistive switching devices with featuring the conductive-point resistive switching mechanism [16, 17].

**Current and Future Challenges**

In these prototype demonstrations of in-sensor computing, the fabricated device arrays are limited due to the challenge in large area synthesis of 2D materials and vdW heterostructures. However, all these works no doubt show that the unique properties of 2D materials and vdW heterostructures can be utilized to achieve ultralow-latency and reconfigurable in-sensor computing [2, 3]. To eventually realize practical applications of in-sensor computing, innovation is demanded to address issues associated with materials, device physics, array size and controlling peripheral circuit. The challenges that must be overcome in the future include growing large-area single crystal materials, exploiting suitable sensing device structures and mechanisms to handle sensory information, fabricating large-scale computational sensory device arrays with good uniformity, high yield and reliability, as well as designing peripheral circuits that efficiently control programmable operations of in-sensor computing arrays.

In contrast to in-sensor computing, many distinct operating mechanisms were already explored to realize 2D materials-based memristive devices for in-memory computing. From structural point of view, planar or vertical metal/insulator/metal (MIM) configurations of isolated devices with relatively large area are mainly studied. Vertical MIM devices with small lateral area would enable high integration density and would be considered by the industry.

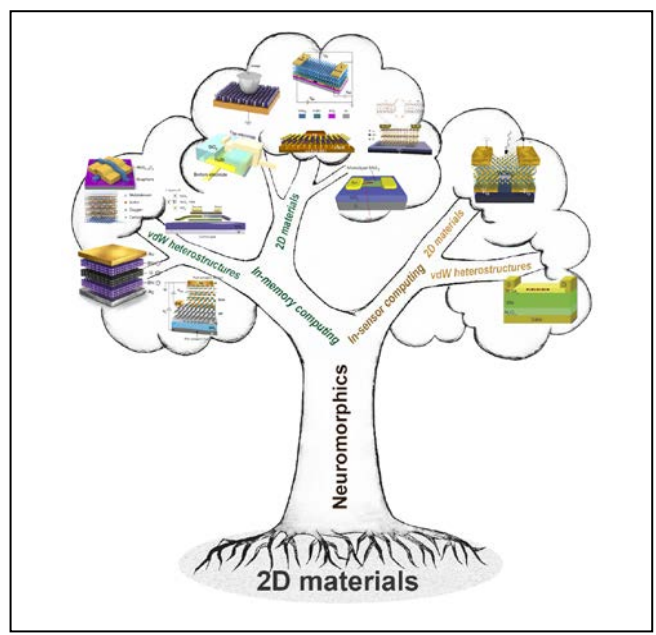

**Figure 1.** 2D and vdW heterostructure materials for neuromorphics. The in-sensor computing devices include $WSe_2$-based homojunction for ultrafast machine vision (Adapted with permission [2], Copyright 2020, Springer Nature) and $WSe_2$/h-BN vdW heterostructure for reconfigurable vision sensor (Adapted with permission [3], Copyright 2020, The American Association for the Advancement of Science); the in-memory computing devices include self-selective vdW memristor (Adapted with permission [4], Copyright 2019, Springer Nature), electrically tuneable homojunction based synaptic circuit (Adapted with permission [5], Copyright 2020, Springer Nature), vdW semi-floating gate memory (Adapted with permission [6], Copyright 2018, Springer Nature), gate-tuneable heterostructure electronic synapse (Adapted with permission [7], Copyright 2017, American Chemistry Society), grain boundary mediated $MoS_2$ planar memristor (Adapted with permission [8], Copyright 2015, Springer Nature), ionic intercalation memristive device (Adapted with permission [10], Copyright 2019, Springer Nature), phase change memristive devices (Adapted with permission [11], Copyright 2019, Springer Nature), robust graphene/$MoS_{2-x}O_x$/graphene vdW memristor (Adapted with permission [13], Copyright 2018, Springer Nature), multilayer h-BN electronic synapse (Adapted with permission [14], Copyright 2018, Springer Nature), atomristor (Adapted with permission [16], Copyright 2018, American Chemistry Society).

However, crossbar study of MIM vertical devices is limited due to the difficulty in synthesis of large area 2D materials with controllable thickness and high quality vdW heterostructure with controllable interface. From electrical point of view, most 2D materials based MIM devices cannot achieve endurances of larger than $10^6$ cycles and stability study of the resistive states was not always demonstrated in multilevel resistive switching devices reported so far. Besides, a unified criteria for yield and variability has been not yet established, which leads to a challenge in evaluating the maturity of 2D materials technology for circuit- and system-level applications of in-memory computing. Clearly stating yield-pass criteria and variability windows of memristive devices is especially important in 2D materials given the large number of local defects intrinsic to scalable synthesis methods as well as other extrinsic defects introduced during integration.

**Advances in Science and Technology to Meet Challenges**

2D materials for in-sensor computing and in-memory computing are emerging research fields and are still in their infancy. The family of 2D materials database is rapidly expanding and a large number of family members already reported in experiments are available for computational sensory and memory devices. In particular, some of air-stable 2D single-crystal materials such as graphene, h-BN and $MoS_2$, etc. can be synthesized directly on metal wafers and transferred to target wafer substrate in a reliable approach [18]. Besides these advances in materials growth, recent advances in device physics and arrays as well as peripheral circuits would offer unprecedented opportunities to realize devices arrays on wafer scale with 2D materials that are suitable for in-sensor and in-memory computing applications.

For practical applications of in-sensor computing, further exploration of novel device physics related to 2D materials is required. For example, in the case of vision sensor, a few distinct types of visual information (*i.e.* orientation, colour, etc.) have to be sensed and processed simultaneously with low power consumption and low latency. Notably, significant progresses have been achieved in anisotropic optoelectronics based on low-lattice symmetry 2D materials and bandgap engineering by electrical field and quantum confinement in 2D materials. This would facilitate the device design with new mechanism that enables to sense and process visual information related to orientation, colour and others. Recently, a $32 \times 32$ optoelectronic machine vision array has been fabricated with a large-area monolayer $MoS_2$ synthesized by metal-organic chemical vapor deposition to propel the functional complexity in visual processing to an unprecedented level [19]. Together with the advance in industrial foundry synthesis of large-area ambipolar $WS_2$ directly on dielectric by plasma enhanced atomic layer deposition, the promising demonstrations of in-sensor computing should be extended to a larger scale array to benchmark against the performance of conventional material-based technology.

Traditionally, in-memory computing is usually implemented in 1T1R crossbar array to avoid the sneak path issue. Similarly, 2D material based resistive switching devices should be organized in such way. To that end, radically new growth processes are desired to achieve all-2D materials 1T1R integrated circuit applications. Furthermore, to fabricate a large-scale crossbar array with high yield and low variance, it is required to spatially engineer the precise atomic vacancy patterns on the surface of wafer-scale single crystal 2D semiconductors or insulators, in particular for monolayer form.

Beyond individual 2D materials, vdW heterostructures by stacking 2D materials with distinct electronic properties can retain the properties of each component and exhibit additional properties inaccessible in individual 2D materials. With the breakthrough in material synthesis and fabrication of large-scale integrated arrays as well as peripheral circuits, use of 2D vdW heterostructures in in-sensor and in-memory computing would provide a disruptive technology to solve the challenges of traditional electronics based on von Neumann architecture.

**Concluding Remarks**

In conclusion, more exploration of 2D vdW heterostructures and continued effort in exploiting novel device physics will offer more possibilities for energy-efficient and low-latency in-sensor and in-memory computing, respectively. For example, vdW heterostructure device with high photoresponsivity over broadband spectrum is expected to deal with visual information in an ultra-wide dynamic range close or beyond human retina. By sensing visual information encoded with other degrees of freedom, *e.g.* multi-coloured wavelengths, polarization, phase, etc., it would further enhance the information processing capability of in-sensor computing chip. The unique properties of 2D vdW heterostructures are not limited to in-sensor computing applications, but also show promise in in-memory computing. Exploring spin-orbit torque and ferroelectric polarization in vdW heterostructure for energy-efficient in-memory computing would be a case in point. The current and future challenges of achieving computing in sensory and non-volatile devices mainly arise from materials synthesis, device physics and array integration. However, all recent advances have indicated that 2D vdW heterostructures can provide numerous opportunities for exploration and hold promise for innovation in material growth, device physics, array integration and peripheral circuit for desirable in-sensor and in-memory computing devices as well as their fusion for real-time and high energy-efficiency data analysis applications [20].

**Acknowledgements**

This work was supported in part by the National Natural Science Foundation of China (61625402, 62034004, 61921005, 61974176), and the Collaborative Innovation Center of Advanced Microstructures. F. M. would like to acknowledge the support from AIQ foundation.

## 1.7 – Organic materials

Tyler J. Quill[1], Scott T. Keene[2] and Alberto Salleo[1]

[1] Department of Materials Science and Engineering, Stanford University, Stanford, CA 94305, United States of America

[2] Department of Engineering, University of Cambridge, Cambridge CB2 1PZ, United Kingdom

**1. Status**

Organic semiconductors (OSCs) have emerged as candidate materials for artificial synaptic devices owing to their low switching energies, wide-range of tunability, and facile ion-migration due to the large free volume within the material. OSCs emulate neuroplasticity at the single unit level with a wide range of synaptic switching mechanisms demonstrated for both two-terminal devices, which utilize filament formation,[1] charge trapping,[2] and ion migration, as well as three-terminal transistor-like architectures such as ion-gated electrochemical transistors[3] and charge trapping transistors. In most cases, the resistive switching of polymers is either via metal-ion migration to form conductive pathways (Figure 1a-b) or by reversible doping where the oxidation state of the OSC is modulated via charge trapping on defect sites (such as implanted nanoparticles), redox reactions (*e.g.* protonation/deprotonation), or ion intercalation (Figure 1c-d).

The ability to tailor the properties of OSCs makes them a particularly promising class of materials for neuromorphic devices since both chemical and microstructural control over the materials can dramatically influence device performance (Figure 1e). Side-chain engineering of OSCs can enhance ionic mobility in the materials, enabling relatively high-speed device operation,[4] whereas modification of chemical moieties on the polymer backbone can be used to tune of energy levels and electronic conductivity.[5] The crystallinity and microstructure of these materials allow for yet another degree of freedom which can be exploited to further optimize them to emulate synaptic behavior.[6] Lastly, the relatively low-cost and solution processability makes OSCs particularly attractive where large-area or printable devices are desired, such as when interfacing with biological systems.

Thus far, OSC neuromorphic devices have demonstrated a variety of synaptic functionality, including the representation of synaptic weight as electrical resistance,[3] excitatory postsynaptic potential (EPSC), global connectivity,[7] and pulse shaping.[8] This broad functionality makes OSCs promising for applications ranging from high-performance computing to biological interfacing of neuromorphic systems. Recently, three-terminal electrochemical devices with low switching energy have been demonstrated which can overcome several challenges associated with parallel operation of a hardware neural network in a crossbar architecture,[9] showing the promise for organic materials in neuromorphic engineering. In this work, however, we will discuss the general challenges and outlook for using OSCs in neuromorphic computing without focusing on any single device, application, or architecture.

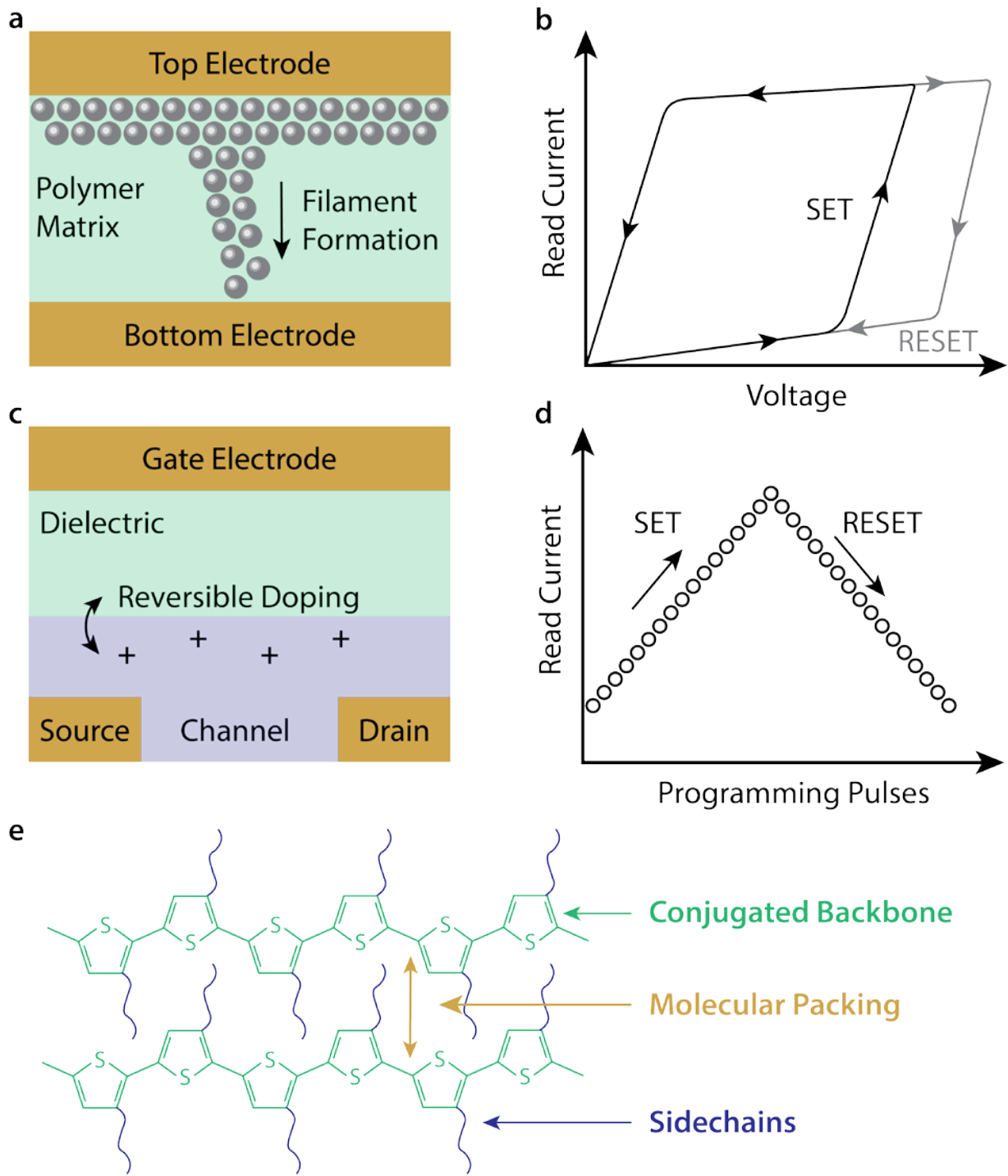

**Figure 1. Organic neuromorphic device operation.**
**a.** Schematic of filament formation and **b.** corresponding read current *vs.* voltage response. **c.** Schematic of three-terminal neuromorphic device based on modulating the channel carrier concentration and **d.** the corresponding programming curve. **e.** Schematic of organic semiconductor structure showing backbone represented by a conjugated thiophene (green), the molecular packing distance (gold), and the tunable sidechains (purple).

**Current and Future Challenges**

*Speed.* Increasing the speed of organic devices has long been a central goal in materials engineering. In OSCs, device speed can be limited by electronic mobilities, ionic mobilities, defects, or stray capacitances. Recent advances in side-chain engineering of mixed ionic/electronic conducting OSCs have notably improved the speed at which organic devices operate, but their speeds still lag behind their inorganic counterparts. Furthermore, the electronic mobilities of OSCs are typically lower than their inorganic/crystalline counterparts, also limiting speed.

*Density.* Patterning of OSCs presents a fundamental challenge for increasing device density due to the incompatibility between OSCs and many of the solvents and photon wavelengths used in photolithography. Consequently, outside of additive manufacturing methods such as printing, the most widespread methods of patterning OSCs rely on either sacrificial hard masks to protect the OSC from solvents,[10] or the use of orthogonal photoresist solvents which do not damage the OSC.[11] Additionally, deposition of highly uniform OSC films is challenging due to complex microstructures, and nontraditional fabrication techniques are required to enable vertical architectures to reduce the individual device footprint (3-terminal) and interconnect complexity.

*Integration.* An additional challenge revolves around incorporating organic neuromorphic devices/systems with traditional digital systems while avoiding damage during back-end-of-the-line (BEOL) processing. The electronic properties of OSCs typically degrade at elevated temperatures (typically >150 $^{o}$C) due to phase transitions (*e.g.* $T_g$, $T_m$) or temperature-induced morphological changes (*e.g.* thermal expansion, backbone twisting). This temperature sensitivity is problematic due to required processing temperatures of ~400 $^{o}$C used to anneal Cu interconnects in inorganic device stacks. Although crossbar architectures offer a straightforward method of replicating the vector-matrix multiplication desired in ANNs, the need to control sneak currents often requires an access device, increasing the complexity of the array and providing an additional integration challenge.

*Environmental and Electronic Stability.* A final remaining challenge for OSCs is to achieve long-term device stability and resistance state retention. Interfaces of OSCs and dielectrics are susceptible to formation of traps resulting from exposure to oxygen or moisture, leading to irreversible changes in device performance. Additionally, because of the inherently low switching energy found in many organic neuromorphic devices, "SET" OSCs are susceptible to leakage due to parasitic reactions with the surrounding atmosphere.[12] Finally, both the charge transport and doping reactions in OSCs must be stable at the typical operating temperatures of computers (~85 $^{o}$C) without suffering from changes in morphology due to thermal annealing.

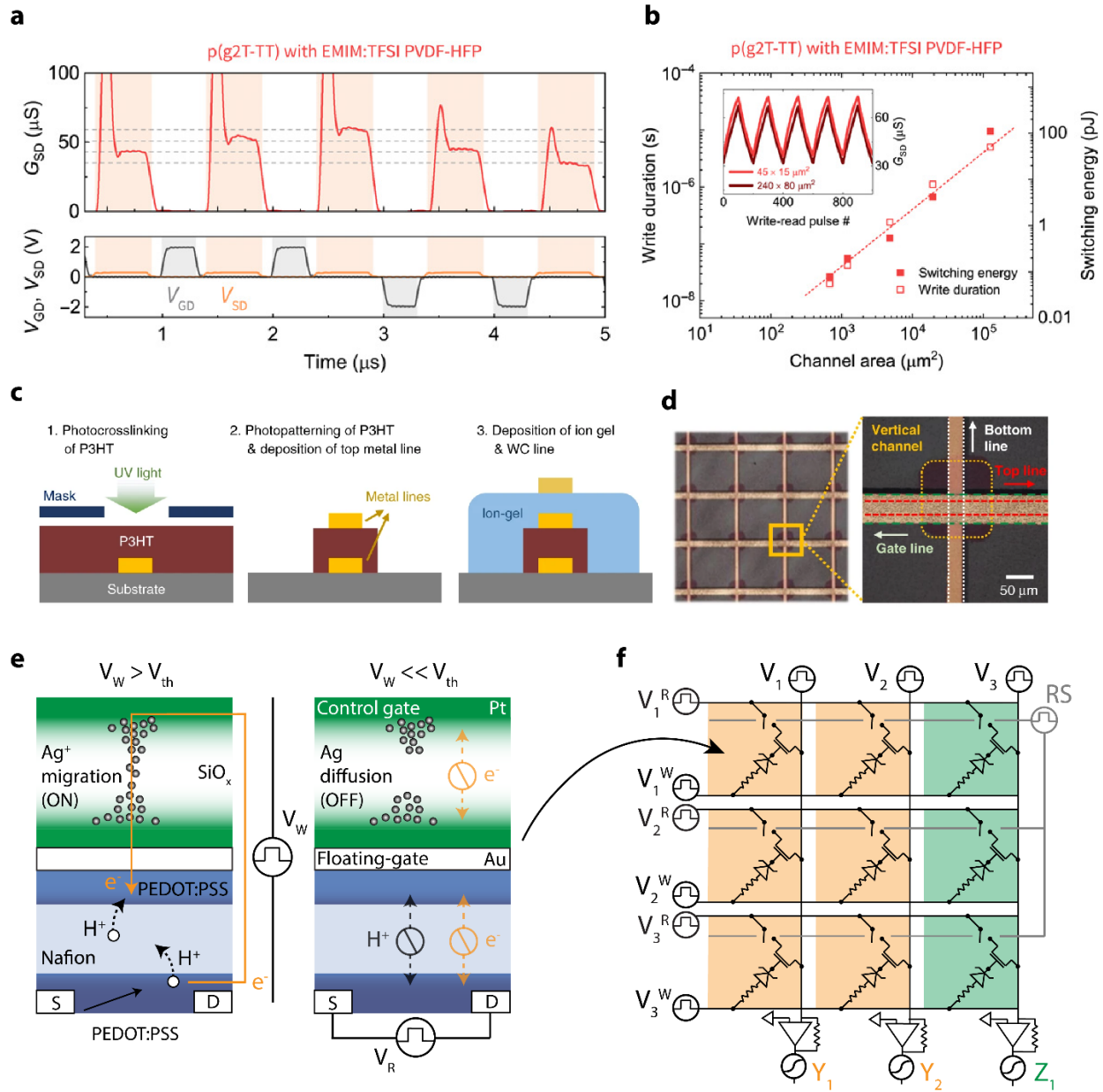

**Figure 2. State-of-the-art organic neuromorphic devices.**
**a.** Analog resistance tuning of an electrochemical neuromorphic device under ±2-V 200-ns write pulses (gray shaded area), followed by 100-ns write-read delay and +0.3-V 500-ns readout (orange shaded area). The horizontal dashed lines are a guide to the eye to represent tunable conductance states. **b.** The volumetric scaling of electrochemical doping enables channel conductance of devices to be tuned with increasingly lower write energies and shorter write pulses as device sizes are reduced. **c.** Cross-sectional schematic of fabrication procedure of densely packed ion-gel-gated vertical P3HT synapses and **d.** optical microscopy images of a crossbar array. **e** A non-volatile ionic floating gate (IFG) memory consisting of a filament forming access device (green) attached to a PEDOT:PSS organic synapse (blue). **f** Schematic of parallel-programmable neuromorphic array using IFG memory divided into a two-layer neural network, as indicated by orange and green. Analog network inputs $V_i^R$ are applied across the source-drain rows, while programming inputs $V_i^W$ and $V_j^W$ are applied along the gate row and drain column, respectively. Adapted from ref.[13], AAAS, (**a,b**); Reproduced from ref. [14], Springer Nature Ltd, (**c,d**); Adapted from ref. [9], AAAS, (**e,f**).

**Advances in science and technology to meet challenges**

*Speed.* To improve organic device speed, it is first essential to identify the rate limitations. For example, in ion-gated devices, both the ion and electronic mobilities can dictate the switching speed. Spyropoulos *et. al.* have shown that for ion-gated devices, presence of the electrolyte ions within the OSC can improve device response time.[15] Additionally, strain-relaxation following the insertion or removal of species may also limit the time to reach a stable resistance state.[16] Once the fundamental speed limits are identified, polymers and devices can be engineered to optimize for the critical parameters (*e.g.* ion mobility, parasitic capacitances, strain response). In a recent example, selection of the electrolyte and OSC materials allowed for electrochemical neuromorphic devices to operate with 200 ns write operations and <1 μs read-write cycles (Figure 2 a-b).[13]

*Density.* Novel nanofabrication processes which can accurately define OSCs in vertical architectures can reduce device footprints, increasing device density. Strategies such as utilizing metal contacts as hard masks can enable nanopatterning of OSC channels with resolutions limited by conventional lithographic techniques,[17] but defining gate and electrolyte geometries with similar precision for complete 3-terminal devices introduces additional complexity. Choi *et. al.* recently demonstrated vertical 3-terminal electrochemical neuromorphic devices which reduced the single cell footprint to *ca.* 100 μm by 100 μm in a crossbar architecture using photo-crosslinked P3HT as the channel material (Figure 2c-d). In principle, this cell could be reduced significantly using the same general technique with the use of advanced photolithography.

*Integration.* Advancements in non-traditional chip manufacturing (BEOL alternatives)[18] are necessary for seamless integration of OSCs with silicon technology. Sneak currents in neuromorphic arrays can be avoided by using filament-forming access devices coupled to three-terminal memories, as shown by Fuller *et. al.* (Figure 2e-f).[9] Increasing the temperature stability of OSCs also helps enable complete integration with conventional BEOL processing. Recently, Gumyusenge *et. al.* demonstrated that nanoconfined OSCs in high-performance polymer blends exhibit robust temperature-independent mobilities up to 220$^o$C,[19] a notable step towards integration.

*Environmental and Electronic Stability.* Although the stability of OSCs presents challenges for developing neuromorphic devices, recent design strategies provide promise. The modularity of OSCs enables tuning of both molecular orbital energies as well as morphology. For example, engineering OSCs with high ionization potentials can eliminate cross-reactions with moisture or oxygen.[5] Further optimization of OSC crystallinity[6] and encapsulation methods[12], which shield devices from the ambient atmosphere, could further improve stability.

**Concluding remarks**

Organic materials have rapidly grown into a promising class of materials for neuromorphic systems and could be harbingers of other unconventional semiconductors for these applications. While there are great challenges facing organics before they are suitable for commercial neuromorphic computing systems, including significant improvements to speed, density, integration, and stability, there are no fundamental barriers preventing OSCs from satisfying these metrics.

Owing to their biocompatibility and softer mechanical properties, organics are of interest for direct connections between biological systems and neuromorphic computers, such as in brain-machine interfaces and adaptive prosthetic devices. Inspired by the biomimicking nature of neuromorphic systems, there is a strong push towards direct integration with prosthetics to match the low power computation already found in the human brain. These systems and devices would form direct interfaces with tissue, repairing augmenting functionality or act as "smart system" for wearable electronics. Recently, organic neuromorphic devices have been tuned to match either the output signals for "talking" to[8] or by responding to neurotransmitter signals for "listening" to[20] the biological domain. We postulate that organic materials will shine as neuromorphic devices in bioelectronic interfaces due to the relative maturity of the materials class in the bioelectronic space.

**Acknowledgements**

*AS and STK acknowledge support from the National Science Foundation and the Semiconductor Research Corporation (Award NSF E2CDA #1507826). TJQ acknowledges support from the National Science Foundation Graduate Research Fellowship Program under grant DGE-1656518.*

## 1.8 – Spintronics

J. Grollier, D. Marković, A. Mizrahi

Unité Mixte de Physique, CNRS, Thales, Université Paris-Saclay, 91767 Palaiseau, France

**Status**

Spintronics, or spin electronics, manipulates the spin of electrons in addition to their charge. This brings multiple interesting features for neuromorphic computing: the non-volatile memory provided by nanomagnets and the non-linear dynamics of magnetization induced by fields or currents (*1*). These two aspects allow the same materials to be used to mimic the essential operations of synapses and neurons. Important experimental results have thus been obtained in recent years.

Synapses - The first way to realize spintronic synapses is to store the weights in digital Spin Torque-Magnetic Random Access Memories (ST-MRAMs) (*2*). Gigabit devices from the latter are now commercially available in several large foundries. They consist of magnetic tunnel junctions, formed by an ultra-thin (~1 nm) insulator sandwiched between magnetic layers, integrated in the CMOS process. The main advantage of ST-MRAMs over their competitors is their endurance, which is more than two orders of magnitude higher, a very important factor for the chips dedicated to learning, that will require very many read/write cycles. Indeed, the resistance change mechanism comes from a reversal of magnetization by current pulses of the order of nanoseconds and a hundred millivolts, a purely electronic phenomenon that does not require the movement of ions or atoms in a nanostructure as in ReRAMs or PCMs. Moreover, they are non-volatile, retaining information even when the power is switched off. Associative memories integrating ST-MRAMs (Fig.1a) have enabled significant gains in power consumption, with only 600µW per recognition operation, i.e. a 91.2% reduction compared to a twin chip using conventional Static Random Access Memory (*2*).

The second way to realize spintronic synapses is to directly imitate a synapse with a magnetic tunnel junction. In this case, the junction acts as a memristor device, which takes as input a current and multiplies it by its resistance, which thus plays the role of the synaptic weight. The stability of magnetization in magnetic tunnel junctions allows them to retain the value of the weight. Since magnetization is naturally bistable, magnetic tunnel junctions are very good candidates for neural networks with binary weights (*3*). It is also possible to modify the materials or geometry so that the magnetization changes orientation via non-uniform states. This has allowed to experimentally realize analog synapses (Fig. 1b) (*4–6*), as well as to train a small neural network with magnetic multi-state synapses (Fig. 1c) (*7*).

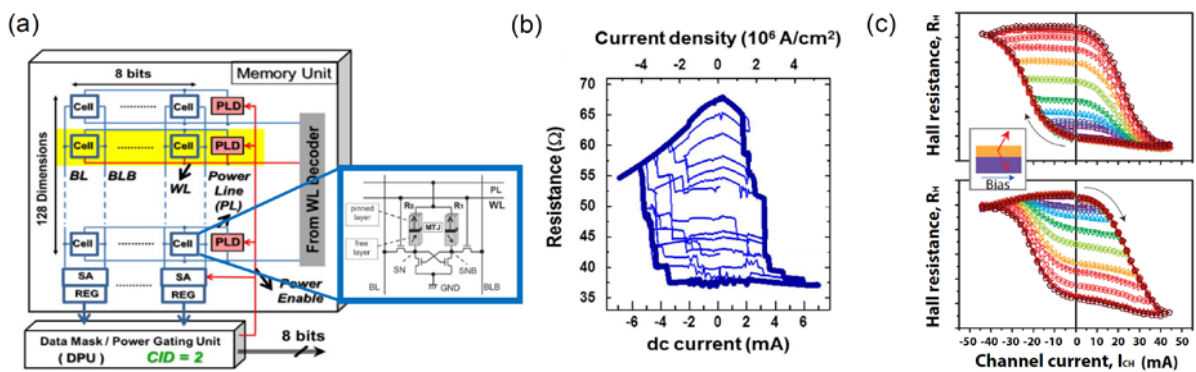

Fig.1 - Spintronic synapses. (a) Schematic of an associative memory circuit with ST-MRAM cell, reproduced from (*2*) (b) R-I hysteresis loop of a spintronic memristor based on current-induced domain wall displacement in a magnetic tunnel junction, reproduced from (*4*) (c) ) R-I hysteresis loop of a spintronic memristor exploiting spin-orbit torques in a ferromagnetic/antiferromagnetic bilayer, reproduced from (*6*).

Neurons - In most neural network algorithms, neurons simply apply a non-linear function to the real-valued synaptic inputs they receive. The characteristics of the nonlinear dynamics of spintronics can be exploited to mimic biology more closely, which could lead to increased computing functionalities such as local and unsupervised learning. Biological neurons transform the voltage on their membrane into electrical spike trains, with a mean frequency that is non-linearly dependent on the voltage. Magnetic tunnel junctions transform DC inputs into an oscillating voltage with a frequency that depends non-linearly on the injected current. This property can be used to imitate neurons. In stable junctions such as those used for ST-MRAMs, the spin torque can induce oscillations between about ten MHz and ten GHz depending on the materials and geometry. These oscillations have been used with a single device to recognize pronounced digits with a time-multiplexed reservoir (*8*). Four coupled spintronic nano-oscillators were also trained to recognize vowels via their synchronization patterns to RF inputs (Fig. 2a) (*9*). In unstable junctions, thermal fluctuations may be sufficient to induce telegraphic voltage behavior, allowing the mimicking of stochastic neurons with minimal energy consumption. Neuromorphic tasks have been performed by small experimental systems composed of such junctions, using neural networks (*10*, *11*) or probabilistic algorithms (*12*).

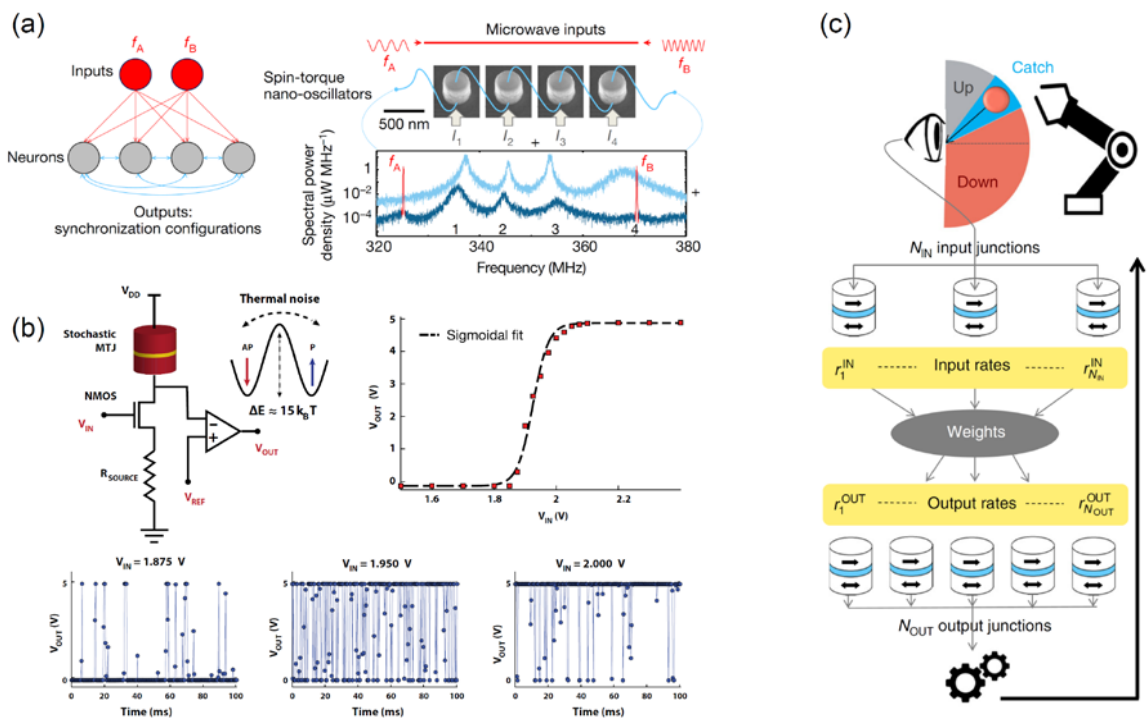

Fig.2 - Spintronic Neurons. (a) Principle of vowel recognition with four coupled spintronic nano-oscillators, reproduced from (*9*). Left: schematic of the implemented neural network. Right: schematic of the experimental set-up and associated microwave emissions in free (light blue) and phase-locked (navy) states. (b) Superparamagnetic tunnel junction behaviour under different input voltage (time traces at the bottom, average resistance top right) and circuit implementing a probabilistic bit (top

left) (*12*). (c) Schematic of a population of superparamagnetic tunnel junctions assembled in a neural network reproduced from (*10*).

**Current and Future Challenges**

The challenge is now to create large-scale spintronic neuromorphic circuits capable of solving useful tasks at low energy consumption, especially for embedded artificial intelligence. This requires the integration of layers of synapses and spintronic neurons interfaced in deep networks. The major disadvantage of spintronics compared to other technologies is the low resistance ratio between the OFF and ON states of the magnetic tunnel junctions, of the order of 2-3 compared to more than 10 in other resistive memory technologies. These small variations in resistance make the reading of the state of the junctions more complex, and have so far prevented the development of magnetic tunnel junction crossbar arrays as realized in other resistive technologies. It is therefore necessary to continue the effort on the material side to push the resistance variations towards their theoretical $R_{off}/R_{on}$ value > 100 (*13*, *14*). On the CMOS design side, the development of low-power circuits allowing efficient reading of the state of the junctions, such as sense-amplifier, is crucial. The first demonstrations will certainly rely on binarization of resistance values for the inference phase and implementation of hardware Binary Neural Networks, before end-to-end on-chip learning solutions are developed.

Combining ionic and spintronic effects will be one of the keys to efficient learning of neuromorphic chips. It was recently demonstrated that strong magnetoelectric effects enable control of magnetic dynamics by the electric field created by the interface, more efficiently than previous methods (*15*).

A critical challenge for the development of hardware neural networks is to achieve a high density of connections. Spintronics offers several opportunities to tackle this issue. Long-range connections can be implemented via spin currents and magnetic waves or by physically moving magnetic textures such as skyrmions and solitons (*1*, *16*). Furthermore, the multilayer nature of spintronic devices allows them to naturally stack in three dimensions, opening the path to vertical communication (*17*).

Spintronic neuromorphic chips will be able to receive as inputs fast signals compatible with digital electronics (classical binary junctions), radio-frequency inputs (GHz oscillator), as well as inputs varying at the speed of the living world, thanks to superparamagnetic junctions or magneto-electric effects that can operate at timescales between seconds and milliseconds. There is active research on developing spintronic devices for on-chip communication (using their capability to emit and receive microwaves), magnetic sensing (with promising biomedical applications) and energy harvesting, all of which could benefit neuromorphic chips (*1*).

Taking full advantage of the dynamical behavior of spintronic devices will require the development of dedicated learning algorithms, inspired by advances in both machine learning and computational neuroscience. The fact that the behavior of spintronic devices relies on purely physical phenomena that can be predictively described and integrated into neural network programming libraries is a key enabler for this task (*18*).

**Advances in Science and Technology to Meet Challenges**

Spintronics is undergoing promising new developments from which neuromorphic chips could benefit. Antiferromagnetic materials and interfaces with optics bring the possibility of information processing and transmission at THz speed (*19*). Spin/charge conversions are increasingly efficient thanks to new materials such as topological insulators.

Finally, the multifunctionality of spintronics also makes it possible to train complex physical systems that do not exactly reproduce the synapse/neuron structure, such as, for example, arrays of spin wave transmitters/receivers, or fixed or mobile magnetic particles, such as skyrmions and domain walls (*20*). Micromagnetic simulations with predictive power, coupled with gradient descent, have modeled learning tasks (*18*). Experimental demonstrations with these complex physical systems remain to be carried out.

**Concluding Remarks**

In the short term, neuromorphic spintronics should see the commercialization of artificial intelligence chips storing synaptic weights into current and future generations of ST-MRAMs. This should be followed by the development hardware neuron circuits leveraging the dynamical properties of magnetic tunnel junctions to implement synapses and neurons for inference and learning. In the longer term, more exotic materials and textures offer the fascinating prospect of in-materio computation based on complex physical effects.

## 2.1 Deep Learning

Peng Yao, J. Joshua Yang

Electrical and Computer Engineering Department, University of Southern California, Los Angeles, CA, USA

**Status**

The development of Deep Learning (DL) has brought Artificial Intelligence (AI) to the spotlight of broad research communities. The brain-inspired neural network models with different structures and configurations have made significant progress in a variety of complex tasks [1]. However, in conventional von Neumann architecture, the physically separated computing unit and memory unit require frequent data shuttling between them, which results in considerable power consumption and latency cost. One promising approach to tackle this issue is to realize in-memory computing (IMC) paradigm where each underlying device component functions as memory and computation elements simultaneously. Non-volatile devices based on resistive switching phenomena [2-3], such as redox memristor, phase change, magnetic and ferroelectric devices, could support such computing system and show greatly improved performance in data centric computation tasks.

Analogue resistive-switching memory based in-memory computing is promising to bring orders of magnitudes improvement in energy efficiency compared to the conventional von Neumann hardware. The devices are assembled in a crossbar structure to conduct vector-matrix multiplication (VMM) operations, where the input vectors are encoded as voltage amplitude, pulse widths, pulse numbers, or sequential pulses with different significances, and the matrix elements are mapped to tunable cell conductance where each cell is often represented in the differential form of a pair of devices. Thanking to Ohm's law for multiplication and Kirchhoff's Current Law for accumulation, the dense crossbar could conduct multiplication-accumulation (MAC) fully in parallel and the computation occurs at the data location. Since VMM calculation accounts for the majority of computation during inference and training of deep learning algorithms, this in-memory computing paradigm could help the hardware to meet stringent requests of low power dissipation and high computing throughput.

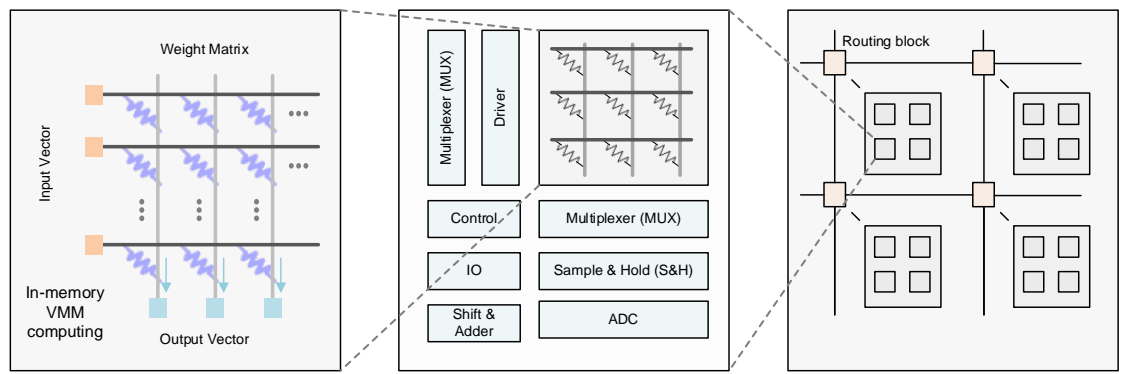

Figure 1. Schematic of underlying in-memory computing hardware for deep learning acceleration, presented from crossbar level, macro circuit level, and monolithic system level, respectively.

Major progresses have been made in this area, spanned from device optimization to system demonstration [2-4]. The oxide-memristor devices have been scaled down to 2 nm in an array [5] and 3D stacked architecture has been fabricated in laboratory to enhance the network connectivity [6]. In addition, various deep neural network models, including perceptron [7-8], Multiple Layer Perceptron (MLPs) [9], Long Short Term Memory (LSTM) [10] based Recurrent Neural Networks (RNNs), and Convolutional Neural Networks (CNNs) [11], have been demonstrated based on non-volatile resistive-switching crossbars or macro circuits. These demonstrations have covered the typical learning algorithms for supervised learning, unsupervised learning and reinforcement learning. More recently, a multiple-array based memristor system [11] and some monolithically integrated memristor chips have been demonstrated [12-13], and it is encouraging to see that this kind of in-memory computing system could achieve an accuracy comparable to software results and reach >10 TOPS/W energy efficiency using 8 bit input precision [11]. However, despite the fast development of hardware prototypes and demonstrations, a monolithically integrated IMC chip with large and tiled crossbars (shown in Fig. 1) for practical and sophisticated DL models (e.g. ResNET50) is still under-explored, and the accomplished tasks are limited to relatively small dataset (e.g. MNIST, CIFAR10) rather than handling large workloads (e.g. ImageNet).

**Current and Future Challenges**

One fundamental challenge of the IMC system originates from the device non-ideal characteristics and process issue towards large-scale heterogeneous integration. No matter what the specific switching mechanism is, the analogue-switching non-volatile memory device show some inherent device-wise and cycle-wise variance, reading fluctuations, state drift or stuck, or limited on/off ratio. Although the rich dynamics could be explored for neuromorphic systems in attempt to resemble biological phenomenon more faithfully, these uncontrolled non-ideal behaviors would inevitably affect the accuracy of the DL model and deteriorate the system performance in most cases. The high-level system researches often unrealistically assume ideal device features, resulting in some significant discrepancy from practical applications. Moreover, the forward propagation and backward propagation pose stringent requirement for linear I-V behavior to conduct accurate MAC operations, while the update procedure demands the device showing state-independent conductance tuning curves with sufficient linearity and symmetry with respect to the programming pulses to insure training convergence and efficiency. In spite of the undergoing advances in device optimization and exploration, ideal device is still missing. Thus, the IMC systems need various compensating strategies, which can substantially degrade system efficiency.

In terms of heterogeneous integration, the advanced foundry process for large-scale fabrication of analogue-switching devices is absent, regardless of the specific device category. Although NOR Flash is commercially available, it's large operation voltages and slow speeds together with its limited endurance and scalability make it at best an interim solution, to be replaced by emerging devices. Oxide memristor is promising in dense integration given the demonstration of 2nm feature size and 3D stacking ability at lab. However, only 130nm analog-switching technology [12-13] and 22nm digital-switching technology [14] in foundry have been reported. Many other kinds of devices require back-end process with high temperature, complex layer deposition or special handling process, which present obstacles for them to be monolithically integrated with the mainstream

CMOS technology. The absence of high-uniformity and high-yield process in mainstream foundries for large-scale and small-footprint integration of analogue-switching devices has been slowing down the development of IMC circuits.

Errors in analog IMC and inefficiency of periphery circuits also imposes serious challenges for practical hardware. Analog computing directly utilizing physical laws is superior in energy efficiency, whereas it only suits for low-precision tasks so far. Although DL algorithms put loose constraints on parameter precisions (such as 4bit weights for regular inference tasks), state-of-the-art models still demands accurate digitalized value representations. However, the conductance states of analog devices always follow a certain distribution and deviate from the target mapping values, which would bring in weight representing errors. In addition, at the array/crossbar level, the parasitic effects along the metal wires would lead to inevitable IR drop and result in inaccurate programming and computing. This effect becomes more severe if the array size is increased for higher performance. Such systematic errors may be mitigated through some algorithm and architecture co-design, such as compensations in the mapping algorithms. The periphery circuits would also introduce computing errors due to the voltage loss on analogue switches, transistor mismatch, unfixed clamping voltage and environmental fluctuations. All these together would substantially lower analogue computing accuracy and prevent IMC system from realistic applications if not appropriately addressed.

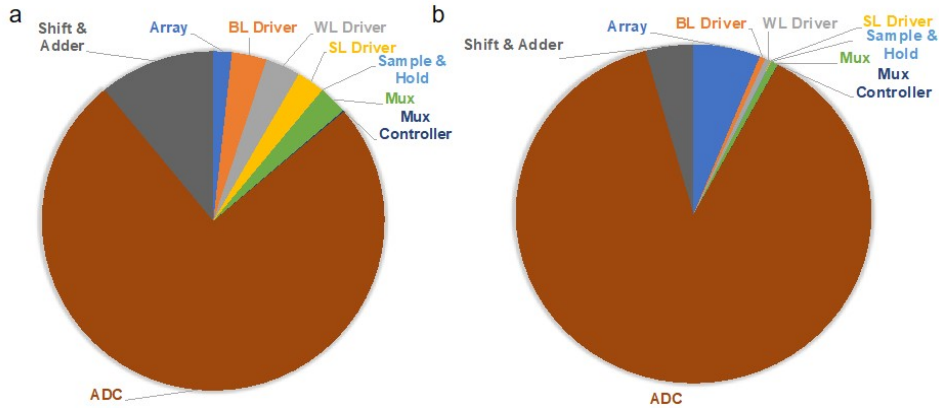

Figure 2. The breakdown of area and power consumption in a macro-circuitry instance [11]. (a) Area overhead. (b) Power overhead.

To take the full advantage of IMC features, all necessary functional blocks should be integrated monolithically with device crossbars (as shown in Fig.1), including buffers, interfacial circuits (mainly ADCs, converting the accumulated current to digital signals), routing units, control logic and digital processing. These circuits are expected to match the device operating requirements, such as programming voltage and driving currents. In such a complete on-chip system with tiled crossbars, the auxiliary periphery circuits might consume much more power, area and latency than the analog-domain VMM calculation. Although the IMC paradigm eliminate the movement of DL weights, it still needs data flowing between different layers and requires on-chip memory accessing. Meanwhile, the parallel MAC calculations desire multiple ADCs locating at the end of each column to carry out fast conversations frequently. According to the profile of a designing instance, the ADCs account for the majority of power and area overhead (shown in Fig.2) [11]. Exploiting a larger crossbar to conduct VMM is beneficial to boost system performance by amortizing the periphery circuit overhead in the whole system, which, however, would lead to larger parasitic capacitance and resistance, higher dynamic range of the output current and lower device utilization ratio. The inefficiency of periphery circuits, especially the ADCs, is becoming the system bottleneck of IMC hardware, where innovations are needed in the co-design of device and architecture.

**Advance in Science and Technology to Meet Challenges**

Extensive multi-disciplinary efforts are needed in order to advance the development of IMC-based deep learning accelerators, as co-designs of device, circuit, architecture and algorithm are required to build practical IMC system [15].

First of all, researches in material engineering and device optimization should be conducted either based on present analogue-switching non-volatile devices or for the exploration of novel devices, aiming at enhanced reliability, improved programming linearity and symmetry while maintaining high switching speed, low programming power and intensive scaling potential. In addition, stable stack process for large-scale heterogeneous integration of highly uniform crossbar is needed for practical applications. The development of 3D process could drive the device density to next level and bring in extra dimension to explore more efficient system. Even more importantly, 3D structures enable the massive connectivity and low-loss communications required for complex neural networks.

Second, at the macro-circuit level, there is plenty of room to optimize the crossbar structure and the periphery circuits. For example, basic 2-transistor-2-memristor (2T2M) configuration [12] could be utilized as a signed-weight unit to construct IMC arrays, where in-situ subtraction operation is conducted in analog domain with the differential current being accumulated subsequently. Such configuration reduces total flowing currents to mitigate IR drop effect, which makes it available to build larger crossbar. Apart from this, encoding input signal by pulse width or low voltage amplitude range might bypass the nonlinear current-voltage characteristic issue, at the expense of increasing system latency or circuitry complexity. On the other hand, novel periphery circuitry design customized for IMC is required, including fast, low-power ADC and high-throughput routing scheme with little on-chip memory. For example, time-domain interfaces could be used to replace conventional ADC-based interfaces [16]. Furthermore, some emerging devices with rich nonlinearities [17] could potentially replace circuitry blocks directly, such as implementation of device-wise ReLU function [18-19].

Finally, system-level innovations are critical to expedite the development of IMC hardware. From architecture perspective, time division multiplexing of ADCs and replicating same weights to different crossbars are key technologies in order to optimize the system dataflow and boost the computing parallelism. In addition, despite the difficulties in data storing and transmission in analogue domain, interfacing, transferring and processing the information in analogue format is intriguing due to the potential of huge efficiency benefits. From the algorithmic point of view, configuring and optimizing the DL models to fit IMC device features and reduce hardware cost is demanded. On-chip learning, hardware-aware learning and hybrid learning are some representative works to mitigate device non-ideal characteristics and computing errors.

**Concluding Remarks**

In-memory computing based on analogue-switching non-volatile device shows exceptional superiority regarding computing throughput and energy efficiency than the conventional von Neumann hardware, suited for dealing with data centric problems and brain-inspired deep learning algorithms. In spite of the significant advancements in device explorations and system demonstrations, device non-ideal behaviors, difficulties in large-scale heterogeneous integration, inaccuracies of analog computing and inefficiency of periphery circuits pose great challenges to promoting the in-memory computing technologies for practical application. Monolithic integration of a complete system that unleash the full potential of the in-memory computing features with tiled crossbar architecture and smooth dataflow is still missing. Consequently, extensive co-design efforts from device optimization, circuitry design, architecture exploration and algorithm tailoring are consistently needed. With the utilization of more emerging devices and advanced 3D integration process, the in-memory computing promises bright future of deep learning hardware.

## 2.2 Spiking neural networks

Giacomo Indiveri, Institute of Neuroinformatics, University of Zurich and ETH Zurich, Switzerland

**Status**

The design of neuromorphic circuits for implementing spiking neural networks represents one of the main activities of Neuromorphic Computing and Engineering. Currently, these activities can be divided into two main classes: (i) the design of large-scale general-purpose spiking neural network *simulation* platforms using digital circuits and advanced Complementary Metal-Oxide Semiconductor (CMOS) fabrication processes [1–3], and (ii) the design of analog biophysically realistic synaptic and neural processing circuits for the real-time *emulation* of neural dynamics applied to specific sensory-motor online processing tasks [4–10]. This latter effort pursues the original goal of Neuromorphic Engineering, set forth over thirty years ago by Carver Mead and colleagues [11, 12], to use the physics of electronic devices for understanding the principles of computation used by neural processing systems. While the strategy of building artificial neural processing systems using CMOS technologies to physically emulate cortical structures and neural processing systems was mainly restricted to academic investigations for basic research in the past, the recent advent of emerging memory technologies based on memristive devices spurred renewed interest in this approach, also for applied research and practical applications. One of the main reasons is that the analog and mixed-signal analog/digital neuromorphic processing architectures that implement adaptation, learning, and homeostatic mechanisms are, by construction, robust to device variability [13, 14]. This is a very appealing feature that enables the exploitation of the intricate physics of nanoscale memristive devices, which have a high degree of variability, for carrying out complex sensory processing, pattern recognition, and computing tasks. Another appealing feature of these mixed-signal neuromorphic computing architectures, that enables a perfect symbiosis with memristive devices, is their "in-memory computing" nature: these architectures are typically implemented as large crossbar arrays of synapse circuits that represent at the same time the site of memory and of computation. The synapses in each row of these arrays are connected to Integrate-and-Fire (I&F) soma circuits, located on the side of the array. The soma circuits sum spatially all the weighted currents produced by the synapses, integrate them over time, and produce an output pulse (spikes) when the integrated signal crosses a set threshold. In turn the synapses are typically stimulated with input spikes (e.g., arriving from other soma circuits in the network), and convert the digital pulse into a weighted analog current [6, 13]. Depending on the complexity of the synapse and soma circuits, it is possible to design systems that can exhibit complex temporal dynamics, for example to create spatio-temporal filters matched to the signals and patterns of interest, or to implement adaptive and learning mechanisms that can be used to "train" the network to carry out specific tasks.

**Current and Future Challenges**

Artificial neural networks can achieve impressive performance in solving an incredible amount of problems, thanks to their *learning* abilities. The backbone of learning in neural networks simulated on standard computing technologies is the "backpropagation through time" (BPTT) algorithm [15]. So an important challenge for spiking neural networks is to understand how to implement learning rules as powerful as BPTT, but using plasticity mechanisms that only have access to local signals, due to their "in-memory computing" nature. This goal is particularly challenging for hardware spiking neural networks, because their synaptic circuits have limited resolution, and because the devices used to store their weights are often affected by cycle-to-cycle and device-to-device variability.

In addition, neuromorphic spiking neural architectures do not have access to external memory blocks where to store or retrieve information, like in von Neumann architectures. On one hand this has the advantage of removing the infamous von Neumann memory bottleneck [16, 17], which refers to the problem of having to transfer information from storage areas to computing ones at very high rates, and which accounts for the vast majority of the exceedingly large power consumption figures of standard computers. On the other however, this introduces the problem that spiking neural computing systems cannot arbitrarily choose the resolution of the data they need to store, or the times at which they access it during the course of their computations.

This adds an even more complicated challenge of understanding how to manage memory and time in such architectures: spiking neural networks can be seen as non-linear filters that process information online, as data is flowing through them. To carry out real-time computation on a sensory input stream, these networks must retain a short-term memory trace of their recent inputs. Without learning, there are fundamental limits on the lifetimes of these memory traces that depend on both the network size and the longest time-scales supported by the elements used in the network [18]. So an important requirement for enabling the construction of hardware spiking neural networks that can be deployed in a wide range of real-world applications is to develop volatile memristive technologies that have a large distribution of time scales, ranging from micro-seconds to hours and days [14].

To exploit the features of neuromorphic spiking hardware to their fullest extent, there are therefore two tightly interlinked critical challenges that need to be addressed in conjunction: (i) the development of a radically different theory of computation that combines the use of fading memory traces and non-linear dynamics with local spike-based learning mechanisms, and (ii) the development of both volatile and non-volatile memory technologies, compatible with CMOS analog circuits, that support the theories developed.

**Advances in Science and Technology to Meet Challenges**

The challenges that the Neuromorphic Computing and Engineering field faces, for understanding how to best design signal processing and online learning mechanisms in hardware spiking neural networks for solving complex problems might seem insurmountable. However, biological brains are an existence proof that robust and stable computation can be achieved, using a computing substrate that is analog, and that uses inhomogeneous and imprecise signal processing elements. So understanding how animal brains, even small insect brains with fewer than one million neurons, manage to achieve this tasks will be key for making progress also in this domain. Specifically, the advances in science that are required to meet these challenges need a strong interdisciplinary approach. Advances in theoretical and computational neuroscience will provide a core component. But these will need to be complemented with notions and results from multiple sub-fields of electrical engineering, such as information theory, signal processing, and control theory, as well as other disciplines such as mathematics, computer science, and robotics.

In parallel, it is clear that the technology used today in conventional computing systems is not ideally suited for building brain-inspired neuromorphic hardware. Emerging nanoscale memristive memory technologies represent a promising development that can provide solid-state electronic elements able to emulate different properties of biological synapses and neurons. Important advances in technology required to meet the challenges outlined above can indeed be provided by the development of both volatile and non-volatile memristive devices with characteristics that are compatible with the specifications provided by the theory. However, to mass-produce neuromorphic computing systems at scale, it is important that these device are compatible with standard CMOS fabrication processes. As one of the most important features of existing CMOS circuits used for implementing spiking neural network is their ability to perform computations using extremely small amounts of power [6], it will be important that the memristive devices designed to be co-integrated with these circuits do not require large amounts of currents to change conductance.

As the development of neuromorphic circuits and memristive devices for building hardware spiking neural networks to carry out computation is a very recent phenomenon, there is a unique opportunity for making concrete progress toward meeting the challenges faced by following, a co-design approach that drives both advances in science and in technology together.

**Concluding Remarks**

Implementing spiking neural networks computing systems with analog CMOS circuits and memristive devices is hard. Besides few examples of proof-of-concept systems that have been applied to very specific tasks, such as sensory processing or spatio-temporal pattern recognition for bio-medical applications [19–24], no general purpose solution exists yet. More importantly, no well established formal methodology exists for automatically designing or programming them.

However we are witnessing incredible progress being made independently in artificial neural networks, in machine learning, in neuroscience, and in memory technology developments. In addition, there is a large demand for the development of novel low-power computing technologies for applications "at the edge", i.e., applications that need to process data measured locally, without connecting to remote servers on the internet, often with low latency and in compact packages.

So brain-inspired approaches for building such technologies are extremely promising, and the potential of research and development in memristive/CMOS spiking neural networks computing systems is extremely high.

**Acknowledgements**

This paper is supported in part by the European Union's Horizon 2020 ERC project NeuroAgents (Grant No. 724295), and in part by the European Union's Horizon 2020 research and innovation programme under grant agreement No.871371 (project MeMScales).

## 2.3 – Emerging Hardware Approaches for Optimization

John Paul Strachan[1*], Suman Datta[2]
[1]Hewlett Packard Laboratories, Hewlett Packard Enterprise, San Jose, CA, USA
[2]Department of Electrical Engineering, University of Notre Dame, Notre Dame, IN, USA
*Currently at Peter Grünberg Institute (PGI-14), Forschungszentrum Jülich GmbH, Jülich, Germany and RWTH Aachen University, Aachen, Germany
Email: j.strachan@fz-juelich.de, sdatta@nd.edu

**Status**

This perspective outlines a roadmap of emerging hardware approaches that utilize neuromorphic and physics-inspired principles to solve combinatorial optimization problems faster and more efficiently than traditional CMOS in von Neumann architectures. Optimization problems are ubiquitous in modern society, needed in training artificial neural networks, building optimal schedules (e.g., airlines), allocating finite resources, drug discovery, path planning (VLSI and shipping), cryptography, and graph analytics problems (social networks, internet search). Such problems are often extremely challenging, requiring compute resources that scale exponentially with the problem size (i.e., NP-complete or NP-hard complexity). Mathematically, in a combinatorial optimization problem [1] one has a pre-defined cost function, $c(x)$, that maps from a discrete domain $\vec{X}$ (nodes, vectors, graph objects) to $\mathbb{R}$, the real number space, and the goal is to find the $\vec{x}_{opt}$ that achieves the globally optimum cost value $c_{min}(\vec{x}_{opt})$.

While exact methods for solving optimization problems have been developed, these can be too time-consuming for challenging or even modest-sized instances. Instead, there is steadily rising popularity for faster meta-heuristic approaches, such as simulated annealing [2] and evolutionary algorithms [3], computing models such as Boltzmann machines [4], Ising models [5][6], and variations of Hopfield networks [7]. These take inspiration from physical and biological systems which solve optimization problems (Fig 1) spontaneously. Many naturally-occurring phenomena, including the trajectories of baseballs and shapes taken by amoeba, are driven to extrema of objective functions by following simple principles (e.g., *least action* or *minimum power dissipation* [8]). In one example, proteins, which are long chains of amino acids, can contort into an exponentially large number of different shapes, yet they repeatably stabilize into a fixed shape on the time-scale of milliseconds. For a protein composed of only 100 peptide bonds, it is estimated that there are over $10^{300}$ different conformal shapes. Even exploring one every picosecond ($10^{-12}$ sec), takes more than the age of the universe to explore them all. Instead, nature uses efficient dynamics to arrive at a solution in less than a second.

**Current and Future Challenges**

Neuromorphic and physics-based hardware complements the meta-heuristic algorithms and models described above. Emerging hardware approaches include Quantum-based Annealers [9][10], Optical or Coherent Ising Machines [11][12][13], CMOS-based digital annealers [14][15][16], analog resistive memory-based systems [17][18], coupled oscillators [8][19][20][21], and probabilistic bit logic [22][23]. Some of these techniques are illustrated in **Figure 1**, and **Table 1** highlights some of their respective features, strengths, and challenges.

We highlight and compare some of today's emerging approaches on the Max-cut benchmark (partitioning a graph). D-Wave's quantum annealer uses low-temperature superconducting devices comprising 5,000 bits in a sparsely coupled network. Prior work shows solving a 200 node cubic graph in 11ms, consuming around 25kW power to operate at cryogenic temperature. The coherent Ising machine (CIM) based on optical parametric oscillators is another approach, where a fully connected 2,000 node CIM uses a kilometer long fiber cavity to accommodate the degenerate optical parametric oscillator (DOPO) pulses and takes about 50ms to solve Max-Cut problem on a 200 node cubic graph. CIM utilizes measurement-and-feedback schemes for coupling the spins, that is provided by traditional CMOS based field-programmable-gate-arrays (FPGAs). Several CMOS-based digital or mixed-signal hardware accelerators have also been developed. The Ising chip demonstrated by Hitachi uses CMOS static random access memory (SRAM) cells as spins while the coupling is realized using digital logic gates, used to implement the simulated annealing algorithm. Due to its non-von Neumann architecture, it exhibits a 50x lower energy-to-solution over that of a CPU running a greedy algorithm to find Max-Cut of a 200 node random cubic graph. The energy dissipation is still orders of magnitude more than other Ising solvers recently implemented with emerging devices such as resistive RAM-based cross-bar arrays and insulator-to-metal phase transition (IMT)-based coupled oscillator arrays. Analog in-memory computing using RRAM crossbar arrays has been utilized to demonstrate four orders of magnitude improvement in energy over CPUs [18]. IMT nano-oscillator-based Ising solvers may exhibit even lower energy consumption, primarily due to the ultra-low power dissipation of the IMT oscillators. The IMT oscillator network implements bi-directional ferromagnetic and anti-ferromagnetic coupling using simple electrical elements such as resistance and capacitance, and can achieve highly parallelized all-to-all connectivity. The analog or continuous-time dynamics of these Ising solvers has an inherent advantage of parallelism which lowers the time to solution compared to CMOS annealers and CPUs operating in discrete time. The time-to-solution (or cycles-to-solutions) remains similar for both the IMT solver and RRAM-based hardware accelerator.

**Advances in Science and Technology to Meet Challenges**

**Table 1** summarizes some of the current capabilities, strengths, and challenges to the highlighted emerging hardware approaches for optimization. In this section, we consider challenges shared by all approaches and the advances needed to address them.

At the lowest level, there is substantial room to improve both the devices used in the solution state (or neuron) representation and in the connections between them. Whether a magnetic, electronic, or optical medium is used, these provide the core computational elements. Desired properties include multi-bit levels (many problem beyond Max-cut are no longer binary), high endurance and robustness, low variability, and rapid re-programmability. Material engineering targeting these properties is needed, and ideally kept compatible with today's CMOS technology for future integration and mass production.

To address important applications of the future, all hardware solutions will need to be able to scale to many thousands of variables and constraints. For example, railway companies operate tens of thousands of trains per week for which optimal crew schedules are desired [24]. Supporting this necessitates breaking large problems across many processing units that must communicate at low latency, low energy, and high bandwidth. While the efficiency of the processing sub-units may be very high, this communication quickly becomes the bottleneck and Amdahl's law limits the benefits of the core processing circuits if there is a high communication overhead. Related to this challenge is the inherent connectivity for the processing units. With limited connections between the nodes (e.g., only nearest or next-nearest neighbours), "embeddings" are used to solve problems with higher density connections, and these come with exponential penalties [25]. Maintaining efficient connectivity for large problems that must span many processing units will challenge all of today's emerging approaches. Indeed, the massive fan-out capabilities in biological nervous systems averaging >10,000 connections between neurons shows how nature has addressed this problem and future neuromorphic systems tackling optimization may need to mimic these designs.

With increasing scale and fan-out there arises the inevitable challenge of significant device parasitics and variability. Non-idealities include interconnect/wire parasitics in terms of line-to-ground capacitance, line-to-line capacitance and frequency variability for the oscillator approaches. With increasing problem size and the concurrent increase in the size of the network, it will be increasingly difficult to find the globally optimal solution. The reduction in success probability can be mitigated by increasing the number of anneal cycles and/or executing larger trial batches, but only at the expense of time-to-solution. An alternate approach could be to exploit emerging monolithic three-dimensional integration technology that provides multiple tiers of interconnect that can be dynamically configured to provide an efficient, scalable and dense network on chip. This promising direction will provide new architectural opportunities for on-chip implementation of large dense networks with programmable connections that are beyond the capabilities of existing process and packaging technologies today.

We stress that optimization problems are highly diverse, and even within a problem category (e.g., scheduling) specific instances can have different traits and levels of difficulty, such as the characteristic scale of barriers between minima, the density of saddle points, or the relative closeness in value between local and global minima. Consequently, domain experts have developed techniques highly tailored to their problem class. This could entail parameter choices such as using different noise distributions or cooling schedules (simulated annealing), to algorithmic variations such as ensembles of models exchanging temperatures (parallel tempering) or populations exchanging and mutating characteristics (genetic algorithms). Thus, it is desired for any emerging hardware to support these rich variations as much as possible, exposing internal parameters to the user for control, as well as provisioning the architecture to efficiently realize the more promising algorithmic variations.

Many optimization problems may also involve substantial pre- and post-processing computations. For example, transforming a practical airline crew scheduling problem into the prototypical NP-hard "set-cover" problem first involves constructing sub-sets from viable rotations. Such pre- and post-processing, let alone mid-stream processing (replica exchange in parallel tempering), requires flexible and complex architectures that include traditional digital units in addition to neuromorphic and physics-based optimization solvers.

The above challenges highlight the need for hardware designs to be algorithm and "software aware." Equally important is the development of algorithms and tools that are strongly "hardware aware." These must be designed to exploit the strengths of the underlying processing units—such as cheap stochasticity or certain types of parallelism [26] —while, simultaneously mitigating their respective weaknesses—such as reduced precision. Thus, constructing successful systems for optimization solving will require a deep co-design from materials, devices, and packaging, all the way up to algorithms and software tools.

**Concluding Remarks**

This perspective has highlighted the promise in leveraging physics- and brain-inspired principles to tackle today's most intractable computational problems. There are future challenges spanning all levels of the computing stack. Highly diverse approaches are being explored leveraging electronic, magnetic, optical, or quantum systems. Ultimately, we expect the winning approach will be the one that gathers an excited community of users by building flexible, performant, and reliable optimization solvers, and thus begin the virtuous relationships between users, software, and hardware researchers.

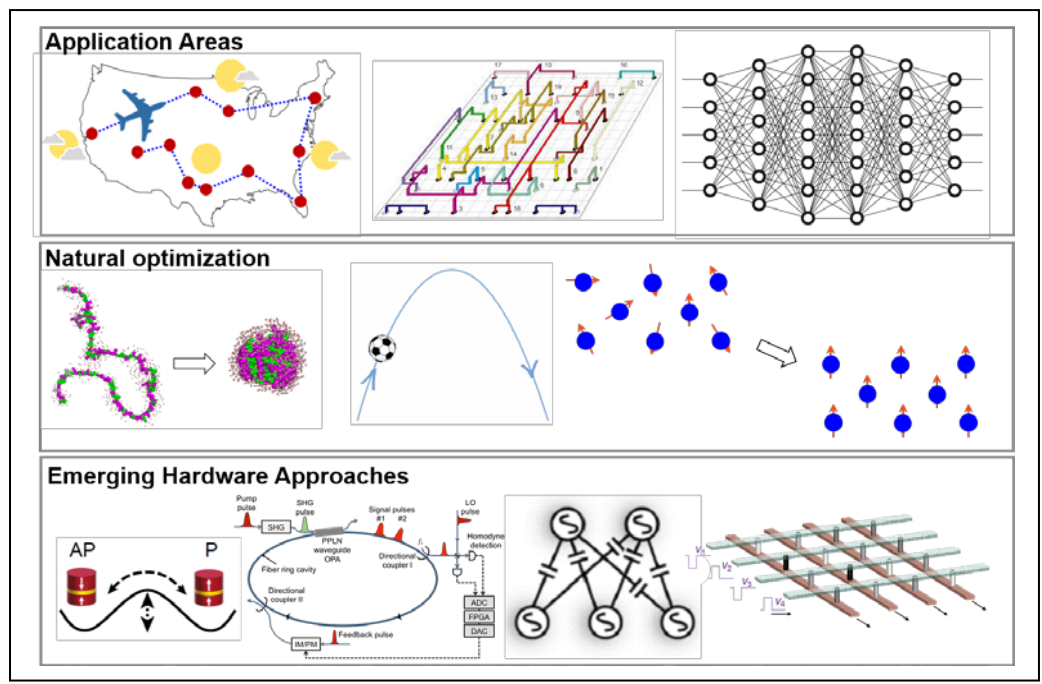

**Figure 1.** Optimization in society and nature. **Top row**: important application areas include flight scheduling, VLSI routing, training artificial neural networks. **Middle row**: optimization in nature includes protein folding (see text), object motion obeying the principle of least action, and the orientation of magnetic spins in a crystal. **Bottom row**: some highlighted emerging hardware approaches include probabilistic logic bits implemented with Magnetic Tunnel Junctions (MTJ) [21], Coherent Ising Machine [12], coupled oscillators [19], and analog in-memory computing [17].

|  | Digital CMOS Annealer | Optical Ising Machine | Analog Resistive Crossbar | Coupled Oscillators | Probabilistic Logic Bits (P-bit) | Quantum Annealing |
|---|---|---|---|---|---|---|
| **Bit Precision** | Up to 64 bit | Binary (>5 bit with Integ. photonics) | Analog, or >5 bit | Binary | Binary | Analog |
| **Problem Re-Programmability** | Easy | Hard | Moderate | Hard | N/A | Moderate |
| **Dense problem sizes demonstrated** | 8192 | 2000 | 60 | 8 | 6 | 60 |
| **Performant Architecture Developed for Large-scale** | Yes | No | No | No | No | No |
| **Room Temperature Operation** | Yes | Yes | Yes | Yes | Yes | No |
| **Strengths** | Mature technology, flexibility | Low latency, high speed, wavelength multiplexing capability | High density, continuous value and time capability | Low energy, continuous-time capability | Inherent stochasticity | Leverage quantum resources, sophisticated control, software, and tools developed |
| **Main Challenges Short Term** | Noise Injection | Power consumption of control system, photo-detectors, nonlinear operations | Device yield, repeatability, analog device drift | Frequency variability and drift | Integrating magnetic materials; multi-bit | Embeddings for dense connectivity |

| Main Challenges Long Term | Interconnect and comm. bottleneck, CMOS scaling, 3D integration | Component sizes, silicon integration, non-volatile storage of weights | Fully passive arrays, interconnect and parasitics, Digital conversion | Continuous time stability, interconnect and parasitics | Only stochastic neurons, requires architectural integration with another approach | Cryogenic requirement, scaling of superconducting integrated circuits |
|---|---|---|---|---|---|---|

Table 1. State-of-the-art and opportunities for emerging hardware approaches to optimization.

**Acknowledgements**

The authors acknowledge helpful input for figure and table content from Thomas Van Vaerenbergh and Suhas Kumar.

## 2.4 – Enabling technologies for the future heterogeneous neuromorphic accelerators

Elisa Vianello, CEA, LETI, Université Grenoble Alpes, Grenoble, France
Alexandre Valentian, CEA, LIST, Université Grenoble Alpes, Grenoble, France

**Status**

Artificial Intelligence (AI) and in particular Artificial Neural Networks (ANNs) have demonstrated amazing results in a wide range of pattern recognition tasks including machine vision, natural language processing, and speech recognition. ANN hardware accelerators place significant demands on both storage and computation. Today's computing architectures cannot efficiently handle AI tasks: the energy costs of transferring data between memory and processor at the highest possible rates are unsustainably high. As a result, the development of radically different chip architectures and device technologies is fundamental to bring AI to power-constrained applications, such as Data of Things, combining Edge analytics with IOT.

**Current and Future Challenges**

Most of the AI applications today are still running in the Cloud, *i.e.* in data centers which offer large storage capacity and processing power [1]. The learning phase, for which you need large datasets, is done in the Cloud, and inference tasks are performed on the same assets. Such a scheme is not sustainable in the long run: with the increasing demand for intelligent devices, the data centers will not be able to sustain the load. Part of it will have to be offloaded to the devices themselves. This is the **current challenge** that companies and research teams are taking on: to enable running inference tasks at the edge, thanks to dedicated hardware accelerators. The main pathway for the implementation of such constrained systems is the reduction of the power consumption. This also relates to longer battery life, and to heat dissipation, which cannot be afforded. Since the complexity of neural networks tends to grow over time, this problem will only become more acute. In data-centric application such as AI, the main source of power consumption is data movement [2]. It costs hundreds of times more energy to move data to/from an external memory than to compute on it. One of the solutions is to handle lightweight data, which is why quantization of weights and activations is a field of active research, down to the extreme of binary neural networks. Second, the integration of dense memories, as close as possible to the processing engines (PE) exploiting new memory technologies and 3D integration schemes is mandatory. Some promising solutions are presented in the next section. The ultimate evolution consists in removing the PEs altogether and computing directly inside the memory: this is the In-Memory-Compute paradigm.

Having accelerators dedicated to inference tasks at the edge is only the first step. The **future challenge** will be to perform the learning phase locally as well. Autonomous agents will need to learn from experience, in order to adapt to their environment, learn basic principles and rules, infer common sense. They will thus have to exhibit lifelong learning abilities. Such accelerators will have additional challenges, linked to the learning phase itself. For instance, the state-of-the-art method to train Artificial Neural Networks is the back-propagation algorithm, which minimizes a given loss function based on gradient descent [3]. However, its implementation on GPUs is energy-consuming and it does not satisfy edge requirements.

**Advances in Science and Technology to Meet Challenges**

Artificial Intelligent (AI) systems are data-hungry: first, large amounts of memory are required for storing network states and parameters; second, the energy cost associated to data transfer between the memory and the processor is the major source of energy dissipation. Resistive memory devices, also referred as memristors, can provide massive on-chip data storage with low voltage and low-latency accessibility. Their basic working principle relies on the modification of the material at atomic level causing a change of resistance. These memories include resistive-switching random access memory (RRAM), phase-change memory (PCM), magnetic random-access memory (MRAM), and the ferroelectric random access memory (FeRAM). They are currently implemented as a 1T1R structure, *i.e.* with one MOS transistor (1T) used for accessing one resistor (1R). The memory cell footprint is around $40F^2$ and is limited by the access transistor. Promising results have been recently demonstrated to increase the memory density: using one transistor to access multiple RRAMs (1T4R) [4], stacking multiple 1T1R thanks to monolithic 3D technology [5], replacing the MOS transistor by a stacked nanowire transistor [6] or by a backend selector [7]. This last technology option enables efficient crossbar arrays with low leakage currents, leading to the highest density, with a footprint of $4F^2$. Another promising approach to achieve high density is to increase the number of resistance levels in a single cell, for storing multiple bits. RRAM arrays storing up to three bits were demonstrated [5, 8]. Those multiple level cells not only increase the density, they are also a great mean for efficiently implementing In-Memory-Compute functions: provided that the resistance levels are linearly allocated, matrix-vector multiplications and accumulation (MAC) of inputs can be done in an analog manner, simply exploiting Ohm's and Kirchhoff's laws [9, 10].

However, on-chip learning based on back-propagation remains a challenge on memristor-based hardware, because of non-ideal device properties: limited endurance, non-linear conductance modulation, as well as device variability [11]. Moreover, back-propagation does not allow continuous, incremental learning. This is tackled at technology level, but also at algorithm level. To achieve unsupervised learning, there have been multiple efforts implementing biologically plausible Spike Timing Dependent Plasticity (STDP)-variants and Hebbian learning using neuromorphic processors [12, 13]. However, they are local learning rules and they do not provide any guarantee that network performance will improve in multilayer or recurrent networks. Novel algorithms where both inference and learning could fully be achieved out of core physics have been recently proposed such as, three-factor spike-based learning rules [14, 15], Direct-Feedback Alignment [16], Equilibrium Propagation (*Eq-prop*) [17]. Next, these algorithms have to be mapped on real hardware tacking into account device non-idealities. Some recent works explored this direction. For instance, it has been demonstrated that drift behavior of PCM devices (generally considered a device non-ideality) can be exploited to implement long lasting eligibility traces, a critical ingredient of three-factor learning rules [18]. A machine learning scheme that leverages the RRAM variability to implement Markov Chain Monte Carlo (MCMC) sampling techniques to enable on-chip learning in Bayesian neural network has been demonstrated [19].

As can be imagined, many actors will provide accelerators for running AI workloads at the edge. They will use a combination of technologies, and different coding strategies: classical coding (FP32, BF16, INT8), spike coding (Rate, Temporal, Time-To-First-Spike), a combination of both. Since there is never one-size-fits-all solution, it is strongly believed that heterogeneous AI systems will play an important role for delivering products tailored to different application needs, as illustrated in Figure 1. Such systems will leverage IP reuse thanks to 2.5D and 3D integration technologies, thus reducing non-recurring engineering (NRE) costs and time to market. Those technologies enable the integration of different chips, also known as chiplet [20], built with different manufacturing processes. Those chiplets are co-integrated on silicon or glass substrates, or an organic interposers. Communication with the outside world is ensured thanks to through-silicon-vias (TSV) or through-mold-vias (TMV) and redistribution layers.

For fulfilling that vision, issues such as interoperability and application mapping need to be addressed. A high-level AI system description language will be specified, for expressing the respective hardware resources and their interactions. Deep learning frameworks will be complemented with export capabilities on those heterogeneous platforms.

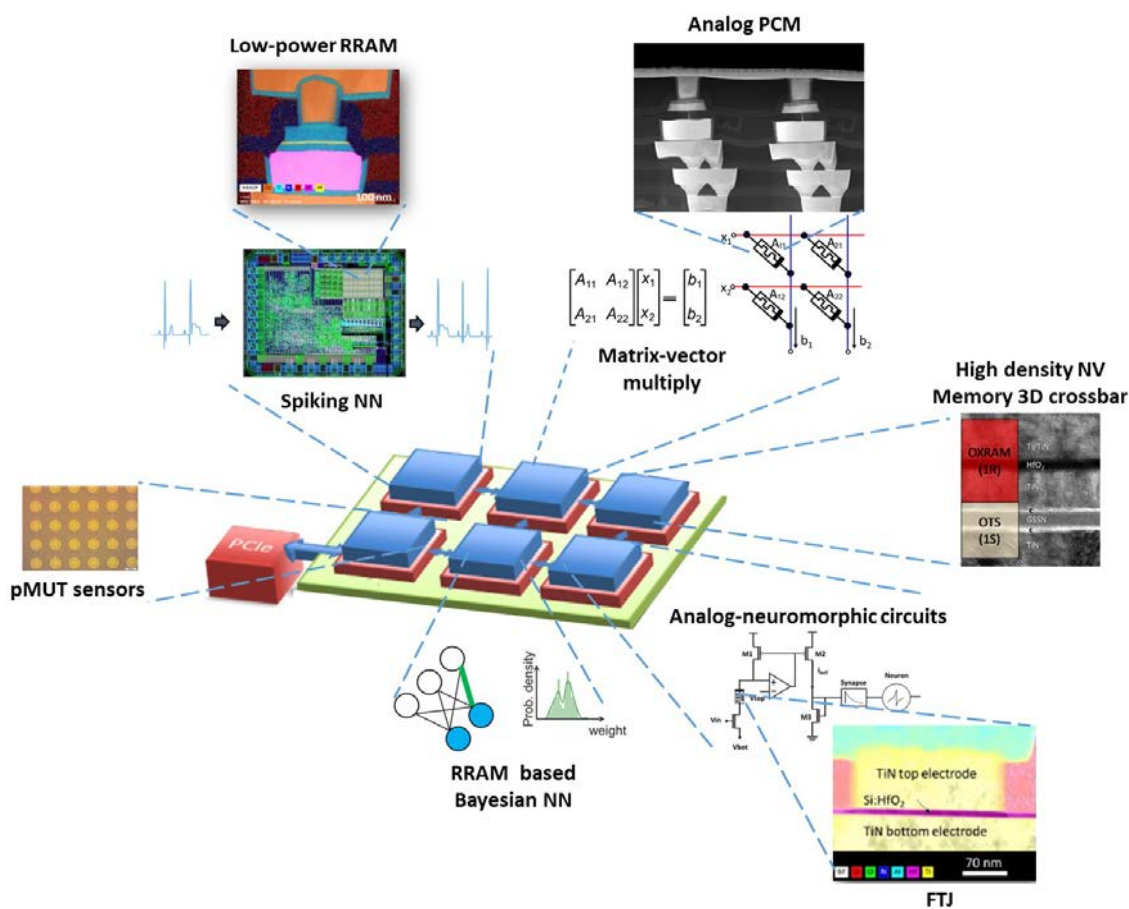

Figure 1. Modular AI systems composed of heterogeneous components – each of which being optimized for a specific task and exploiting different technology solutions.

**Concluding Remarks**

Artificial Intelligence and in particular machine learning have made tremendous progress in image, video, sound and speech recognition tasks. The next challenge is enable unsupervised and continuous learning at the edge. Innovations ranging from single memory devices to full-scale architectures exploiting heterogeneous integration are required to meet the computational needs of such applications.

**Acknowledgements**

*We acknowledge funding support from the H2020 MeM-Scales project (871371), the ECSEL TEMPO project (826655) and the ECSEL ANDANTE project (876925).*

## 2.5 - Photonics

*J. Feldmann[1], X. Li[1], W.H.P. Pernice[2,3] and H. Bhaskaran[1]*

[1] Department of Materials, University of Oxford, Parks Road, OX1 3PH Oxford, UK
[2] Institute of Physics, University of Münster, Heisenbergstr. 11, 48149 Münster, Germany
[3] Center for Soft Nanoscience, University of Münster, 48149 Münster, Germany

**Status**

The field of optical computing began with the development of the laser in 1960 and has since been followed by many inventions especially from the 1980s demonstrating optical pattern recognition and optical Fourier-transform processing [1]. Although these optical processors never evolved to commercial products due to a limited application space and the high competition with emerging electronic computers, photonic computing again gained much interest in recent years to overcome the bottlenecks of electronic computing in the field of artificial intelligence, where large datasets must be processed energy efficiently and at high speeds [2]. Optical computers are able to seriously challenge electronic implementations in these domains, particularly in throughput. Photonics further has allowed one to integrate optics on-chip enabling such optical neuromorphic processors to have several advantages compared to their electronic counterparts. One of them is based on the fact that photons are bosons and are able to occupy the same physical location (i.e. not subject to the Pauli exclusion principle). Thus, many can be transmitted through the same channel without mutual interference. This offers an intrinsically high degree of parallelization by wavelength and mode multiplexing techniques, enabling the use of the same physical processor to carry out multiple operations in parallel leading to high computing densities. Additionally, the data transport problem that is apparent in electronics at high signal speeds is easily addressed using photonic waveguides that serve as low power data links. Taken together with the fact that linear operations can be implemented in the optical domain with very high energy efficiency [3], photonics offers a promising platform for high speed and highly parallelised neuromorphic computing [4].

Many non-von Neumann photonic computing techniques have been demonstrated using integrated, fibre-based and free-space optics [3], showing a large variety of different approaches ranging from coherent neural networks [5], reservoir computing [6] and phase-change photonics [7]–[9] to hardware accelerators for the main computational bottlenecks (usually matrix multiplications) in conventional AI solutions [10], [11]. Most of these are analogous to in-memory computing that has most prominently been developed by IBM [12], [13]. Further advances in photonic computing might first lead to optical co-processors that accelerate specific operations such as vector-matrix multiplications and are implemented together with conventional electronic processors. The next step could be photonic neuromorphic computers avoiding electro-optic conversions.

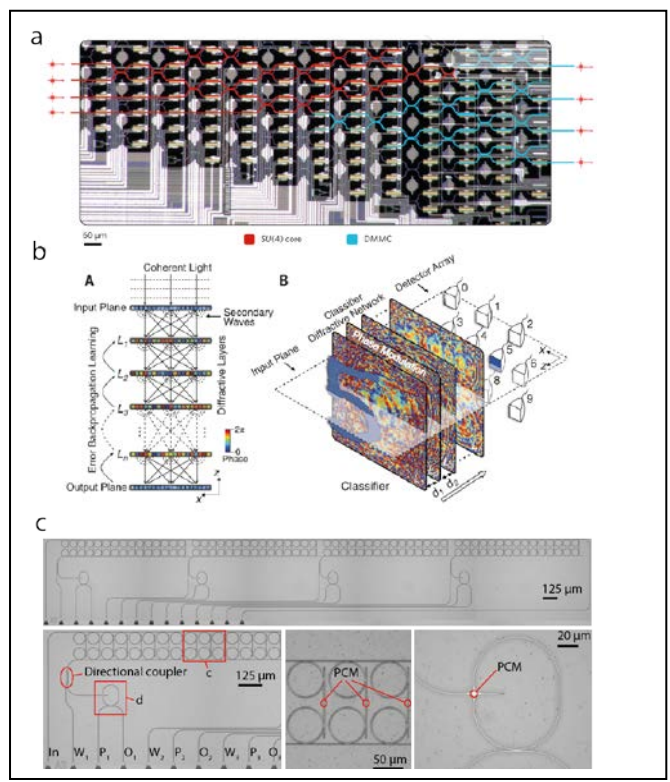

**Figure 1.** Different implementations of neuromorphic photonic circuits. a) Coherent matrix multiplication unit based on MZIs [5]. Diffractive deep neural network [3]. c) All-optical neural network using phase-change materials [7].

**Current and Future Challenges**

The current main limitations of neuromorphic photonics lie in scalability, system stability and interfacing with electronic systems. As most approaches that rely on analogue computing, the precision of photonic approaches depends on the noise accumulated in the processor.

Whereas free-space implementations are often bulky and therefore difficult to scale, the challenges for integrated circuits are optical loss and the reliable fabrication of the individual components. Especially tuning the fabrication of resonant elements as ring resonators that are used for fast signal modulation and multiplexing require thermal tuning in order to operate properly. To gain from wavelength multiplexing and achieve highly parallel processing, the fabrication techniques need to provide sufficient reproducibility of the wavelength channels of the multiplexers.

In addition to fabrication challenges, many neuromorphic photonic processors rely on coherent light and require precise control of the optical phase. Because the phase is strongly temperature dependent, excellent thermal stability of the system is necessary, which in itself can become a significant component of the overall power consumption of the system. Another challenge for integrated photonic circuits is the integration of light sources on CMOS compatible silicon platforms.

Whereas linear operations can be carried out in photonic circuits intrinsically very well, all-optical non-linear elements working at sufficiently low power and high speeds are more challenging to develop. However, from a strictly neuromorphic computing standpoint this is crucial, as non-linear elements determine spiking functionality in neurosynaptic networks, at least at the device level.

From the overall system architecture standpoint for neuromorphic photonic computing, additional challenges regarding integration with conventional technologies exist. Given that most information processing is performed in the electronic domain, photonic neuromorphic systems usually have to be interfaced with electronic processors, e.g. to input the data or store the results of the computation. This imposes at least two electro-optic conversions, which decrease the overall energy efficiency of the system. As most neuromorphic processors are a type of analogue computing, additionally conversions between the digital and analogue domain have to be performed. Especially analogue to digital converters (ADC) can make up a huge part of the power budget and scale badly in terms of energy with the number of bits and operation speed [14]. To be able to use the high modulation speeds accessible in modulating and detecting optical signals, significant improvements in digitizing the results of the computation have to be made.

**Advances in Science and Technology to Meet Challenges**

Like their electronic counterparts, photonic neuromorphic processors require precise fabrication. With increased interest in photonic information processing, several photonic foundry services are emerging and are continuing to enhance their capabilities. These will be crucial to progress this field. Although vastly improved with good performance, the existing capabilities in photonic foundries are well behind those that exist in the more mature and established electronics. Improvements in fabrication techniques will give way to less variation in device specifications, e.g. in the wavelength specification of certain components like resonators or multiplexers and reduction of optical loss. This will be important to improve on the parameters that make photonic neuromorphic processors more advantageous, specifically the ability to wavelength multiplex. A useful tool in the fabrication process could be an additional tuning step after fabrication, to match the designed specifications such as measuring the resonance wavelength of a resonator and adjusting it to the desired wavelength as a post-processing correction. Advances in the standard components as modulators and detectors as well as the addition of new components to the libraries of photonic foundries and the development of new materials for non-volatile optical storage will enhance this field and bring these circuits closer to commercialization.

Yet another crucial component is efficient light sources that can be integrated on photonics alongside reliable many-channel on-chip multiplexers. Integrated optical frequency combs that provide a wide optical spectrum with a fixed channel spacing that can be exploited for computing as a coherent light source are a prime example of this [15]. Photonic neuromorphic circuits rely on electronic control and therefore improvements in high-speed electronic components such as digital-to analogue (DAC) and analogue-to digital converters (ADC) are also very important. Further research could also lead to all-optical DACs and ADCs circumventing the need for electro-optic conversions. In general, photonic neuromorphic processors that minimize conversions between the digital and analogue domains are preferable. A specific class of neural networks that could prove especially suitable for low power photonic processing are spiking neural networks, that reduce digital-to-analogue conversions by using binary spikes and their time dependence as information carriers.

As the non-linear optical coefficients for silicon are small, functional materials that allow for such non-linearity or other added functionality are also important [16]. A promising class of materials are phase-change materials (PCMs) that switch their optical properties upon excitation and therefore effectively resemble a non-linear element [17], [18]. Although PCMs can be switched with low optical powers, significant improvements have to be made in increasing the switching speed in order to keep up with high modulation speeds enabled by photonics. Another class of materials considered for low power optical non-linearities are epsilon-near-zero materials [19].

Operating with analogue signals results in a higher sensitivity to noise; recent advances in reducing the precision of neural networks to lower numbers of bits with low loss in prediction accuracy is one step to overcome this challenge [20] and further research in this area is also required.

As photonic integrated circuits become more and more complex, similar to electronics, a three-dimensional implementation seems necessary to avoid crosstalk and loss when routing the signals and avoid waveguide crossings, which also requires investigation.

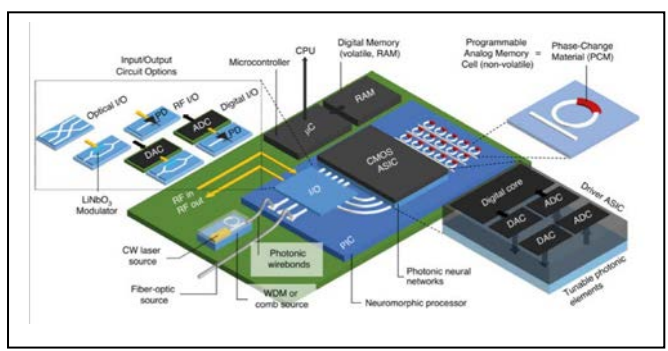

Figure 2. Potential architecture of a photonic neuromorphic architecture (from [4])

**Concluding Remarks**

Photonic computing has been a promising field of research over the last decade as photonics matures. New challenges in handling huge amounts of data in the fields of artificial intelligence and machine learning that bring conventional electronic processors to their limits have resulted in a surge interest in photonic computing. This is because of their inherent advantage that enables high throughput, a high degree of parallelization together with the ability to carry out linear operations at very low energies. This makes photonic neuromorphic processors a very promising route to tackle the upcoming challenges in AI applications.

In spite of the challenges photonic computing concepts that can overcome the limitations of electronic processors have been demonstrated in the recent years, and a roadmap to address their march into commercialization would be a huge benefit to society.

**Acknowledgements**

This research was supported by EPSRC via grants EP/J018694/1, EP/M015173/1, and EP/M015130/1 in the United Kingdom and the Deutsche Forschungsgemeinschaft (DFG) grant PE 1832/5-1 in Germany. WHPP gratefully acknowledges support by the European Research Council through grant 724707. We further acknowledge funding for this work from the European Union's Horizon 2020 Research and Innovation Program (Fun-COMP project, #780848).

## 2.6 – Large-Scale Neuromorphic Computing Platforms
Steve Furber, The University of Manchester, UK

**Status**

The last decade has seen the development of a number of large-scale neuromorphic computing platforms. Notable among these are the SpiNNaker [1] and BrainScaleS [2] systems, developed prior to, but supported under the auspices of, the EU Flagship Human Brain Project, and somewhat later the Intel Loihi [3] system. These systems all have large-scale implementations and widespread user communities.

All three systems are based upon conventional CMOS technology but with different architectural approaches. SpiNNaker uses a large array of conventional small, embedded processors connected through a bespoke packet-switched fabric designed to support large-scale spiking neural networks in biological real time and optimised for brain modelling applications. BrainScaleS uses above threshold analogue circuits to model neurons running 10,000 times faster than biology, implemented on a wafer-scale substrate, optimised for experiments involving accelerated learning. Loihi sits somewhere between these two, using a large array of asynchronous digital hardware engines for modelling and generally running somewhat faster than biological real time, with the primary purpose of accelerating research to enable the commercial adaptation of future neuromorphic technology.

In order to support their respective user communities these systems have extensive software stacks, allowing users to describe their models in a high-level neural modelling language such as PyNN [4] (used for both SpiNNaker and BrainScaleS) so that straightforward applications can be developed without a detailed understanding of the underlying hardware.

These large-scale systems have been up and running reliably for some time, supporting large user communities, and offer readily accessible platforms for experiments in neuromorphic computing. Access to neuromorphic technology is no longer a limiting factor for those who wish to explore its potential and capabilities, including using these existing platforms to model future technologies.

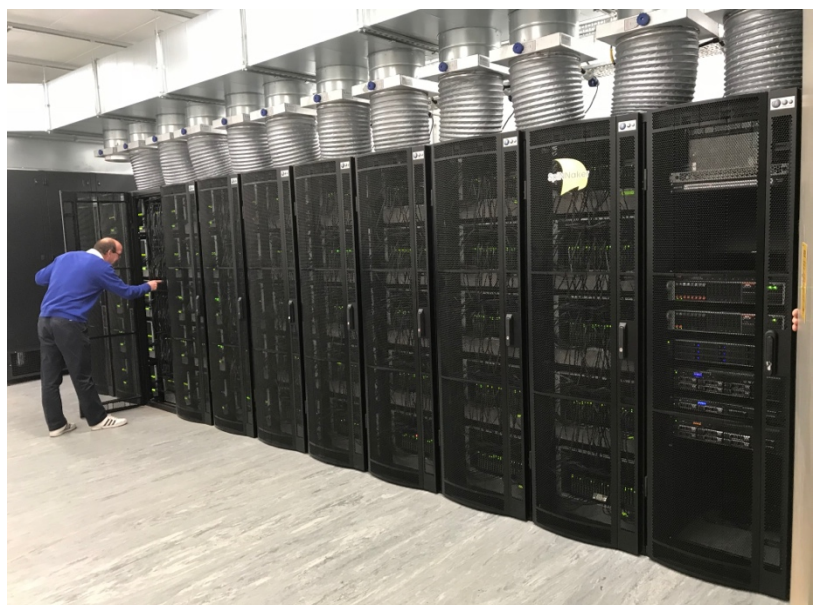

**Figure 1.** The million-core SpiNNaker machine at the University of Manchester, occupying 10 rack cabinets with an 11[th] cabinet (shown on the right) containing the associated servers.

**Current and Future Challenges**

The major challenge for neuromorphic computing technology, if it is to gain widespread adoption, is to offer a compelling demonstration of its commercial viability. It must demonstrate some significant advantage over competing technologies in terms of its capabilities, performance and/or energy-efficiency. The applicability of neuromorphic technologies to brain science is easier to justify, though even there it is competing with more conventional computing technologies such as GPUs [5] and HPC [6], and clear wins [7] are hard to come by. But brain science is a niche research application, and something with much broader applicability is required.

Although there have been a few demonstrations of neuromorphic superiority in application domains such as keyword recognition [8] these are still rather thin on the ground, but there is growing confidence that something will emerge in the near future. This confidence is underlined by the growing number of venture capital backed start-up companies in the neuromorphic arena. Most of these are focussed on small-scale AI applications in edge computing, though some such as SpiNNcloud Systems GmbH have an eye on data centre scale neuromorphic technology, in this case based on the second-generation SpiNNaker2 chip [9].

Large-scale systems of spiking neurons amplify many of the challenges faced by smaller-scale systems in areas such as the design of homeostatic mechanisms that control the operating point of a neural population, adaptation mechanisms that compensate for local failures, unsupervised learning mechanisms that allow the system to focus resources around the statistical distribution of observed inputs, synaptogenesis, neurogenesis, on-line learning, and so on. But as the system scales up, the need for built-in automatic adaptation only increases as the time and cost of optimising by hand increases rapidly beyond reasonable bounds.

A challenge for the future will be to combine the scalability demonstrated by the current large-scale neuromorphic systems with the enhanced characteristics offered by the advanced device technologies described elsewhere in this review.

**Advances in Science and Technology to Meet Challenges**

Given the challenges that face neuromorphic technologies at all scales, where should we look for inspiration for the solutions to those challenges? There are two obvious places to seek such inspiration: biology, and mainstream AI.

Biological brains, from small insects up to human scale, demonstrate the flexibility, adaptability and energy-efficiency that we aspire to for our engineered systems, if only we could understand how they work. This is why neuromorphic engineers track advances in brain science, exploring the potential of advances such as new insights into dendritic computation to improve the capabilities of engineered systems.

Similarly, the explosion over the last decade of applications of AI based upon artificial neural networks offers insights into the effective organisation of neurons, whether spiking or not. There is a strong sense that the success of artificial neural networks must be telling us something about how brains work, despite the absence of evidence in biology for, for example, the error backpropagation learning mechanism that is so effective in artificial neural networks. Some form of gradient descent (the principle underlying backprop) must be at work in biological learning, and recent developments in algorithms such as e-prop [10] offer a glimpse of how that could work.

The prospect of the convergence of neuromorphic engineering with brain science and mainstream AI is tantalising for all three branches of science/engineering.

**Concluding Remarks**

Large-scale neuromorphic computing platforms offer reliable, well-supported services that can be used to research neuromorphic technologies at the system level at minimal cost. They are accompanied by software stacks that abstract users away from low-level hardware details, allowing them to focus on higher-level issues such as network architectures, learning rules, brain modelling, etc. As such they play an important role in training the next generation in neuromorphic technology and in providing flexible development platforms for exploring neuromorphic system development without the constraints imposed by smaller research prototypes.

**Acknowledgements**

The design and construction of the SpiNNaker machine was supported by EPSRC (the UK Engineering and Physical Sciences Research Council) under grants EP/D07908X/1 and EP/G015740/1. Ongoing development of the software is supported by the EU ICT Flagship Human Brain Project (FP7-604102, H2020-720270, H2020-785907, H2020-945539)

## 3.1 - Learning in Spiking Neural Networks

Emre Neftci,
Peter Grünberg Institute – Neuromorphic Software Ecosystems
Forschungszentrum Jülich, Germany
Faculty of Electrical Engineering and Information Technology
RWTH Aachen, Germany
Department of Cognitive Sciences, Department of Computer Science University of California Irvine

### Status

The dynamical nature of Spiking Neural Networks (SNN) circuits and their spatiotemporal sparsity supported by asynchronous technologies makes them particularly promising for fast and efficient processing of dynamical signals (Sec. 2.2). Here, we discuss learning in SNNs, which refers to the tuning of their states and parameters to learn new behaviors, achieve homeostasis and other basic computations. In biology, this is achieved via local plasticity mechanisms that operate at various spatial and temporal scales. While several neural and synaptic plasticity rules investigated in neurosciences have been implemented in neuromorphic hardware [19], recent work has shown that many of these rules can be captured through three factor rules (3F) of the type [13, 19, 10]

$\Delta W = F_{pre}F_{post}M_{post}$ where factors $F_{pre}$ and $F_{post}$ correspond to functions over presynaptic and postsynaptic states, respectively, and the factor $M_{post}$ is a post-synaptic modulation term (See also sec. 3.4 for a specific example). The modulation is a task-dependent function, which can for example represent error in supervised learning task, surprise in an unsupervised learning task, or reward in reinforcement learning. Given the generality of the three-factor rule in representing existing learning rules and paradigms, this section focuses on the requirements for implementing 3F plasticity in neuromorphic hardware. By analogy to the brain, the most intuitive implementation of synaptic plasticity is on-chip, *i.e.* plasticity is achieved at or the synapse circuit or equivalent circuit near the SNN (Fig. 1, top). Neuromorphic engineers have extensively implemented learning dynamics derived from computational neurosciences, such as Spike Time Driven Plasticity (STDP) variants [1, 5, 22] and more recently, 3F rules [7]. On-chip learning requires precious memory and routing resources [21], which hinders scalability. On digital technologies, this problem can be sidestepped by time-multiplexing a dedicated local plasticity processor [6, 12]. The time-multiplexing approach however suffers from the same caveats as a Von Neumann computer due to the separation between the SNN and the associated plasticity processor. Other promising alternatives for true local plasticity are emerging devices (Sec. 1) and related architectures (Sec. 2.1), which allow storage and computation for plasticity to occur at the same place.

A more practical approach to learning in SNNs is off-chip (Fig. 1, Bottom), which relies on a separate general purpose computer to train a model of the SNN, where memory and computational are potentially more abundant. In this approach, once the SNN is trained, the parameters are then mapped to the hardware. Provided a suitable model of the hardware substrate or a method to convert parameters from a conventional network to an SNN, off-chip learning generally achieves the best inference accuracy on practical tasks [9, 28]. Heterogenous approaches combining on-chip and off-chip approaches (also called chip-in-the-loop, Fig. 1, middle) have been successful at smaller scales [11], although scalability there remains hindered by the access to the local states necessary for plasticity in the chip. The suitability of on-chip or off-chip learning is highly dependent on the task. The former is best for continual learning (Sec 3.4) and the latter is best when a large dataset is already available and the SNN model and parameter conversion are near-exact. If the model is not exact, hybrid learning is often the most suitable method. On-chip and hybrid learning also have the advantage that learning can occur online, *i.e.* during task performance.

Although Hebbian STDP variants have been instrumental for modeling in neuroscience, mathematically rigorous rules derived from task objectives such as 3F rules have a clear upper hand in terms of practical performance [28, 36, 16]. This is arguably because some forms of spatial credit assignment are necessary to learn in non-shallow networks [2]. Thus, we anticipate that mathematically motivated (top-down driven) rules grounded in neuroscience are likely to drive the majority of future research in SNN learning and their neuromorphic implementation. Already today, the success of top-down modeling of learning to efficiently train SNNs ushered in a new wave of inspiration from Machine Learning (ML) [25], and accelerated the quest to build neuromorphic learning machines. In the following, we focus on specific challenges of 3F learning approaches.

## Current and Future Challenges

While the technological advantages of spike-based sensing are now increasingly evident (Sec. 2.2 and Sec. 4), it is natural to question the advantages of SNNs and their related learning algorithms. In search of an answer, it is fruitful to consider some outstanding challenges of state-of-the-art artificial intelligence technologies, which are largely based on deep learning: the mismatch between the training dataset and the real world (generalizability), and their high power requirements [31]. Learning online can overcome the former by learning continuously during task performance, so as to correct errors at the moment when they occur (provided issues related to sequential learning, *e.g.* catastrophic interference, are avoided [17]). On-chip learning on the other hand lends itself to distributing the plasticity processes across the hardware substrate, where it occurs, as is the case for memristor crossbar arrays [27]. Learning then becomes a local process, which can reduce power requirements, provided the necessary learning factors of the learning rule can be computed efficiently. This is possible for example by communicating errors using spikes [32, 20].

Because the challenges of on-chip learning form a superset of those faced by off-chip and hybrid learning strategies, and can solve key problems in state-of-the-art AI, we focus our challenge description on the on-chip case and the 3F rule. SNN learning requires the computation and storage of suitable learning factors, which is dictated by the requirements of space-time credit assignment, *i.e.* how blame or credit should be assigned given the performance on a given task. The form of the 3F rule described above is compatible with this requirement by virtue of the third factor $M_{post}$, which modulates the learning rule according to a (potentially global) learning signal. Phenomenological rules hypothesize the nature of these factors from neuroscientific evidence. However, by taking inspiration from ML, these terms can be derived from first principles, such as gradient descent on suitable loss functions. For example, in the gradient backpropagation (BP) rule, a form of gradient descent applied to differentiable networks, $M_{post}$ can be identified with the backpropagated errors. However, as BP requires non-local signals, it is difficult to implement on a physical substrate and it is inconsistent with the brain [2]. As such, BP has significant scaling problems [14]. This could be circumvented if the errors can be estimated using local signals [18]. Thus, achieving local forms of BP is an active field of research that is now steadily advancing towards viable BP alternatives in SNNs [16, 36, 4]. Interestingly, the terms $F_{pre}$ and $F_{post}$ of the 3F rules correspond to neuron and synapse-specific traces of postsynaptic and presynaptic activity, respectively. An important challenge is the computation and storage of synaptic-specific traces locally, which are distinct from weight and parameter storage. Emerging volatile devices are central to this effort [33, 7]. Among the various non-idealities of emerging materials and devices, device-to-device variability remains a major unsolved problem for successful learning in SNNs [23]. As no device will be perfect, the community is now striving to co-design neuron, synapse, plasticity and algorithms that are tailored to the device and the application [19].

**Advances in Science and Technology to Meet Challenges**

State-of-the-art learning algorithms are specifically designed for conventional Von Neumann computers. While relying on deep learning technologies for SNN learning via off-chip learning will help reap early results, real breakthroughs will take place when software and algorithms are designed specifically for the constraints and dynamics of neuromorphic hardware. This involves moving beyond the concept of mapping conventional neural network architectures and operations to SNNs, and instead modeling and exploiting the computational properties of biological neurons and circuits. Beyond advances in the emerging devices themselves (Sec. 1), one key enabler of such breakthroughs will be a differentiable programming library (e.g. Tensorflow) operating at the level of spike-events and temporal dynamics that facilitates the scalable composition and tracing of operations [3]. Such a framework can in turn facilitate the computation of the necessary learning factors. While recent work demonstrated SNN learning with ML frameworks [29, 26], the mapping of the 3F computations on dedicated hardware does not yet exist. This is due to a lack of applications and the more stringent requirements for learning. Additionally, current technologies are not optimized to training large-scale SNNs, which remains today very slow and memory-intensive due to the high complexity of the underlying dynamics and gradients [34, 35]. However, provided that SNN models capture key features of the brain, namely that average spike rate of neurons in the brain is at most 2Hz and that connectivity is locally dense but globally sparse [8], specialized computers capable of sparse matrix operations can greatly accelerate offline training compared to conventional computers. This is because a neuron that does not spike does not elicit any additional computations or learning at the afferent neurons. Spurred by the hardware efficiency of binarized neural networks [24], some ML hardware now support efficient sparse operations [15] which could be exploited in SNN computations. A community-wide effort in these directions (software and general-purpose hardware) are likely to boost several research areas, including the discovery of new (spatial) credit assignment solutions, the identification and control of the distinctive dynamics of SNNs (multiple compartments, dendrites, feedback dynamics, reward circuits etc.), and the evaluation of new materials and devices, all in the light of community-accepted benchmarks. Undertaking such device evaluations prior to the design and fabrication cycle, for instance via a suitable surrogate model of the device, can save precious resources and dramatically accelerate the development of emerging devices.

The ability to cross-compile models in a software library can blur the line between hardware and software. This resonates well with the idea of on-chip and off-chip learning working in concert. That approach is attractive because the difficulties of online learning can be mitigated in hardware with multiple stages of training, for example by first training offline and then fine-tuning online [30]. Furthermore, fewer learning cycles entail fewer perturbations of the network, and thus mitigating the problems of sequential learning. At the same time, learning is achieved after a much smaller number of observations (*e.g.* few-shot learning), which is essential in continual learning tasks (Sec. 3.4). The success of such *meta*-learning hinges on a good task set definition and is compute- and memory-intensive. Once again, general-purpose computers supporting sparse matrix operations, associated ML libraries and community-wide efforts are essential to achieve this at scale. Although ML is not the only approach for SNN learning, the tools developed to enable ML-style learning algorithms are central to other learning models and approaches. These include hyperdimensional computing, variational inference algorithms, and neural Monte Carlo sampling, all of which rely on well-controlled models and stochasticity that can be supported by such tools.

## Concluding Remarks

Beyond hardware developments, the hidden figure of deep learning's success has been software libraries. It enabled the composition and differentiation of hardware-optimized routines for inference and learning. Neuromorphic computing, and more specifically learning in neuromorphic hardware is arguably at a similar standpoint today. Neuromorphic computing entails a different set of constraints (distributed, sparsity, locality) that are determined by the physics of the underlying hardware. The next step in exploring this new set of constraints requires a software library tailored to SNN dynamics and sparsity capable processors, which in turn can greatly accelerate the discovery of new materials, architectures, and learning algorithms for SNNs.

## Acknowledgements

This work was supported by the National Science Foundation under grant 1652159 and 1823366.

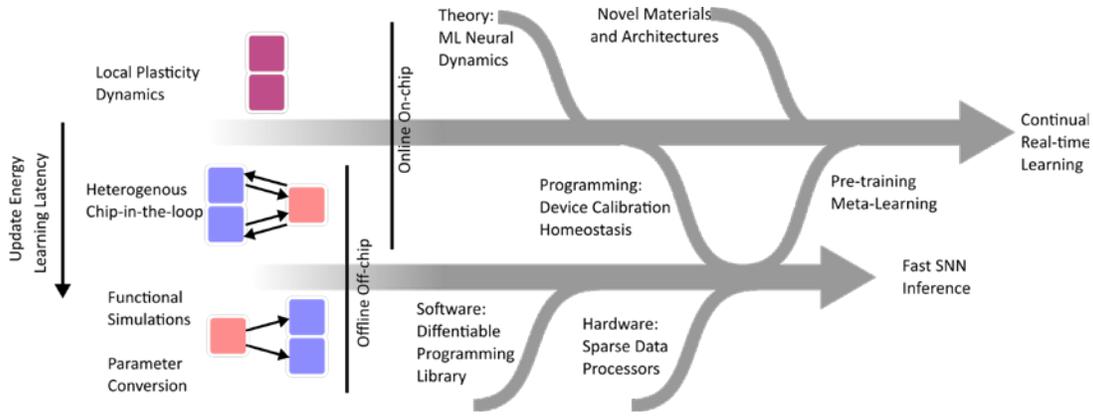

Figure 1: Implementation Strategies and Roadmap of SNN Learning. Learning in SNNs can be achieved on-chip, off-chip or a combination of both (chip-in-the-loop). In off-chip learning, the parameters trained on a general-purpose computer (pink box) are mapped on to the neuromorphic device (blue). In the chip-in-the-loop approach, updates are computed partially off-chip, but using states recorded from the chip. In the on-chip implementation, the updates are computed locally to the neurons or synapses. While brain-like, efficient, continual learning can only be achieved using on-chip learning, off-chip approaches also play an important role in pre-training the model and prototyping new algorithms, circuits and devices, or when learning is not necessary (fast SNN inference).

## 3.2 – Learning-to-Learn for Neuromorphic Hardware
Franz Scherr, Wolfgang Maass
Institute of Theoretical Computer Science, TU Graz

**Status**

An important goal for neuromorphic hardware is to support fast on-chip learning in the hand of a user. Two problems need to be solved for that:
1. A sufficiently powerful learning method has to run on the chip, such as stochastic gradient descent.
2. This on-chip learning needs to converge fast, requiring ideally just a single example (one-shot learning).

Evolution has found methods that enable brains to learn a new class from a single or very few examples. For instance, we can recognize a new face in many orientations, scales, and lighting conditions after seeing it just once, or at least after seeing it a few times. But this fast learning is supported by a long series of prior optimization processes of the neural networks in the brain during evolution, development, and prior learning. In addition, insight from cognitive science suggests that the learning and generalization capability of our brains is supported by innate knowledge, e.g. about basic properties of objects, 3D space, and physics. Hence, in contrast to most prior on-chip learning experiments in neuromorphic engineering, neural networks in the brain do not start from a tabula rasa state when they learn something new.

Learning from few examples has already been addressed in modern machine learning and AI [1]. Of particular interest for neuromorphic applications are methods that enable recurrently connected neural networks (RNNs) to learn from single or few examples. RNNs are usually needed for online temporal processing —an application domain of particular interest for energy-efficient neuromorphic hardware. The gold standard for RNN-learning is backpropagation through time (BPTT). While BPTT is inherently an offline learning method that appears to be off-limit for online on-chip learning, it has recently been shown that BPTT can typically be approximated quite well by computationally efficient online approximations. In particular, one can port the online broadcast alignment heuristic from feedforward to recurrent neural networks [2]. In addition, one can emulate the common LSTM (long short-term memory) units of RNNs in machine learning by neuromorphic hardware-friendly adapting spiking neurons. Finally, a computationally efficient online approximation of BPTT —called e-prop— exists that also works well for recurrent networks of spiking neurons (RSNNs) with such adapting neurons [3]. The resulting algorithm for on-chip training of the weights $W_{ji}$ for neuron $i$ to neuron $j$ of an RSNN —for reducing some arbitrary but differentiable loss function $E$— takes there the form

$$\frac{dE}{dW_{ji}} = \sum_t L_j^t\, e_{ji}^t\,.$$

The so-called learning signal $L_j^t$ at time $t$ is some online available approximation to the derivative of the loss function $E$ with regard to the spike output of neuron $j$, and the eligibility trace $e_{ji}^t$ is an online and locally computable eligibility trace. While this would usually require even more training examples than BPTT, one can speed it up substantially by optimizing the learning signal $L_j^t$ and the initial values of the synaptic weights $W_{ji}$ to enable learning from few examples for a large —in general even infinitely large— family $\mathcal{F}$ of on-chip learning tasks $C$ [4]. This can be achieved through Learning-to-Learn (L2L) [5]. A scheme for the application of L2L to enable fast on-chip learning is shown in Figure 1.

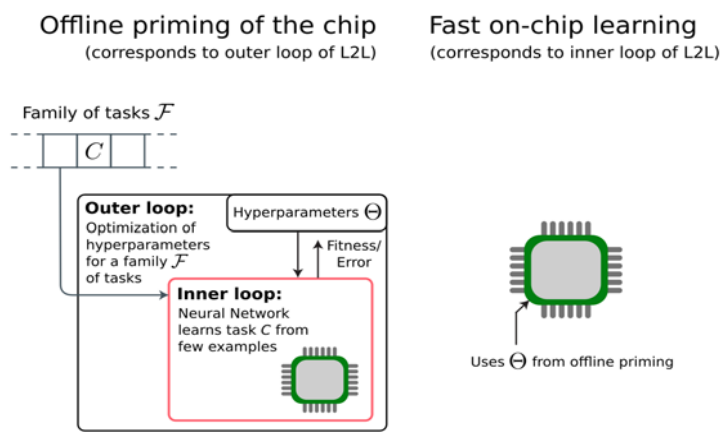

**Figure 1.** Scheme for the application of L2L for offline priming of a neuromorphic chip. Hyperparameters $\Theta$ of the RSNN on the chip are optimized for supporting fast learning of arbitrary tasks $C$ from a family $\mathcal{F}$ that captures learning challenges that may arise in the hands of a user. The resulting hyperparameters are then loaded onto the chip. Note that the desired generalization capability is here more demanding than usually: We want that the chip can also learn tasks $C$ from the family $\mathcal{F}$ very fast that did not occur during offline priming (but share structural properties with other tasks in the family $\mathcal{F}$).

**Current and Future Challenges**

The main choices that have to be made for such realization of fast on-chip learning are the choice of the family $\mathcal{F}$ of tasks, the choice of the optimization method for offline priming through the definition of hyperparameters, and the choice of the hyperparameters. Options for the latter are for example:

1. Just the learning rate parameters of on-chip learning rules are hyperparameters.
2. Also the values of all synaptic weights of the RSNN are hyperparameters.
3. Only the initial values of the synaptic weights of the RSNN are hyperparameters.
4. In addition all parameters of an auxiliary NN —the learning signal generator— that generates online learning signals $L_j^t$ for fast convergence of e-prop on the chip are hyperparameters.

Option 1 has been explored for RSNNs in [6, 7], and with an application to analog neuromorphic hardware in [8]. Option 2 is arguably the most commonly considered application of L2L in machine learning and computational neuroscience models [5, 9-13]. An attractive feature of this option for realizing fast on-chip learning is that it requires no synaptic plasticity for that. Rather, it uses hidden variables of the RNN for storing information from the few training examples that are needed for fast learning. In the case of machine learning, these hidden variables are the values of memory cells of LSTM units. In spiking neural networks these are the current values of firing thresholds of adapting neurons. An alternative is to choose only some synaptic weights to be hyperparameters, and to leave others open for fast on-chip learning [14]. Option 3 is used by the MAML approach of [15], where only very few updates of synaptic weights via BPTT are required in the inner loop of L2L. It also occurs in [4] in conjunction with option 4, see Figure 2 for an illustration.

One common challenge that underlies the success of all mentioned options, is the efficacy of the training algorithm for the offline priming phase, the outer loop of L2L. While option 1 can often be carried out by gradient-free methods, the more demanding network optimizations of the other options tend to require BPTT for offline priming of the RNN.
)

**Advances in Science and Technology to Meet Challenges**
It is quite realistic to enable according to this L2L method fast on-chip learning on neuromorphic hardware. The most demanding aspect for the hardware is to be able to run the on-chip learning algorithm that is required. This can be implemented on most neuromorphic hardware if only simple local rules for synaptic plasticity are required in the inner loop of L2L, as in option 1. In the case of option 2 a spike-based neuromorphic hardware just needs to be able to emulate adapting spiking neurons. This can be done for example on SpiNNaker [16] and Intel's Loihi chip [17]. Using BPTT for on-chip learning appears to be currently infeasible, but on-chip learning with e-prop is supported by SpiNNaker and the next generation of Loihi. Then option 4 can be used for enabling more powerful fast on-chip learning. The only additional requirement for the hardware is that an offline primed learning signal generator can be downloaded onto the chip (once and for all), and that the chip supports communication of its learning signal for gating local synaptic plasticity rules according to e-prop.

An illustration of a sample application which becomes then realistic is shown in Figure 2: On-chip learning of a new spoken command from a single example in such a way that the same command can then also be recognized under different acoustic conditions and from different speakers.

Future advances need to address the challenge of training extended learning problems during the offline phase. Besides improved gradient-based algorithms, also gradient-free training methods such as Evolution Strategies [18] are attractive for that. In fact, since the latter paradigm allows to employ neuromorphic hardware directly for evaluating the learning performance, this approach can benefit from the speed and efficiency of fast neuromorphic devices, as in [8]. Particularly fast neuromorphic hardware such as Brainscales [19] might support then even more powerful offline priming with training algorithms that could not be carried out on GPU-based hardware, thereby providing the basis for superior hybrid systems.

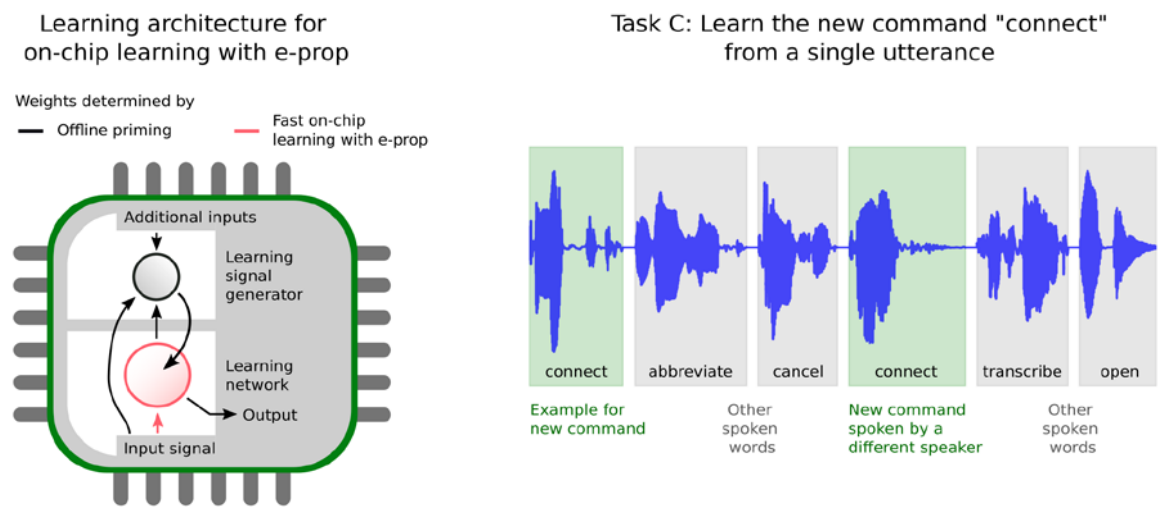

**Figure 2. Left) Learning architecture for fast on-chip learning with e-prop**. A learning signal generator produces online learning signals for fast on-chip learning. The weights of the learning signal generator as well as the initial weights of the learning network result from offline priming. **Right) Example application for fast learning**. In this task $C$, the learning network has to learn the new command "connect" from a single utterance, so that it recognizes it also from other speakers. The learning signal generator is activated when the new command is learnt (leftmost green segment).

**Concluding Remarks**

Learning in neuromorphic hardware is likely to become split into two phases that each have different goals and require different learning methods: An extensive offline priming phase —either on the actual hardware or a software model for it— that optimizes selected hyperparameters but possibly also the network architecture for a large family of potential on-chip learning tasks in the hands of the user. Resulting hyperparameter values and network architectures will be downloaded onto the neuromorphic hardware before it gets into the hands of the user. The hardware is then primed so that remaining open parameters can be learnt on-chip from very few examples, possibly even just one example.

It is conceivable that this method can be expanded to provide another useful property for neuromorphic hardware in the hands of the user: That on-chip learning cannot bring the chip into an operating regime which is unsafe, or undesired for other reasons. It has already been verified that the outer loop of L2L can impose powerful priors for subsequent computing and learning of RSNNs [13].

**Acknowledgements**

This research/project was supported by the Human Brain Project (Grant Agreement number 785907) of the European Union and a grant from Intel.

## 3.3 - Computational neuroscience

Srikanth Ramaswamy, Newcastle University

**Status**

Understanding the brain is probably the final frontier of modern science. Rising to this challenge can provide fundamental insights into what makes us human, develop new therapeutic treatments for brain disorders, and design revolutionary information and communication tools. Recent years have witnessed phenomenal strides in employing mathematical models, theoretical analyses and computer simulations to understand the multi-scale principles governing brain function and dysfunction – a field referred to as "Computational neuroscience"[1], [2]. Computational neuroscience aims at distilling the necessary properties and features of a biological system across multiple spatio-temporal scales – from membrane currents, firing properties, neuronal morphology, synaptic responses, structure and function of microcircuits and brain regions, to higher-order cognitive functions such as memory, learning and behavior. Computational models enable the formulation and testing of hypotheses, which can be validated by further experiments.

The multidisciplinary foundations of computational neuroscience can be broadly attributed to neurophysiology, and the interface of experimental psychology and computer science. The first school of thought, neurophysiology, is exemplified by the model of action potential initiation and propagation proposed by Hodgkin and Huxley [3] and theoretical models of neural population dynamics [4]. Whereas, the second school of thought, at the interface of experimental psychology and computer science focuses on information processing and learning, which could be traced back to models of artificial neural networks that were developed about half a century ago [5]. Computational neuroscience became its own nascent field about three decades ago and has rapidly evolved ever since [6].

In the early stages of its conception, computational neuroscience focused almost entirely on states of sensory processing, mainly due to the fact that studies of cognitive function were restricted to the domain of psychology, which was beyond what empirical neuroscience could offer. However, since then, rapid strides in tools and techniques have enabled tremendous advances in our knowledge of the neural mechanisms brain underlying cognitive states such learning and memory, reward and decision-making, arousal and attention [7]–[9]. Consequently, the dynamic field of neuroscience bestows many opportunities and challenges. A recent development is the symbiosis between computational neuroscience and deep learning [10]. Deep learning models have enabled efficient means to analyze vast amounts of data to catalyze computational modeling in brain research. However, the current framework of deep learning is mostly restricted to object recognition or language translation. Identifying the fundamental mechanisms responsible for the emergence of higher cognitive functions such as attention and decision making, appropriately recapitulated into algorithms by computational models, will influence the next generation of intelligent devices.

**Current and Future Challenges**

Thus far, computational neuroscience has been limited to modelling local circuits – through point neurons or multicompartmental models. However, the advent of "big" spatial and temporal data, from single-cell transcriptomes, micro, meso and macro-scale connectomes, large-scale neurophysiology, and functional brain activity mapping, which is nothing short of an industrial revolution in neuroscience, are rapidly enabling computational models to incorporate greater biological realism and move beyond local circuits. The grand challenge, however, is to assimilate the exponentially increasing datasets into theories and models of global brain function that complement local circuit and large-scale network models. One possibility towards surmounting this challenge is through collaborative endeavors, which have pioneered a high-throughput, team science approach to neuronal modelling. These initiatives have brought together disparate disciplines such as experimental biology, computational neuroscience, high-performance computing and data science that attempt to model neural circuits with thousands of detailed multicompartmental neurons and millions of synapses and simulate these models on supercomputers. The first of these was probably the Blue Brain Project, followed by the Allen Institute for Brain Science, the US Brain Initiative, Japan Brain and Minds, the Human Brain Project, and the International Brain Laboratory among several others [11]–[18].

**Advances in Science and Technology to Meet Challenges**

The brain is undoubtedly the most sophisticated information processing device known to humankind. However, the underlying principles of its function and dysfunction seem to vastly differ from those understood through conventional computing hardware. Although we are far from fully understanding the principles of brain function, which operates by consuming a few watts of power, it still manages to solve complex problems by devising algorithms that appear to be intractable with current computing resources. The brain is robust, reliable and resilient in its operation despite the fact that its building blocks could fail. All of these are highly advantageous features to instruct the better design of the next generation of computing hardware [19].

In the future, computational models and simulations of brain function and dysfunction will be better informed through its unique capabilities to model and predict the outside environment, underlying design principles, their mechanisms and multi-scale organization and operation. The interface of experimental and computational neuroscience will shed new light on the unique biological architecture of the brain and help translate this knowledge into the development of brain-inspired technologies.

We are still at the tip of the iceberg in dealing with and solving diverse challenges. It is possible that "neuromorphic" computing systems of the future will comprise billions of artificial neurons and the development, design, configuration and testing of radically different hardware systems will require new software compatible with the organizing principles of brain function. This will require a deep theoretical understanding of the way the brain implements its computational principles. Knowledge of the cognitive architectures underlying capabilities such as attention, visual and sensory perception can enable us to implement biological features that current computing systems lack.

Recent advances in "connectomics" allow unprecedented reconstructions of biological brain networks. These connectomes display rich structural properties, which include heavy-tailed degree distributions, segregated ensembles and small world networks [20]. Despite these advances, how network structure determines function computation remains unknown. Going forward, approaches from computational modelling and neuroscience could better inform the design of neuromorphic systems to unravel how structure leads to function, how the same network configuration could result in a spectrum of cognitive tasks depending on the network state and how different network architectures could support the emergence of similar cognitive states.

**Concluding Remarks**

The fledgling field of computational neuroscience has now transformed into a dynamic and vibrant global community. Computational models and simulations offer an unprecedented framework to understand brain function and dysfunction. But what could computational neuroscience possibly achieve in the next two or three decades? Can models of neural networks explain any higher-order brain function? For example, visual function involves progressive information processing with across brain regions with similar network architectures but different modalities of representation and computation [21]. Most neural network models to date are exclusively designed with functionally homogeneous populations of neurons, comprising simplified point neuron models or "ball and stick" compartmental models that do not necessarily capture the rich features of brain organization and architecture.

Although computational neuroscience initially took off in the "reductionist" direction that enabled researchers to abstract approaches and insights from disparate fields such as mathematics, physics and engineering it probably resulted in overly simplified assumptions about the unprecedented complexity of the building blocks of the brain. Future models of neural networks and theories would have to take cognizance of the astounding assortment of neuron types, impact of neuromodulatory systems, metabolic factors, co-release of neurotransmitters, neuro-glial-vasculature interactions adaptive mechanisms of neurons and synapses across diverse spatio-temporal scales among many other biological details.

In conclusion, computational neuroscientists that use their preferred level of abstraction and modeling traditions should come together as a community, which surprisingly, is rather difficult. As a famous neuroscientist and Nobel laureate, Sydney Brenner remarked "we are drowning in a sea of data and starving for knowledge". Computational neuroscientists should interact a lot more with experimentalists and learn from each other. More theory is the need of the hour to crack the brain.

**Acknowledgements**

I acknowledge support from the ETH Board to the Blue Brain Project.

# 3.4 – Stochastic Computing
Jonathan Tapson, School of Electrical and Data Engineering, University of Technology, Sydney

**Status**

The human brain is extraordinarily efficient in computation, using at least five orders of magnitude less power than the best neuromorphic silicon circuits [1]. Nonetheless, it still consumes approximately 20-25% of a human's available metabolic energy, and it is safe to assume that the evolutionary pressure to optimize for power efficiency in the brain was extremely severe [2]. It therefore comes as a suprise that the transmission of signals through the brain's synaptic junctions is apparently noisy and inefficient, with probabilities of 0.4-0.8 for transmission of an axonal spike being typical (see [3] for a detailed review). This raises the question: is this transmission variability a bug or a feature? Also, can any brain-inspired computational system which does not include synaptic stochasticity capture the essence of human thought? Perhaps it serves to regularize biological neural networks, in the same way that machine learning techniques such as dropout are used to make artificial neural networks more robust.

This, and many other similar questions, drive the field of stochastic computation [4, 5]. The field covers a large number of techniques in which some kind of probabilistic function, filter or network is used to create a computational output which would not be possible with deterministic systems. For example, in neuromorphic neural networks, the use of nonlinear random projections has become a commonplace method for raising the dimensionality of an input space prior to a learned solution layer. Technologies as diverse as silicon device mismatch, memristor mismatch, and even random networks of conductive fibres have been proposed and tested for this purpose [6].

Generally, stochastic computation methodologies fall into a number of categories:

1. Systems where noise or randomness is used to add energy to a system, enabling it to traverse or establish states which were otherwise inaccessible. Energy in this sense means potential or kinetic energy (in mechanical, electrical or chemical form), rather than general system power consumption. The various phenomena of stochastic resonance and stochastic facilitation [4] are typical examples of these systems.

2. Systems where the data or input streams are intrinsically random or noisy, and rather than filter or otherwise reduce the uncertainty in the signals, a computational system is devised which processes the raw signal to produce a computationally optimal output. Recently, many of these systems apply Bayesian models [7], particularly when derived from biological principles.

3. Systems in which it is required to project the input space nonlinearly to a higher dimension, in order to facilitate linear or other numerical solutions to some regression or classification problem. This obviously includes conventional neural networks; however, there is an increasing body of research in both human neuroscience and machine learning in which random nonlinear projections are found to be optimal for some function.

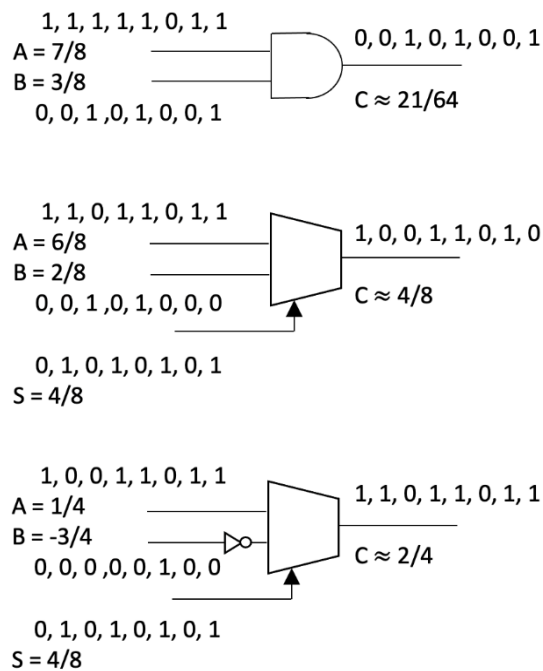

**Figure 1.** Illustration of a simple stochastic computation, using logic gates (AND or MUX) to compute on strings of bits representing probabilistic coding of numbers. Note that single-bit errors in the computation (as a result of noise) will not significantly change the output. After [4].

**Current and Future Challenges**

Perhaps the most exciting theme for the future of stochastic computation is the modelling of random or noisy processes in the brain, and applying these to real world computational problems. For example, there is increasing evidence that random projection layers are used in the brain to improve the versatility and selectivity of neural circuits [8]. This has parallels in random projection artificial neural networks, but the utility of these systems has, until now, been overshadowed by the performance of application specific machine learning systems. Additionally, there is evidence that in spiking neural networks, noise appears to enable probabilistic inference [9]. As these systems move from academia into commercial fields of exploitation, the adapability and robustness of the random networks is likely to shine.

There are some significant issues in exploiting stochastic computation. For example, if randomness is created using intrinsically variable hardware means such as device mismatch, then the randomness is unique to each device, and therefore each device will need to be characterised before it can be used. This is reasonable in laboratory research, but infeasible as part of the workflow in producing high-volume consumer goods.

A further problem for stochastic computation is that many applications – for example, biomedical or aerospace systems – require deterministic and provably failsafe operation. Systems that compute bsed on probability and statistical likelihoods may struggle to meet industrial standards, notwithstanding that their performance may be superior to current methods.

**Advances in Science and Technology to Meet Challenges**

The foundations of stochastic computation are not mature. The field at present consists largely of proofs-of-concept, with few widely-applicable principles that can be applied in a generic way. Noise and randomness are still seen as obstacles to be overcome in signal processing, so unless a practitioner has some prior experience of stochastic computing, there is little likelihood of applying these methods as part of a standard process. Nonetheless, the increasing interest in Baysian methods, and their existence and application in biological neural systems, is generating growing interest in this approach. What the field requires for advancement are the following:

- A foundational theory of probabilistic computing, integrating Bayesian, SR and random projections into a more general framework which answers the question: "I have noisy and nonlinear input data, or a noisy and nonlinear system. How do I take advantage of the energy and nonlinearity rather than simply filtering it out?"
- Current methods of implementing random connective weights and random numbers in silicon are not ideal. In particular, there are few methods which allow the deterministic generation of "random" weights in ways that obviate the requirement to characterise each individual part prior to computation. Similarly, generation of random numbers "on the fly" in computational platforms such as custom silicon or FPGAs requires a significant proportion of hardware and energy to be dedicated to it.

**Concluding Remarks**

The real world and natural organisms are intrinsically noisy and variable, and yet the most efficient computer known – the human brain – computes very successfully using noisy and variable hardware on noisy and variable signals. As our understanding of this phenomenon grows, the desire to apply the principles in artificial computation will grow accordingly. In stochastic computation, we have the beginnings of a model for this process, and it seems likely that in the future we will regard computational systems which do not use probabilistic principles as being as limited as we now regard classification and regression using the purely analytic methods that were conmmonplace before machine learning.

## 3.5 – Convolutional Spiking Neural Networks

Priyadarshini Panda, Youngeun Kim, Department of Electrical Engineering, Yale University, New Haven, USA

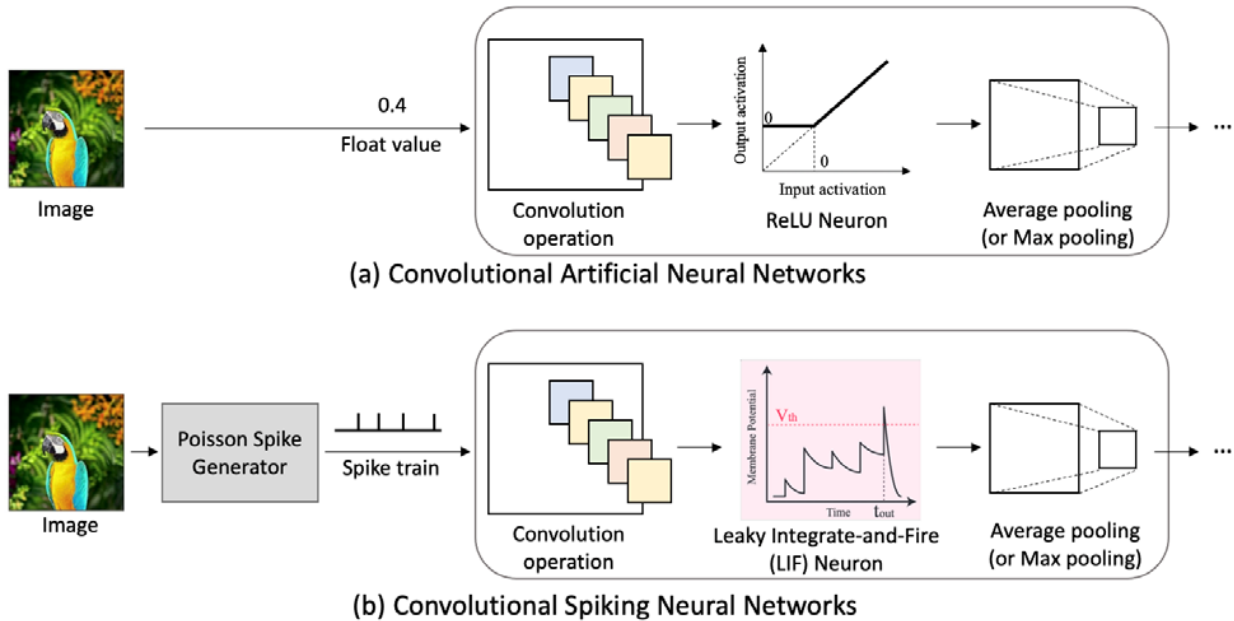

Figure 1. Illustration of the architectural difference between (a) Convolutional Artificial Neural Networks and (b) Convolutional Spiking Neural Networks. Both architectures are based on spatial convolution operation, however, convolutional SNNs convey information using binary spikes across multiple time-steps. To this end, most convolutional SNNs use a Poisson spike generator in order to encode an RGB image into temporal spikes. Also, the Rectified Linear Unit (ReLU) neuron is replaced by Leaky -and-Integrate-Fire (LIF) spiking neuron in which an output spike is generated whenever a membrane potential exceeds a firing threshold (Vth).

## Current Status

Recent machine learning literature show that convolutional neural networks have been adopted successfully in various vision applications [1, 20]. To apply Spiking Neural Networks (SNNs) to diverse vision tasks, convolutional SNNs (Fig. 1) should be explored as a priority. Consequently, there have been many attempts to train a convolutional SNN architecture. Conversion [6] proposes a weight balancing technique in order to convert pre-trained Convolutional Artificial Neural Networks (ANNs) to convolutional SNNs. Note, we make a distinction between convolutional ANN and convolutional SNN based on the type of input processing and the neuronal activation as shown in Fig. 1. The overall architecture still uses convolution and pooling operation suitable for enhanced performance in vision tasks. The conversion method yields competitive accuracy, but requires careful weight scaling to emulate the floating point activation value of ANNs and therefore its application is restricted to simple static image recognition. Surrogate gradient learning circumvents the non-differentiability of a Leaky -and-Integrate-Fire (LIF) neuron by defining an approximate gradient during backward propagation [7]. Training SNNs directly on spike trains can enable parsing various types of input representation beyond static images, and potentially lead to SNN implementation for diverse computer vision applications.

Further, the development of convolutional SNNs is in tandem with a growing interest in Dynamic Vision Sensors (DVS). Different from a standard frame-based camera, a bio-plausible DVS camera emulates the human retina and generates asynchronous and binary spikes according to illumination changes. Therefore, this visual sensor has the advantage of a high frame rate, low energy consumption, and less blurring effect compared to a conventional frame-based camera [8]. A convolutional ANN fails to exploit the advantages of DVS camera inputs since ReLU neurons cannot capture temporal information. On the other hand, owing to temporal spiking neuronal dynamics, convolutional SNNs are a natural fit for processing DVS data. Also, combining DVS camera inputs with SNNs yields a fully-spiking and energy-efficient system. Despite the potential advantages, an algorithm for training convolutional SNNs on DVS data is still underdeveloped.

## Challenges

Majority of previous works on convolutional SNNs classify static images. SNNs require a spike train to process in the temporal domain. Thus, various coding schemes have been proposed for converting static images to spikes [19]. It is important to select a proper coding strategy since the energy consumption on an asynchronous neuromorphic hardware is approximately proportional to the number of spikes. Currently, rate coding is the most widely-used coding scheme since it yields high application-level accuracy. But, rate coding generates spikes, where the number of spikes is proportional to the pixel intensity. This causes multiple (and sometimes redundant) spikes per neuron and therefore reduces the energy-efficiency of the overall system. In order to bring more energy advantages, a new coding scheme with fewer spikes should be explored.

Another challenge is directly training deep convolutional SNNs. From ANN literature, it is well known that network depth is a crucial factor for achieving high accuracy on vision tasks. ANN-to-SNN conversion enables deep SNNs with competitive accuracy. But, emulating float activation with multiple binary spikes requires a large number of time-steps, which in turn increases overall energy and latency. Surrogate gradient learning allows short latency and can be used with flexible input representations, but it suffers convergence issues when we scale up the depth. Therefore, convolutional SNNs with surrogate learning are still restricted to shallow networks on trivial datasets. Overall, effective spike-based training techniques for deep convolutional SNNs is necessary to reap the full energy-efficiency advantages of SNNs.

Finally, there is a need to investigate SNNs beyond the perspective of accuracy and energy-efficiency. Goodfellow et al. [9] showed that unrecognizable noise can induce a significant accuracy drop in ANNs. This questions the reliability of ANNs since humans do not misclassify when presented with such perturbed adversarial inputs. In this light, there is a need to analyze the robustness of SNNs. Furthermore, the internal spike behavior of SNNs still remains to be a "black-box" as that of conventional ANNs. In the ANN domain, several interpretation tools have been proposed and provide cues for advanced computer vision applications such as visual-question answering. In a similar vein, an SNN interpretation tool should be explored because of its potential usage for real-world applications where interpretability in addition to high energy-efficiency is crucial.

**Advances in Algorithms and Hardware to Meet Challenges**

Beyond rate coding, temporal coding generates one spike per one neuron in which spike latency is inversely proportional to the pixel intensity. Mostafa et al. [10] applied an exponential kernel to derive locally exact gradients of spiking neurons. The authors trains SNNs on XOR problem and MNIST dataset with much fewer number of spikes compared to standard rate coding. Han and Roy [11] tackled the memory access issue for tracking a synapse at every time-step until the synapse receives its first spike. They use two signed spikes that denote the start and finish time-step respectively, improving energy-efficiency. Besides temporal coding, phase coding has also been proposed based on biological observation [12], which encodes temporal information into spike patterns based on a global oscillator in an energy-efficient manner [13].

In order to address the shortcomings of ANN-to-SNN conversion and surrogate gradient learning, Rathi et al. [14] proposed hybrid training- a conversion process followed by surrogate gradient learning. As a result, they achieve competitive performance with significantly lower latency. Wu et al. [15] presented a spatio-temporal backpropagation technique for direct training convolutional SNNs. The authors evaluate their method on static MNIST as well as DVS-based N-MNIST. Recently, Kim and Panda [16] addressed the scalability problem of direct spike-based training by proposing a time-specific batch normalization technique, called Batch Normalization Through Time (BNTT). The authors successfully trained convolutional SNNs from scratch with BNTT spike-based training on complicated datasets such as CIFAR100, Tiny-ImageNet and DVS CIFAR10.

Few recent studies have shown that SNNs have more robustness compared to their ANN counterparts. For the first time, Sharmin et al. [17] observed that SNNs are more robust with respect to adversarial images. They provided a comprehensive analysis on VGG9 networks trained on CIFAR10 dataset with various attack scenarios including black-box and white-box attacks. Also, the authors of [18] asserted that a Poisson coding generator and non-differentiable neuronal dynamics are the main reasons for SNN robustness. Recently, the authors of [3] presented a heatmap visualization tool that shows where SNN focuses on during classification. The results support that SNNs show more robust attention on discriminative regions with respect to adversarial samples. We believe that research on interpretability is essential to open up the possibility of an interpretable and reliable neuromorphic system.

Finally, from a hardware implementation perspective, memristive crossbar array based neural network accelerators have been shown to be area and energy efficient by many orders of magnitude than fully digital CMOS implementations [19]. Thus, it is natural to explore the possibility of implementing convolutional SNNs on such crossbars. But, the convolutional architecture requires a careful weight mapping process. This is because the Matrix-Vector-Multiplication operation of the crossbar arrays are suitable for a fully-connected layer. On the other hand, a convolutional layer uses spatially shared weight kernel where in, same weight kernel interact with different input patches to produce different outputs. Thus, weight sharing can lead to ineffective crossbar utilization (see [19] for more details). To address the mapping problem due to the weight sharing, the authors in [4, 5] present a mapping protocol for convolutional ANNs. They also provide a simulation tool for crossbar implementation, which evaluates the energy and performance of a network during inference on crossbars. We believe that similar mapping protocols can be extended to convolutional SNNs.

## Concluding Remarks

Convolutional architecture plays a key role in artificial intelligence. Therefore, this architecture is also important in the neuromorphic community because of its practical usage for expanding SNN applicability for diverse vision scenarios. Moreover, from a neuroscientific perspective, convolutional architecture is an interesting research topic itself. For example, Hubel and Wiesel [1] discovered a hierarchical model like convolutional architectures in the primary visual cortex (V1) of cats. After two decades, Fukushima and Miyake [2] suggested a more structured neural system, which consists of S-cells extracting visual features and C-cells working as a non-linear function (like ReLU). Today, convolutional ANNs have found use in neuroscience studies to interpret the activity recorded from different brain regions and explain about the functionality of the region. Since convolutional SNNs are more bio-plausible given the temporal spike processing, we believe convolutional SNNs alongside suitable neuromorphic hardware can be used as interpretation tools for neuroscience. Finally, in order to bring the convolutional architecture to a neuromorphic system, recent works have focused on energy-efficient coding schemes and training techniques. In addition to the development of SNNs with higher energy-efficiency and accuracy, robustness and interpretability of convolutional SNNs are significant research directions. Overall, many technical challenges, such as training, coding, and interpretability need to be addressed to build reliable and accurate convolutional SNNs compatible with neuromorphic hardware.

## Acknowledgements

The research was funded in part by C-BRIC, one of six centres in JUMP, a Semiconductor Research Corporation (SRC) program sponsored by DARPA, the National Science Foundation, the Technology Innovation Institute (Abu Dhabi) and the Amazon Research Award.

## 3.6 Reservoir computing
Gouhei Tanaka, The University of Tokyo

**Status**

Reservoir computing (RC) is a machine learning framework capable of fast learning, suited mainly for temporal/sequential information processing [1]. The general concept of RC is to transform sequential input data into a high-dimensional dynamical state using a "reservoir" and then perform a pattern analysis for the reservoir state in a "readout". This concept was originally conceived with a special class of recurrent neural network (RNN) models (see Fig. 1(a)), such as echo state networks (ESNs) [2] and liquid state machines (LSMs) [3]. The main characteristic is that the reservoir is fixed and only the readout is adapted or optimized using a simple (mostly linear) learning algorithm, thereby enabling fast model training. Owing to the computational efficiency, software-based RC on general-purpose digital computers has been widely applied to pattern recognition, such as classification, prediction, system control, and anomaly detection, for various time series data. To improve computational performance, many variants of RC models have been actively studied [4].

On the other hand, hardware-based RC is an attracting option for realizing efficient machine learning devices. A reservoir can be constructed not only with RNNs but also with other nonlinear systems. In fact, a rich variety of physical reservoirs have been demonstrated using electrical, photonic, spintronic, mechanical, material, biological, and many other systems (see Fig. 1(b)) [5]. Such physical RC is promising for developing novel machine learning devices as well as for finding unconventional physical substrates available for computation. The system architectures of hardware-based reservoir can be mainly classified into several types, including network-type reservoirs consisting of nonlinear nodes, single-nonlinear-node reservoirs with time-delayed feedback [6], and continuous medium reservoirs [7]. Many efforts are currently underway to improve computational performance, enhance energy efficiency, reduce computational cost, and promote implementation efficiency of the physical reservoirs. They are often combined with a software-based readout or a readout device based on reconfigurable hardware capable of multiply-accumulate operation.

Further advances in physical RC would contribute to realizing novel artificial intelligence (AI) chips, which are distinguished from AI chips for deep learning. One of their potential targets is edge computing [8]. High-speed machine learning computation for data stream obtained from sensors and terminal devices would lead to data traffic reduction and data security enhancement in the Internet of Things (IoT) society.

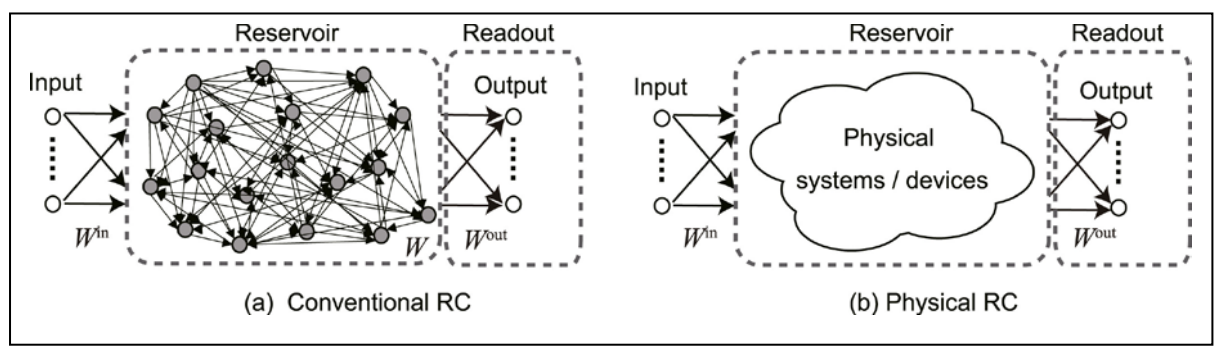

**Figure 1.** RC frameworks where the reservoir is fixed and only the readout weights $W^{\text{out}}$ is trained. (a) A conventional RC system with an RNN-based reservoir as in ESNs and LSMs. (b) A physical RC system in which the reservoir is realized using a physical system or device. Figure reproduced from [5]. CC BY 4.0.

**Current and Future Challenges**

In exchange for the merit that speedy training is possible for RC systems with simple readouts, the fixed reservoir needs to be well designed for high computational performance. For ESNs and LSMs, there are plenty of theoretical and experimental results helpful in designing "good" RNN-based reservoirs [9]-[12]. So far, the potential of RC framework has been shown for many benchmark tasks and practical applications (see Table 1) [5]. Current challenges on software-based RC include enhancing the utility of RC models for industrial applications, comparing their cost-performance ratio with that of other machine learning methods, and developing extremely efficient RC models through various extensions.

For hardware-based RC, however, design principles have yet to be fully established. Therefore, a fundamental challenge is to make recipes for constructing "good" physical reservoirs as well as identifying the main components governing the computational performance. The computational performance of RC systems relies on dynamical states of reservoirs when driven by input signals, which should have nonlinearity and memory (or history dependency) for temporal pattern recognition. The response characteristics of reservoirs can differ depending on the type of reservoir architecture, the physical property of the media/substrates in the reservoir, and the pre- and post-processing of signals to and from the reservoir [5,13]. It is not straightforward to clarify how each of these factors influences the computational ability of the whole RC system, because they are not independent of each other in many cases [14]. In hardware implementation of physical reservoirs, it is needed to consider physical constraints, such as possible time resolution and measurement precision in signal processing, which could also affect the computational property [15]. Therefore, more numerical and experimental studies are necessary for elucidating the relationship between physical and computational properties of RC systems [16].

Toward practical applications of hardware-based RC systems, it is demanded to evaluate each physical reservoir in terms of different aspects such as computational performance, processing speed, power efficiency, scalability, miniaturizability, and robustness. After an accumulation of these evaluations for various pattern recognition tasks, it would be possible to make a comparison between different RC systems and determine which physical RC system meets a specific purpose. It is also significant to promote an integration of RC-based machine learning devices with IoT devices.

**Advances in Science and Technology to Meet Challenges**

RC models can be more efficient by using additional techniques and extended architectures. For instance, a hybrid of an ESN and a knowledge-based model enables better prediction of spatiotemporal chaotic behaviour in high-dimensional dynamical systems [17]. RC models can also be combined with some other machine learning techniques found in this roadmap. Another extended RC model is the deep RC model consisting of multiple reservoirs [18], which is effective for diversifying temporal representations and generating rich dynamical behaviour.

Methodologies for designing and implementing hardware-based RC systems are currently under intensive investigation. For instance, substantial progress has been made for electronic RC [19] and photonic RC [20]. Compared with them, other physical RC systems are still in the initial stage of development [5,13]. More studies are required to harness nonlinearity and memory of physical reservoirs for efficient computation.

On the one hand, it is significant to improve computational performance of physically implemented RC hardware such that it is competitive with other machine learning hardware. On the other hand, mathematical modelling and simulations of physical reservoirs are useful for proof-of-concept of novel hardware-based RC systems and analyses of their mechanisms. Both approaches are complementary for addressing the challenges mentioned above.

Spiking neural networks (SNNs), which is one of the central topics in neuromorphic computing, have often been used for implementing a reservoir, mostly in the context of LSMs. Neuromorphic chips based on SNNs (e.g. BrainScaleS 2 prototype system) are available for emulating LSM models and thereby useful for exploring optimal setting of SNN-based reservoirs [21]. In the future, efficient SNN-based machine learning hardware for real-time computation could be realized based on the RC framework with biologically plausible mechanisms.

| Category | Examples |
|---|---|
| Biomedical | EEG, fMRI, ECG, EMG, heart rates, biomarkers, BMI, eye movement, mammogram, lung images. |
| Visual | Images, videos. |
| Audio | Speech, sounds, music, bird calls. |
| Machinery | Vehicles, robots, sensors, motors, compressors, controllers, actuators. |
| Engineering | Power plants, power lines, renewable energy, engines, fuel cells, batteries, gas flows, diesel oil, coal mines, hydraulic excavators, steam generators, roller mills, footbridges, air conditioners. |
| Communication | Radio waves, telephone calls, Internet traffic. |
| Environmental | Wind power and speed, ozone concentration, PM2.5, wastewater, rainfall, seismicity. |
| Security | Cryptography. |
| Financial | Stock price, stock index, exchange rate. |
| Social | Language, grammar, syntax, smart phone. |

Table 1. Examples of subjects in RC applications. Table reproduced from [Tanaka et al. 2019]. CC BY 4.0.

**Concluding Remarks**
Toward realizing extremely efficient RC systems and hardware for temporal pattern recognition, both algorithmic and implementation efforts are necessary [22]. Moreover, both physical and computational viewpoints are required for development of physical RC. Therefore, the progress of physical RC would be accelerated by interdisciplinary collaborations between experts in different research areas.

**Acknowledgements**

This work was partially based on results obtained from a project, JPNP16007, commissioned by the New Energy and Industrial Technology Development Organization (NEDO), and supported in part by JSPS KAKENHI Grant Number 20K11882, JST CREST Grant Number JPMJCR19K2, and JST-Mirai Program Grant Number JPMJMI19B1.

## 3.7 – Computing with Spikes
Simon Thorpe, CerCo-CNRS

**Status**

Deep learning architectures now dominate artificial intelligence. But although they are superficially neurally-inspired, there are significant differences with biology. The "neurons" in such systems typically send floating-point numbers, whereas real neurons send spikes. Attempting to model a system with the complexity of the human brain with floating-point numbers seems doomed to failure. The brain has 86 billion neurons, with around 7000 synapses each on average. Real-time simulation of such a system with a resolution of 1 millisecond would require $(8.6E+10)*(7.0E+3)*(1.0E+3)$ floating-point operations a second – over 600 PetaFLOPS, even without worrying about the details of individual neurons. This would saturate the most powerful supercomputer on the planet and require 30 Megawatts of power – over one million times the brain's remarkable 20W budget. How does the brain achieve such a low energy budget? It seems very likely that spikes could be a key to this efficiency and a reason why, since the late 1990s, spiking neural networks have attracted increasing interest [1-3].

A first critical advantage is that computation only occurs when there are spikes to process. The AER protocol (Address Event Representation), first proposed in 1992 [4], communicates by sending lists of spikes. It is used in many Neuromorphic systems, including Dynamic Vision Sensors (see section 3.7) and the multi-million processor SpiNNaker project [5, 6]. An early event-driven spiking neuron simulator was the original version of SpikeNet [7, 8]. At the time, the joke was that such a system could simulate the entire human brain in real-time – as long as none of the neurons spiked!

Second, spikes allow the development of far more efficient coding schemes. Researchers in both neuroscience and neural networks typically assume that neurons send information using a firing rate code. And yet, the very first recording of responses of the optic nerve by Lord Adrian in Cambridge in the 1920s demonstrated that while increasing the luminosity of a flashed stimuli increased both the peak firing rate and maintained firing rate of fibres, there was also a striking reduction in latency [9]. Thus, even with a flashed stimulus, response latency is not fixed. Sensory neurons effectively act as intensity-to-delay convertors - a fact effectively ignored for over six decades. But in 1990, it was proposed that spike-arrival times across a population of neurons could be a highly efficient code [10], an idea confirmed experimentally for the retina in 2008 [11].

**Coding with Spiking Neurons**

If we accept that sensory neurons can act as intensity-to-latency convertors, it opens a whole range of interesting options. Figure 1 compares various spike coding schemes for transmitting the activation levels for 16 input channels labelled A-P with 4-bit resolution. The "standard" method would be rate coding, which counts the number of spikes in a fixed period (e.g. 100 ms). But there are numerous drawbacks of such an approach. First, how do you determine the length of the observation window? Second, rate coding is exceptionally vulnerable to changes in global intensity or contrast of the input profile that completely disrupt firing rates.

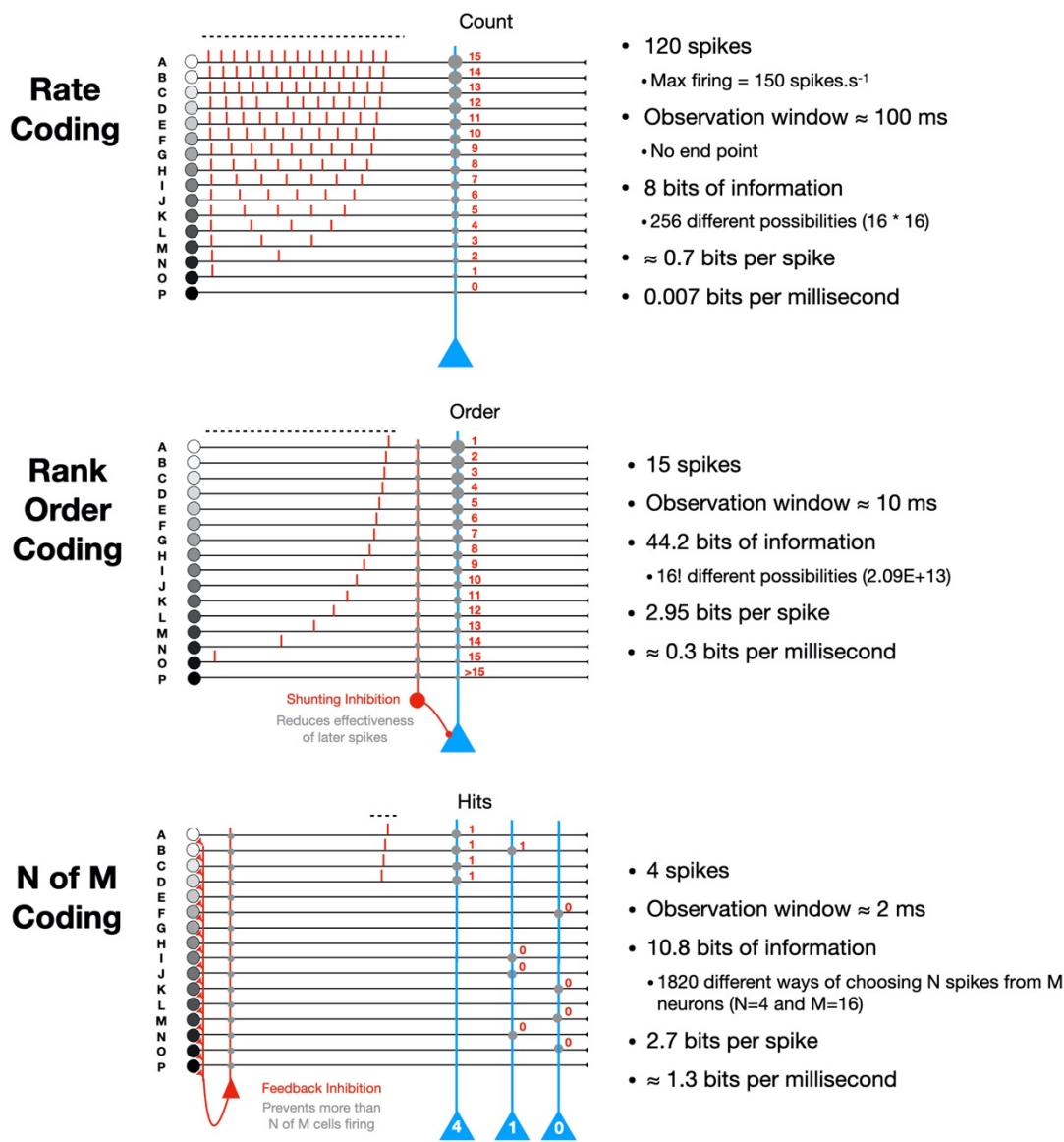

**Figure 1.** Comparison of three different spike-based coding strategies. **Top**: Conventional Rate Coding using counts of spikes in an relatively long observation window. **Middle**: Rank Order Coding uses the order of firing of just the first spike in a shorter window. **Bottom**: N of M Coding which limits the number of spikes that a transmitted allows very rapid and efficient transmission of large amounts of data.

Using a temporal coding scheme avoids these problems. One option uses the order of firing as a code - Rank Order Coding [12]. Synaptic weights are set to values that depend on the input's rank. Then a simple feedforward shunting inhibition mechanism makes neurons selective by progressively reducing the effectiveness of later spikes. Although feasible, Rank Order Coding is perhaps too powerful. Even with just 16 inputs, there are factorial 16 possible firing orders– nearly 21 trillion. Each pattern could be detected by a neuron with the appropriate set of weights.

Another option, N-of-M coding [13], is much simpler. It uses a feedback inhibition circuit to count input spikes and block further firing once a given number of inputs have fired - an inexpensive way to implement a k-WTA (Winner-Take-All circuit). If a target neuron has W connections, it can detect specific combinations of inputs. Effectively, each neuron gets a level of excitation that depends on how well the N inputs spikes match their set of weights. Crucially, this can be done with binary synapses, greatly simplifying the implementation.

Interestingly, this sort of coding had already been implemented in a commercial image processing package developed by SpikeNet Technology around two decades ago. The one published paper on the technology [14] explained that synapses between the eight orientation maps and neurons in the recognition maps were very sparse (less than 1% of possible connections). But it failed to mention that all the connections were binary - simply because the company wanted to keep its "secret sauce" secret!

**Learning with Spiking Neurons**

Switching to temporal coding with spiking neurons also makes it possible to envisage much more efficient learning algorithms. Standard Backpropagation can undoubtedly train systems to solve particular problems but requires huge numbers of training trials with labelled data. Fortunately, a human infant's brain does not need to be trained with millions of images of dogs and cats to categorize new images correctly! Instead, they can learn about new objects very rapidly. There is now good evidence that humans learn to detect virtually anything new that repeats, with no need for labelled data. If humans listen to meaningless Gaussian noise containing sections that repeat, they rapidly notice the repeating structure and form memories lasting for weeks [15]. And in experiments where random images from the ImageNet database are flashed at rates of up to 120 frames per second, humans notice images that repeat, even with only 2-5 presentations [16]. None of the existing floating-point (or rate-based) supervised learning schemes could explain such learning. In contrast, a simple Spike-Time Dependent Plasticity rule that reinforces synapses activated before the target neuron fires makes them develop selectivity to patterns of input spikes that repeat[17], and will even find the start of the pattern [18]. Similar methods have also been used to generate selectivity to repeating patterns in the output of a Dynamic Vision Sensor corresponding to cars going by on a freeway – again in a totally unsupervised way [19].

These STDP based learning rules use continuously variable synaptic weights and typically require tens of repeats to find the repeating pattern, even when all parameters are optimized. But we have recently developed a new learning rule called JAST using binary weights [20] that can match our ability to spot repeating patterns in as few as 2-5 presentations. The target neuron starts with a fixed number of binary connections. Then, instead of varying the strength of the synapses (as in conventional STDP learning), the algorithm effectively swaps the locations of the connections to match the repeating input pattern.

The algorithm was originally implemented on a low-cost Spartan-6 FPGA, already capable of implementing a network with 4096 inputs and 1024 output neurons, and calculating the activation level and updating all the outputs 100,000 times a second. The circuit also included the learning algorithm on-chip.

**Concluding Remarks**

Although the Deep Learning revolution has transformed the landscape and allowed the development of artificial systems that rival or surpass human levels of performance, current systems lag behind their biological equivalents on several fronts. They are several orders of magnitude less energy efficient than biological systems, and the mode of learning is highly non-biological.

Spiking neural networks could overcome many of these limitations in the next few years. Firstly, current systems are handicapped by very inefficient rate-based coding schemes or their floating-point equivalents. Spikes allow the use of temporal coding schemes such as Rank Order Coding and N-of-M coding that are way more efficient. Second, spikes allow the use of Spike-Time Dependent Plasticity rules, some of which operate with binary connections and can learn to recognize repeating patterns very efficiently. Such coding schemes avoid the internal noise problems associated with Poisson rate coding, since they are effectively completely determinist. Of course, there will always be some degree of noise in the sensory data and in the mechanism leading to spike initiation in the sensors themselves.

Finally, sending information as spikes will allow the implementation of vast networks with billions of neurons and trillions of connections even with currently available technology. The critical question is the level of sparseness that can be achieved. It is quite possible that many cortical neurons are silent a lot of the time ("Neocortical Dark Matter"), and that each neuron may only need a small number of (binary) connections. If so, the future impact of spike-based computation could be enormous.

**Acknowledgements**
*The author thanks the CNRS for support.*

## 4.1 - Robotics

Chiara Bartolozzi, Istituto Italiano di Tecnologia (IIT)

**Status**

Neuromorphic systems, being inspired on how the brain computes, are a key technology for the implementation of artificial systems that solve problems that the brain solves, under very similar constraints and challenges. As such, they hold the promise to efficiently implement autonomous systems capable of robustly understanding the external world in relation to themselves, and plan and execute appropriate actions.

The first neuromorphic robots were proof of concepts based on ad hoc hardware devices that emulated biological motion perception [1]. They relied on the know-how of chip designers, who had to manually turn knobs to tune the chip behaviour. That seed could grow into a mature field thanks to the availability of hardware that could be more easily tuned by non-experts with standard software tools [2] and of quality dynamic vision sensors [3] and neuromorphic computing chips and systems [4,5] featuring many instances of neurons and (learning) synapses that could be used as computational primitives for perception and decision making. Since then, neuromorphic robotics followed three main paths, with the development of visual perception for robots using event-driven (dynamic) vision sensors [6,7], proof-of-concept systems linking sensing to control [8] and spiking neural networks for the control of motors [9,10]. At the same time, the neurorobotics community started developing models of perception, cognition and behaviour based on spiking neural networks, with recent attempts to implement those on neuromorphic platforms [11,12,13]. Finally, the computational neuroscience community has developed learning theories to reconcile deep neural networks with biologically inspired spike-based learning and to directly develop spiking neural models for motor control that in the future could be implemented on neuromorphic hardware [14,15,16].

In this rich and lively scenario, the multiple contributing communities and research fields have the potential to lead to the next breakthrough, whereby neuromorphic sensing and computing support the development of smart, efficient and robust robots. This research is timely and necessary, as robots are moving from extremely controlled environments, to spaces where they collaborate with humans, where they must dynamically adapt, borrowing from neural computational principles.

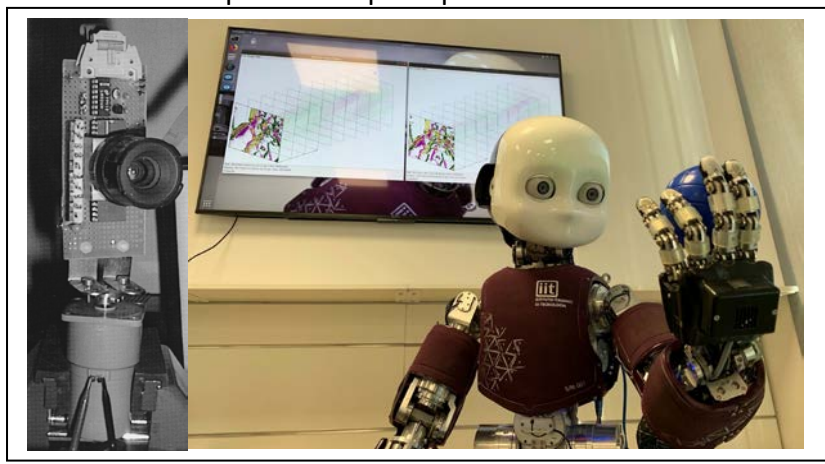

**Figure 1.** Neuromorphic robots: on the left the tracker chip mounted on a pan-tilt unit [1], on the right the iCub humanoid platform featuring event-driven vision sensors.

**Current and Future Challenges**

The complexity of robotics poses a number of challenges for the neuromorphic community, that needs to demonstrate scalable methods for sensing, perception, decision making and control, that can be deployed in complex applications.

The event-driven readout should be applied to the majority of sensors on the platforms, including those that convey information about the robot's state (inertial, F/T sensors, encoders, etc.). Additionally, more sophisticated event-driven encoding could be implemented, by using local computation that emulates the neural encoding of specific sensory modalities.

Understanding the neural code relative to stimuli properties up to decision making [17] will guide the design of spiking neural networks to extract meaningful information and take appropriate decisions.

Controlling a robot requires translating information from the sensory to the joint domain, coordinating different joints to implement stable and smooth trajectories, and counteracting actuation errors. Also in this domain, it is crucial to understand principles of biological control and actuation that lead from the cortical motor commands, down to muscles' recruitment and the relative feedback signals.

At the same time, the big question is how much these principles are shaped on the properties of muscles, and how they have to change to be applied to any type of physical (artificial) actuation. Hidden in this research domain is the intrinsic link between the neural code for actuation and the sensory feedback code that interacts with control signals to effectively drive the system.

Given the number of components that need to be developed and orchestrated, a possible path is the progressive integration of specific neuromorphic modules within a "traditional" robotic architecture. However, the overall system's architectures might be based on very different assumptions and this hybrid architecture might not fully exploit the advantages of the neuromorphic paradigm. A fully neuromorphic system should be the final goal, requiring an exponential growth of the community working on all the bits and pieces that compose a robotic system.

The signature of neuromorphic robots will be continuous learning and adaptation to different environments, different tasks, changes in the robot plant, different collaborators. This must be supported by hardware capable of handling plasticity at multiple temporal scales and a strong knowledge of how the brain implements such mechanisms.

At the technological level, it is paramount to develop neuromorphic devices that can be embedded on robots, increasing the neurons and synapses count and fan-in fan-out capabilities, while maintaining a low power budget. Ideally, those devices have standard interfaces that do not require the use of additional components to be connected to the software infrastructure of the robots. With growing task complexity, and the need of multiple hardware platforms to run different computational modules, the neuromorphic paradigm could take advantage of robotic middlewares, such as ROS, or YARP, that are currently seamlessly integrated with neuromorphic sensors and computing devices.

**Advances in Science and Technology to Meet Challenges**

The main components in which robotics is traditionally categorised – perception, decision-making and control – are strictly connected and influence each other. Studying their interplay and understanding how the neural code reflects sensory information, decision processes and how these are influenced by action and vice-versa, is the future research challenge to bring neuromorphic agents to a level of complexity capable of enabling robots to effectively interact with humans.

An advancement in the understanding of the role of different brain areas, their working principles and their interaction with other areas across different temporal and spatial scales shall guide the design of artificial architectures using spiking neuron models, synaptic plasticity mechanisms, connectivity structures to implement specific functionalities. It is crucial to find the right level of detail and abstraction of each neural computational primitive and develop a principled methodology to combine them. Starting from highly detailed models of brain areas, the community shall find reduced models that can implement their basic functionality and that can be implementable on neuromorphic hardware.

As the community is now developing spiking neural networks to extract information from a single sensory modality, the next step would be to take into account information from other sensory modalities, so that decisions depend on the state of the environment, of the robot and of the ongoing task. Among many others, a key area to take inspiration from is the cerebellum, that supports the acquisition of motor plans and their adaptation to the current (sensed) conditions [18]. The resulting computational frameworks shall therefore include dynamic and continuous learning and adaptation.

On the other hand, progress is necessary in the neuromorphic hardware supporting those new frameworks. New circuits for the emulation of additional computational primitives are needed, as well as the possibility to support dynamic, continuous learning and adaptation at multiple timescales.

Specific to the robotic domain, neuromorphic devices should be truly embeddable. To this aim, standardisation of communication protocols, programming tools, online integration with the robot's middleware must be developed. The necessary miniaturisation to pack more computational resources on a single system that can be mounted on a robot goes through the integration of silicon devices with memristive devices. On a longer term, nanotechnology and flexible electronics could represent a viable solution to further miniaturize, or distribute computational substrates that can de-localise computation to the periphery, or create folded structures similar to the cortex, that through folding increased the surface available for computation, achieving higher computational capabilities.

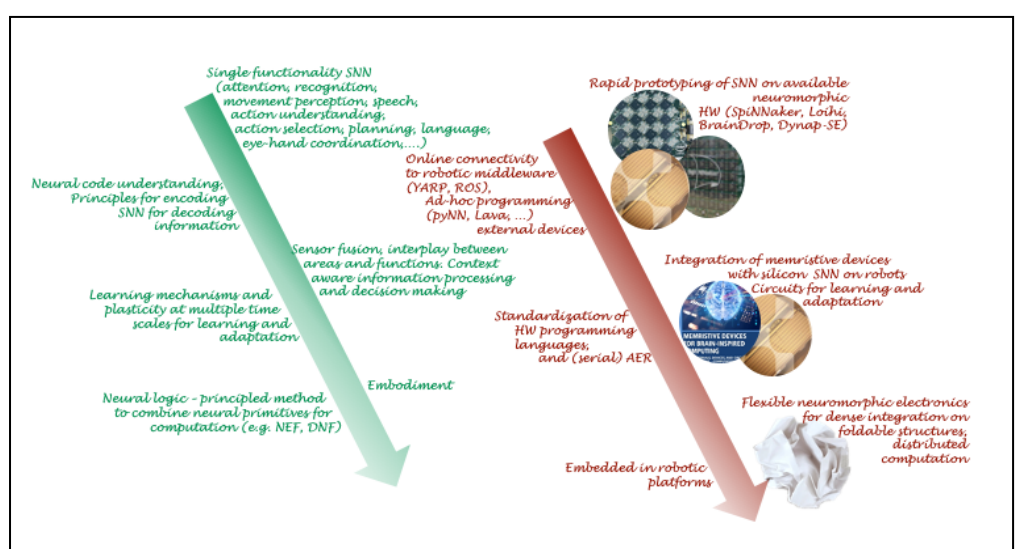

Figure 2. Timeline of a possible development roadmap. In green required theoretical advancements in order of increasing complexity. In red the technological roadmap highlighting the path for new circuit and devices development, as well as the infrastructure needed for the integration on robotic platforms.

**Concluding Remarks**

Neuromorphic sensing, computing and actuation are based on design principles that will make them excel in applications for which animals have evolved. At the same time, animals' behaviour does not only depend on the brain but also on its embodiment, as the shape and distribution of sensors and of muscles, tendons, bones and their interaction influence how the sensory signals are acquired and must be interpreted, and how the control signal has to be delivered. The continuous interplay between brain, body and environment shapes neural computation and this must be taken into account when tailoring neuromorphic computation to robots.

The result of this research path will be crucial and timely for developing the next generation of robots that face the extremely hard challenge of collaborating with humans in human-designed habitat.

The interdisciplinary nature of the neuromorphic approach results in a call for action towards different research communities and different souls within the neuromorphic community. Computational neuroscience and machine learning are called to be the theoretical backbone of neuromorphic computing, micro- and nano-electronics, engineering, physics and material science are called to develop the next generation physical substrate for neuromorphic sensing and computing.

**Acknowledgements**

*The author would like to thank E. Donati for fun and insightful discussions and brainstorming on the topic.*

## 4.2 – Self-Driving Cars

Jonathan Tapson, School of Electrical and Data Engineering, University of Technology, Sydney

**Status**

Self-driving cars have been a staple of science fiction for decades; more recently, they have seemed like an attainable goal in the near future. The machine-learning (ML) boom of the period 2015-2020 gave great cause for optimism, with experts such as the US Secretary of Transport Anthony Foxx declaring in 2016 [1] that "By 2021, we will see autonomous vehicles in operation across the country in ways that we [only] imagine today…My daughter, who will be 16 in 2021, won't have her driver's license. She will be using a service."

This optimism has faded away in the last three years, with the recognition that while it is straightforward to make cars autonomous in simple environments such as freeway driving, there are a multitiude of situations where driving becomes too complex for current solutions to achieve autonomy. It is tempting to refer to these as "corner cases" or "edge cases" – in the sense of being a highly unlikely combination of circumstances, at a "corner" or "edge" of the feature space, which produces a situation where a machine learning algorithms fails to operate correctly – except that, in the real-world of driving, these situations appear to be far more common than was originally expected.

It may be helpful to use the industry terminology when discussing self-driving cars. Self-driving is more formally known as ADAS (Advanced Driver Assistance Systems) and the industry generally uses the Society for Automotive Engineering's (SAE) five-level ADAS model, illustrated below, when discussing autonomous driving capabilities.

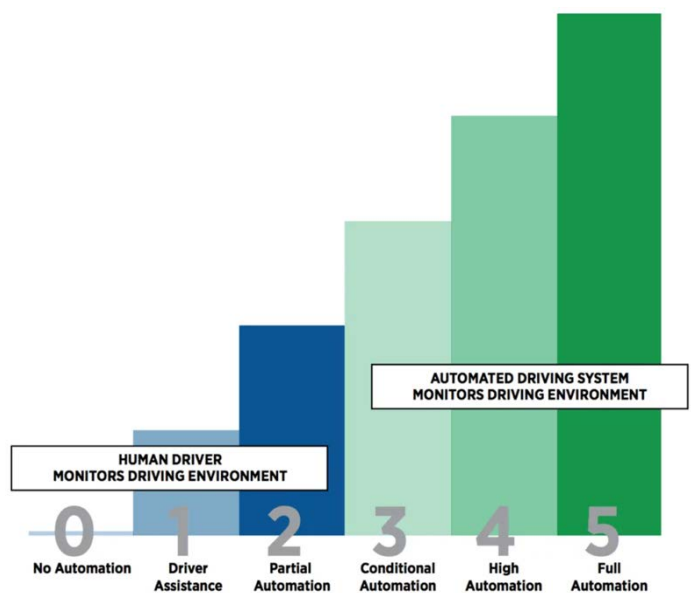

**Figure 1.** The SAE ADAS model. Note that levels 0-2 depend on continuous monitoring by the driver, whereas 3-5 do not. The customary vision of an autonomous car would be ADAS Level 5 – a car which is able to be autonomous in all environments without any human supervision or intervention.

The more recent perception of ADAS progress can be summed up in a quote from Prof. Mary Cummings, Director of Duke University's Humans and Autonomy Laboratory [2]: "There are basically two camps. First are those who understand that full autonomy is not really achievable on any large scale, but are pretending they are still in the game to keep investors happy. Second are those who are in denial and really believe it is going to happen."

Between the optimistic and pessimistic extremes, there is a consensus view amongst ADAS researchers that while full Level 5 ADAS is unlikely to be available in the next five years, Level 4 ADAS is both an attainable and useful target.

**Current and Future Challenges**

One core challenge in achieving high levels of ADAS is perception, and particularly visual perception. It has become apparent that the visual perception of human drivers is extraordinarily hard to reproduce artificially. The problem may be summed up with the following example: most, if not all, human drivers can infer the intent of a pedestrian viewed at a distance of 100m. This is critically important for driving at moderate to high speeds (70-100kmh$^{-1}$) in non-freeway environments, and is as yet impossible to achieve with machine vision [3]. Consider that a pedestrian 1.5m high when viewed at 100m distance subtends an angle of less than 1° vertically and perhaps 0.1° horizontally. When imaged by a video imager through a moderately wide-angle lens, such as is necessary for the forward-facing cameras in an ADAS system, this corresponds to about 3x16 pixels in an HD system. It might be thought that the system is improved by increasing the image resolution (it would be 10x85 pixels in a 4K system) but this also increases the space which must be searched for the pixels of interest.

A second core challenge in ADAS is the complexity of the world that must be modeled by the machine learning systems. Consider the example of a driver who sees a soccer ball bounce across the road in front of their moving car. A human driver is able to draw on an entire world experience concerning balls (and their relationship to perhaps children, or dogs) in reacting to this event. A machine learning system can (generally) only draw upon cases in their training set in determining a response. The effect is that a human is able to respond to events which have never previously occurred, in their or anyone's experience; machine learning systems are, as yet, not reliable in this regard. This is utterly different to the problem faced by, say, a Go-playing ML system – the rules (the world model) of Go can be written in a single page of text. It has taken some time for ML researchers to fully appreciate the difference between these two problems.

A third problem is the power cost of computation required for ADAS. It is obviously difficult to anticipate the power cost of something that hasn't been achieved yet, but the table below gives typical figures. For context, the 2019 Tesla FSD (Full Self Driving) chip is reported to achieve 36 TOPS at 72W power consumption.

| ADAS Capability | Compute Requirements |
| --- | --- |
| L2 | 2 TOPS |
| L3 | 24 TOPS |
| L4 | 320 TOPS |
| L5 | 4000+ TOPS |

**Table 1.** Compute requirements for various ADAS levels (source: Horizon Robotics).

These figures raise questions as to the extent to which ADAS systems will reduce the power and range available, particularly for electric vehicles. One model suggests a range reduction of the order 9-20% for the control systems for Level 4 ADAS [4], which is not insignificant, particularly for electric vehicles where range is a critical issue.

**Advances in Science and Technology to Meet Challenges**

The perceptive reader will see that the problems facing the self-driving car industry are those which are addressed by neuromorphic engineering: modeling and reproducing human-quality perception; building cognitive models of the world; and reducing the power required for real-time perception-cognition-action loop computation.

There has been some interest in using neuromorphic dynamic vision sensors [5] for ADAS vision, but this is hampered by the low spatial resolution of current DVS sensors (which has a complex relationship to the spatio-temporal event processing capacity of the interfaces and downstream processors used) and a perception, perhaps inaccurate, that ADAS will always require conventional (static, framed) sensors, and therefore the inclusion of DVS systems in the senor pack is a costly redundancy. This problem may be solved in due course by hybrid DVS/static imagers. In the bigger picture, the benefits offered by the high dynamic range and spatio-temporal resolution of DVS imagers has led to a general interest in event-based visual processing, which shows some potential to improve ADAS vision processing independently of the use of DVS.

Perhaps the most important feature of human vision which enables driving is the extraordinary resolution of the fovea, coupled with saccadic motion to apply this resolution to small regions of interest (such as pedestrians). There is significant (mostly unpublished) interest in using models of human salience [6] and attention to direct machine vision to areas of the visual (imager) field which require extra or high-resolution processing.

Given the real-time nature of ADAS computation, and the necessity to process correlated streams of visual and 3D point-cloud data (from lidar systems), there is some expectation that event-based neuromorphic computation may be more suitable than current GPU-type computational hardware. At least one neuromorphic event-based hardware startup is focused on real-time vision processing for this purpose [7].

In terms of building cognitive models of the world, we are reaching a point where brute-force approaches to ML are producing diminishing returns. Language models such as GPT-3 [8] can produce impressive passages of text, but it becomes clear that there is no real insight being generated; and these models are trained on orders of magnitude more text than any human could assimilate in a lifetime, suggesting that there is an unfilled deficiency in the model. Neuromorphic approaches such as Legendre Memory Units [9] are offering equal performance to GPT-3 architectures with 10x lower training and memory requirement, suggesting that this may help to close this gap. Similarly, the use of neuromorphic hardware such as Intel's Loihi [10] and GML's GrAIOne chips [7], which are strictly event-based and intrinsically sparse, may provide a computational platform that enables these more biologically realistic machine learning methods.

**Concluding Remarks**

Achieving truly autonomous vehicles is something of a holy grail for engineering in the current era. Recently, there has been a reckoning that it is unlikely to be solved by incremental imporvements based on current technologies. Neuromorphic engineering offers exactly the technological leaps that are required in perception, computation and cognition in order to achieve this goal. In particular, building machine vision systems that acknowledge and model the unique features of human vision, and building computational systems that exploit the event-based sensory flow tht is common to both human and ADAS systems, seem like clear areas for fruitful research.

## 4.3 – Olfaction and chemosensation
Thomas A. Cleland, Cornell University

**Status**

Artificial olfactory systems were early adopters of biologically-inspired design principles. Persaud and Dodd constructed an electronic nose in 1982 based explicitly on the principles of the mammalian olfactory system – specifically, the deployment of a diverse set of broadly-tuned chemosensors, with odorant selectivity arising from a convergent feature detection process based on the pattern of sensor responses to each odorant [1]. Such cross-sensor patterns, inclusive of sampling error and other sources of noise, invite machine learning strategies for classification. Gardner and colleagues subsequently trained artificial neural networks to recognize odorant-associated response patterns from chemosensor arrays [2], and constructed a portable, field-deployable system for this purpose [3].

The biomimetic principle of chemical sensing by arrays of partially selective chemosensors has remained the state of the art [4, 5]. Arrays obviate the need to develop highly selective sensors for analytes of interest, as high specificity can be readily achieved by the deployment of larger numbers of partially selective sensors [6]. Moreover, such systems are responsive to a wide range of chemical diversity (odorant *quality space*; Figure 1), enabling the identification of multiple chemical species and diagnostic odorant mixtures and effectively representing their similarity relationships. The intrinsic redundancy of such chemosensor arrays also renders their responses more robust to contamination or interference, provided the analysis method is able to use the redundant information effectively.

In contrast, strategies for post-sampling signal processing and analysis have varied. Typically, chemosensor array responses are conditioned by electronic preprocessors and then analyzed by one of a range of methods including linear discriminant analysis, principal components analysis, similarity-based cluster analyses, and support vector machines, along with a variety of artificial neural network-based techniques [4, 5, 7, 8]. However, more directly brain-inspired techniques also have been applied to both the conditioning and analysis stages of processing. For example, the biological olfactory bulb (OB) network (Figure 2) decorrelates similar inputs using contrast enhancement [9]. When applied as signal conditioning to artificial sensor data, this operation improved the performance of a naïve Bayes classifier [10]. Similarly, inhibitory circuit elements inspired by the analogous insect antennal lobe (AL) have been deployed to enhance the performance of support vector machines [8, 11]. Finally, fully neuromorphic circuits for analysis and classification have been developed that are based directly on OB/AL circuit architectures (first by [12]; reviewed in [13]; more recently [14, 15]). These approaches are discussed below.

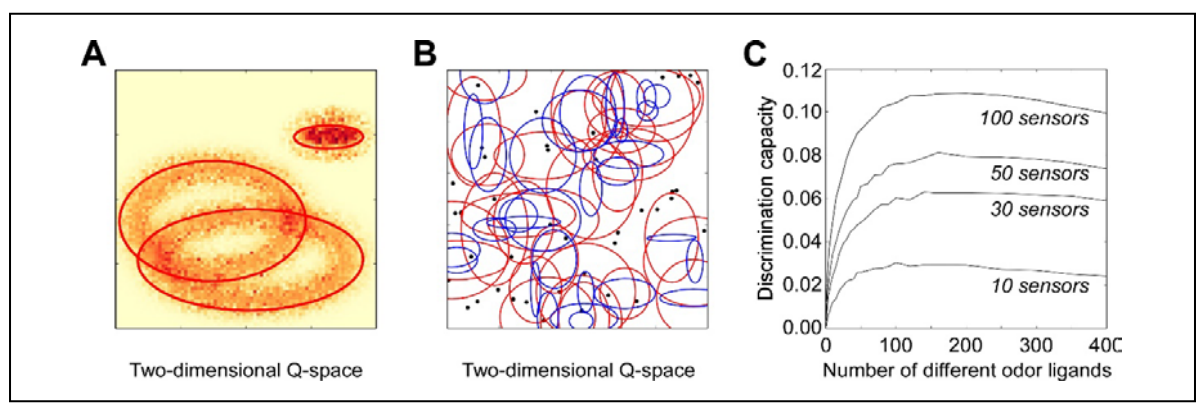

**Figure 1.** Illustration of the capacities of chemosensor arrays to distinguish small changes in odorant quality. [A] Axes denote a two-dimensional quality (Q-) space of physicochemical similarity, ellipses depict the selectivities of three different chemosensors sampling that space. Sensors with broader receptive fields cover a given sensory space more effectively. Discrimination capacity (denoted by *hot colors*) is maximized where the dropoff of sensitivity is steepest, and where the chemoreceptive fields of multiple sensors overlap. [B] Example two-dimensional Q-space with 30 sensors (ellipse pairs, distinguishing selectivity from sensitivity) and 40 chemical ligands (points) deployed. [C] Mean discrimination capacity depends on the number of sensors deployed into a Q-space, shown as a function of the number of competing ligands deployed into the Q-space illustrated in B. Deploying additional chemosensors reliably improves system performance. Adapted from [6].

**Current and Future Challenges**

In many ways, the development of functional/deployable neuromorphic systems is presently limited by algorithms. The energetic and efficiency benefits of neuromorphic engineering – arising from strictly local computation, the colocalization of memory and compute, and spike-mediated communication – are clear in principle, but these properties demand qualitatively new computational strategies to address real-world problems effectively. Developing and optimizing these algorithms for particular tasks of interest is a central challenge for contemporary neuromorphic design.

One strategy for developing effective neuromorphic algorithms is to adapt circuit motifs from the biological brain. Armed with well-developed computational circuit models of olfactory brain structures, neuromorphic olfaction has embraced this strategy, incorporating features such as layered columnar organization, recurrent lateral inhibition, temporal dynamics, spike phase coding, distinct classes of excitatory and inhibitory neurons, custom synaptic learning rules, and adaptive network expansion [9]. Notably, even within these biomimetic constraints, algorithms can differ substantially. For example, lateral inhibition is often applied to decorrelate inputs, but also has served in different implementations to implement attractor dynamics [7, 12, 15], unsupervised clustering [14], and attractor-based denoising [15] in service to different functional goals.

The present challenge is to adapt these biomimetic algorithms to practical applications. Neuromorphic technology is well suited for low-power edge devices, putting a premium on properties like rapid, online learning and adaptation to local conditions that may be partially or wholly unpredictable. Notably, both properties are weaknesses of deep networks [15], which require extensive training with examples of all planned targets including reasonable estimates of predicted future variance. Fundamentally, online learning in neuromorphic olfaction arises from *selective plasticity*, in which the network architecture directs the potentially disruptive effects of new learning to specific circuit elements, coupled with *adaptive network expansion*, which expands the physical network to encompass this new learning and render it accessible. Systems with sufficient capacity for expansion can exhibit *lifelong learning* capabilities. We have referred to this collection of properties as *learning in the wild* [9, 16], and focused on the capacity of such olfaction-inspired algorithms to learn targets from one- or few-shot learning and identify known targets amidst unpredictable interference [15], function in statistically unpredictable environments [17], and mitigate the effects of sensor drift and decay [16]. Notably, working neuromorphic olfaction algorithms have been deployed on diverse edge-compatible hardware platforms including Intel Loihi, IBM TrueNorth, field-programmable gate arrays, and custom neuromorphic devices [7, 12, 15, 18, 19].

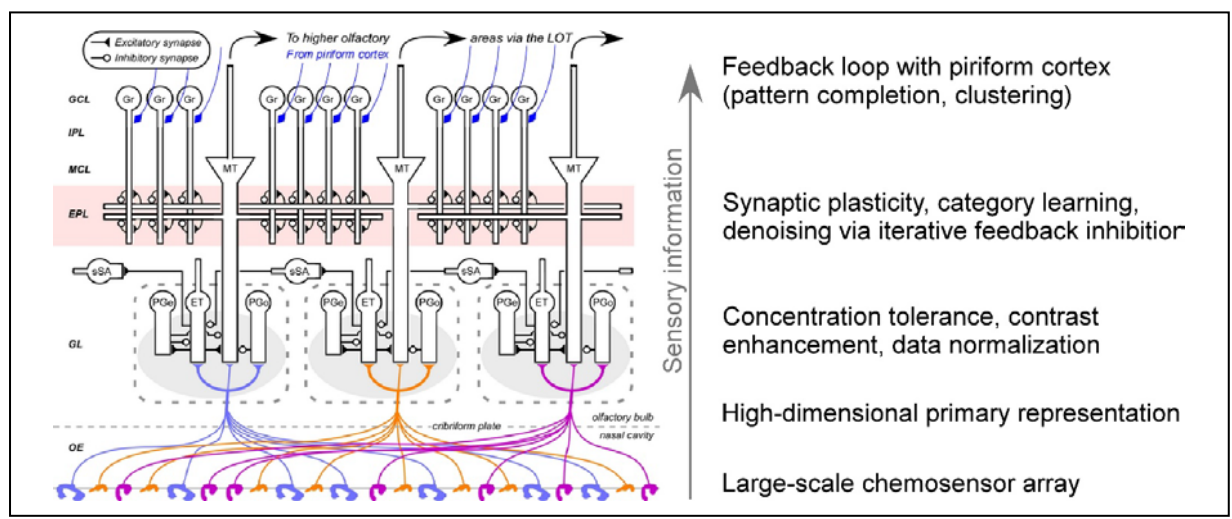

**Figure 2.** Annotated circuit diagram of mammalian olfactory bulb with three sensor classes (denoted by color). Human olfactory bulbs exhibit roughly 400 sensor classes, whereas those of rats and mice express roughly 1200. Glomerular layer (GL) circuitry performs signal conditioning, whereas the formation of target representations depends on synaptic plasticity between principal neurons (MT) and granule cell interneurons (Gr) in the external plexiform layer (EPL). Principal neurons then project to multiple target structures including piriform cortex, which feeds back excitation onto granule cell interneurons. Adapted from [9].

**Advances in Science and Technology to Meet Challenges**

The theoretical performance capacities of machine olfaction are not clearly constrained by anything short of the fundamental signal to noise limits of their deployed environment. The basic capacity to distinguish among similar odorants can be steadily increased by deploying arrays with larger numbers of different chemosensors, provided those chemosensors are responsive to a range of relevant ligands and distinguish between some of them (Figure 1; [6]). The commensurate expansion in network size is tractable on neuromorphic devices in terms of both execution time and energy expenditure (illustrated on Intel Loihi by [15]). The redundancy derived from sensors' overlapping chemoreceptive fields also offers improved resistance to interference from other odorant sources. The actual capacity for such signal restoration under noise depends on the development of circuits that leverage this capacity, and while early efforts are promising [15], there is substantial room for algorithm improvement, such as integrating the pattern completion and clustering capabilities of piriform cortex circuitry, developing cognitive computing methods such as hierarchical category learning to optimize speed-precision tradeoff decisions, and improving performance in the wild. Making these capacities robust requires the development of larger-scale chemosensor arrays, including compact, high-density arrays that can be deployed in the field. Different sensor technologies, optimized both for different sample phases (gas, liquid) and different chemoreceptive ranges of sample quality (e.g., food odors for quality control, toxic gases for safety), will be required. Large libraries of candidate sensors can be screened [4, 20], reducing the need for predictive models of sensors' chemoreceptive fields in this process. However, molecular imprinting technology, developed to produce highly specific chemosensors, now provides this capacity [21], and in principle could be adapted to produce broader receptive fields by imprinting analyte mixtures.

Neuromorphic circuits are not readily adaptable to arbitrary tasks; the domain-specific architectures that underlie their efficient operation also delimit the range of their applications. Olfaction-inspired networks are not limited to chemosensory applications [16], but they are not likely to be effective when tasks do not match their structural priors. However, the characterization and analysis of such fully functional neuromorphic circuits enables the identification and extraction of computational motifs, yielding toolkits that can be intelligently applied to new functional circuits. Moreover, new techniques for spike-based gradient descent learning have successfully demonstrated few-shot learning in neuromorphic circuits preconfigured for the task domain by transfer learning [22]. The design of task-specific neuromorphic circuits in the future is likely to depend on combinations of these strategies, with qualitative circuit elements drawn from theory and generalized domains established therein via emerging optimization strategies.

**Concluding Remarks**

Machine olfaction has been an early and successful adopter of neuromorphic strategies for sampling and computation, in part because of the detailed elucidation of olfactory system networks by experimental and computational neuroscience. Larger-scale sensor arrays and the continued development of postsampling neuromorphic circuitry for signal conditioning and contextually-aware category learning in the wild are critical emphases for near-term progress. More broadly, the benchmarking and analysis of the various computational motifs contained in these functional biomimetic architectures may address broader theoretical and task-dependent questions such as where and when to gate plasticity in selectively plastic architectures, and the transformative capacities of local learning rules deployed within particular network contexts.

**Acknowledgements**
This work was supported by NIH/NIDCD grant R01 DC014701.

## 4.4 – Event Vision Sensors
Christoph Posch, Prophesee

**Status**

Neuromorphic Event-based (EB) vision sensors take inspiration from the functioning of the human retina, trying to recreate its visual information acquisition and processing operations on VLSI silicon chips. The first device of this kind out of C. Mead's group at Caltech, named the "Silicon Retina", made it on the cover of Scientific American in 1991 [1]. In contrast to early more biologically faithful models, often modelling many different cell types and signalling pathways, in turn leading to very complex designs with limited practical usability, in recent years more focus has been put on the creation of practical sensor designs, usable in real-world artificial vision applications. In [2], a comprehensive history and state-of-the-art review of neuromorphic vision sensors is presented.

Today, the majority of EB sensor devices are based on the "temporal contrast" or "change detection" (CD) type of operation, loosely mimicking the transient Magno-cellular pathway of the human visual system (Figure 1). In contrast to conventional image sensors, CD sensors do not use one common sampling rate (=frame rate) for all pixels, but each pixel defines the timing of its own sampling points in response to its visual input by reacting to changes of the amount of incident light [3][6][7]. Consequently, the entire sampling process is no longer governed by an artificial timing source but by the signal to be sampled itself, or more precisely by the variations over time of the signal. The output generated by such a sensor is not a sequence of images but a quasi-time-continuous stream of pixel-individual contrast events, generated and transmitted conditionally, based on the dynamics happening in the scene. Acquired Information is encoded and transmitted in the form of data packets containing the originating pixel's X,Y coordinate, time stamp, and often contrast polarity. Other families of EB devices complement the pure asynchronous temporal contrast function with the additional acquisition of sustained intensity information, either pixel individually [4] or in the form of frames like in conventional image sensors [5].

Due to the high temporal precision of acquired visual dynamics, inherent data sparsity, and robust high dynamic range operation, EB sensors gain increasing prevalence as visual transducer for artificial vision systems in applications where the need for high-speed or low-latency operation, uncontrolled lighting conditions and limited resources in terms of power budget, post-processing capabilities or transmission bandwidth, coincide, e.g. in various automotive, IoT, surveillance, mobile or industrial use cases [8].

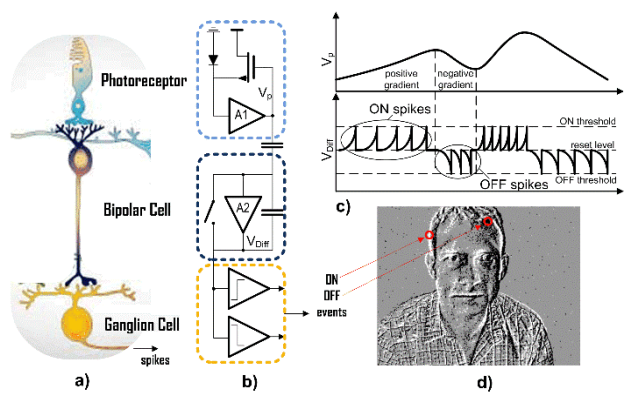

**Figure 1.** a) Simplified three-layer retina model and b) corresponding CD pixel circuitry; in c) typical signal waveforms of the pixel circuit are shown. The upper trace represents an arbitrary voltage waveform at the node Vp tracking the photocurrent through the photoreceptor. The bipolar cell circuit responds with spike events of different polarity to positive and negative gradients of the photocurrent, while being monitored by the ganglion cell circuit that also transports the spikes to the next processing stage; the rate of change is encoded in inter-event intervals; d) shows the response of an array of CD pixels to a natural scene (person moving in the field-of-view of the sensor). Spikes, also called "Events", have been collected for some tens of milliseconds and are displayed as an image with ON (going brighter) and OFF (going darker) events drawn as white and black dots.

**Current and Future Challenges**

Despite their undeniably beneficial characteristics, EB sensors face challenges regarding practical usability and competitiveness in an industrial product. Given the non-deterministic asynchronous nature of the data generation process, leading to e.g. non-constant data rates out of the sensor, the topic of integration of an EB sensor with post-processing into a vision system raises different questions around on-chip event data preparation and interfacing.

Competing technologies for EB vision systems range from conventional computer vision to radar, lidar, ultrasound, passive infrared (PIR) and more. The challenges to make EB a successful technology are related to (1) electro-optical performance of the sensor itself, (2) system integration issues, and (3) challenges around the topic of post-processing for extracting the relevant information from the event data stream.

Pixel size is one of the most important properties for competitiveness with respect to conventional image sensor-based systems. Silicon area, optical format, compatibility with standard optics, camera module dimensions, and form factors dominate cost and applicability of the vision system. Pixel area limits the sensor resolution but also influences the dynamic range of the sensor, in particular its sensitivity at low light levels, which are important factors in key applications such as automotive, mobile or IoT. Thanks to semiconductor technology advances including 3D integration, back-side illumination and improving photodiode efficiency, continuous progress is being made on these issues. With increasing pixel array sizes, the throughput of the sensor readout system and data interface are becoming increasingly relevant to retain the pixel data temporal precision in high-speed/low-latency applications. Power consumption is a key aspect in many artificial vision applications, particularly at the edge. EB vision systems promise to deliver low-power operation thanks to the data sparsity and related efficiency in sensing and processing.

Processing of EB sensor data can be coarsely divided into (1) algorithms that prepare and optimize the raw pixel event data for more efficient transmission and processing on an (external) computation platform, and (2) algorithms that extract higher-level application-specific information out of the event data, to solve a vision task such as e.g. object detection, classification, tracking, optic flow, etc. The first group is preferably implemented close to where the raw data are generated, i.e near-sensor or in-sensor. Typically implemented in an on-chip HW data pipeline, algorithms pre-process the raw pixel data for more efficient transmission and post-processing, also with respect to memory access and processing algorithms requirements. This data conditioning pipeline can include functions such as recoding, formatting, rearranging, compressing, thinning, filtering, binning, histogramming, framing etc. The latter group includes all application-specific vision processing using computer-vision and/or ML-based algorithms and compute models, typically running on some form of application processor. The question of the optimal compute fabric and architecture to be used with EB sensors is unresolved today, and the optimal choice application dependent. However, as discussed widely in other parts of this review, emerging non-Von-Neumann architectures, in particular neuromorphic approaches such as SNN, are better suited to realize an efficient EB system than e.g. general purpose CPUs. Much progress is being made in this area, however challenges remain around the absence of well-established deep learning architectures, including training techniques, for event data, or the lack of largescale datasets.

**Advances in Science and Technology to Meet Challenges**

EB sensors benefit from advances in semiconductor technology. Significant progress was made since early designs with array dimensions of about thousand pixels and pixel sizes of 1000µm$^2$, to today, about 15 years later, where the most advanced EB sensors have 1-megapixel arrays (1000x larger) with 20µm$^2$ pixels (50x smaller) [7] (Figure 2). In contrast to conventional image sensors, typically having one photodiode and 4 or 5 transistors in each pixel, pixels of EB sensors are small analog computers, typically using 50+ transistors for the required signal processing and communication functions. Early EB sensor designs employed standard CMOS processes and later front-side illuminated (FSI) CIS processes. All transistors and other devices needed to be placed next to the photodiode, leading to large pixel sizes and low fill factors. The introduction of back-side illuminated (BSI) CIS processes relaxed the situation and today, 3D wafer stacking combined with small pitch metal bonding allows to place most of the pixel circuitry underneath the photodiode. Latest generation EB sensors today reach pixel pitches below 5µm and fill factors above 75% [7].

Following the CMOS technology and integration roadmaps will yield EB devices with increasing industrial applicability. Further advances in production and packaging technologies like triple wafer stacking, die-stacking system-in-package and wafer-level optics will support the trend to autonomous ultra-low power/small form-factor edge perception devices and artificial intelligence systems where the sensor is highly integrated and tightly packaged with pre-processing and application processing, thereby significantly reducing power consumption and transmission bandwidth requirements of an artificial vision system, e.g. in IoT, mobile or perception networks applications.

A big impact on the usability and competitiveness of EB systems is expected to come from future advances in neuromorphic computing and event-based processing techniques. Spiking neural networks (SNN) are a natural fit for post-processing to the data generated by EB sensors [8][9]. But the sparse data output of EB sensors is also a good match to future hardware accelerators for conventional deep neural networks (DNN) that exploit activation and network sparsity [10].

Recently, new kinds of neuromorphic vision devices beyond CMOS have been demonstrated, exploiting different electro-optical material properties and fabrication techniques to further advance the tight integration of sensing and processing, often combining photon transduction and analog neural network (ANN) functions into a single fabric [11]-[14] . Even though these devices are in their early proof-of-concept phase, interesting and promising results have already been demonstrated.

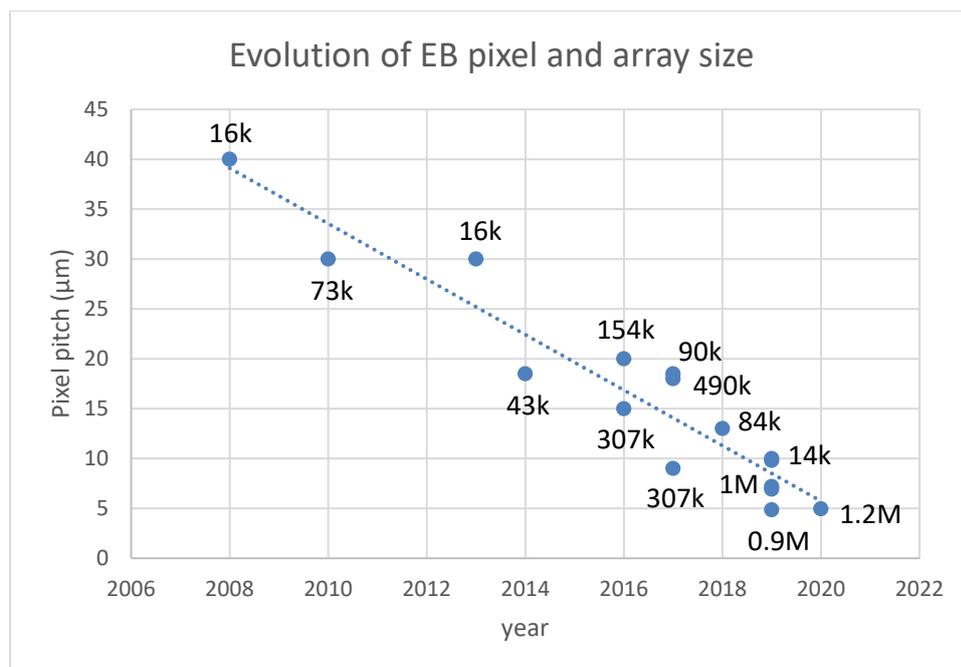

**Figure 2.** Evolution over time of pixel pitch and array size of CD-based EB sensors.

**Concluding Remarks**

Integrating event-based vision sensing and processing with neuromorphic computation techniques is expected to yield solutions that will be able to penetrate the artificial vision market and gain considerable market share in the coming years [16]. This new class of extremely low-power and low-latency artificial intelligence systems could, In a world where power-hungry deep learning techniques are becoming a commodity, and at the same time, environmental concerns are increasingly pressuring our way of life, become an essential component of a sustainable society.

## 4.5 – Neuromorphic Audition
Shih-Chii Liu, Institute of Neuroinformatics, University of Zurich and ETH Zurich

**Status**

Neuromorphic audition technology is inspired by the amazing capability of human hearing. Humans understand speech even in difficult auditory scenarios and using a tiny fraction of the brain's entire 10W. Matching the capability of human hearing is an important goal of the development of algorithms, hardware technology and applications for artificial hearing devices.

**Brief History:** Human hearing starts with the biological cochlea which uses a space-to-rate encoding. The incoming sound is encoded as asynchronous output pulses generated by a set of broadly frequency-selective channels [1]. Fsor frequencies below 3 kHz, these pulses are phase locked to the frequency [2]. This encoding scheme leads to sparser sampling of frequency information from active frequency channels instead of the maximal sampling rate used on a single audio input. The first silicon cochlea designs starting with the work of Lyon and Mead (electronic cochlea) [3], model the basilar membrane (**BM**) of the cochlea by a set of coupled filter stages. Subsequent designs include those with better matching properties for the filter stages and using coupled filter architectures ranging from the originally proposed cascaded type modeling the phenomenological output of the cochlea [3], to a resistively-coupled bank of bandpass filters that models the role of the BM and the cochlear fluid more explicitly [4][5].

Later designs include models of the inner-hair cells (**IHC**s) on the BM, that transduce the BM and fluid vibrations into an electrical signal. They are frequently modelled as half-wave rectifiers (**HWR**s) in silicon designs. Some designs include the automatic gain control mechanism of outer hair cells (**OHC**s) that are useful for dealing with larger sound volume ranges from 60 – 120 dB. Cochlea designs starting from the early 2000s include circuits that generate asynchronous binary outputs (or spikes) encoded using the address-event representation (**AER**). Details and historical evolution of these VLSI designs are described in [4][5]. Recent spiking cochlea designs in more advanced technologies such as 65nm and 180nm CMOS demonstrate better power-efficiency (e.g., < 1 uW/channel in [6]). These new designs show competitive power efficiency compared to other audio front end designs that compute spectrogram features from regular samples of a single audio source.

**Importance of field:** In the early 2000s, cochlea circuits were developed for audio bionic applications [7] and models of biological auditory localization circuits [8]. With increasing prevalence of voice-controlled devices in everyday life, neuromorphic and bio-inspired solutions can potentially be interesting because of the need for low-latency and energy-efficient design solutions in audio edge application domains.

**Current and Future Challenges**
**Big research issues:** Neuromorphic bio-inspired features such as the sparse sampling (e.g., the non-sampling during silent pauses), the event-driven form of brain computing, the natural temporal encoding carried by the asynchronous events (pulses, or sets of pulses, within a specified time window), can enable more energy-efficient solutions for hardware-friendly models that solve an auditory task. The timing information from the asynchronous events of the spiking silicon cochleas is ideal for extracting interaural time differences (ITD), which is useful for spatial audition at lower latencies [9][10]. Speech recognition examples that use spiking cochlea inputs include the reservoir networks of spiking neurons applied to digit recognition [11].

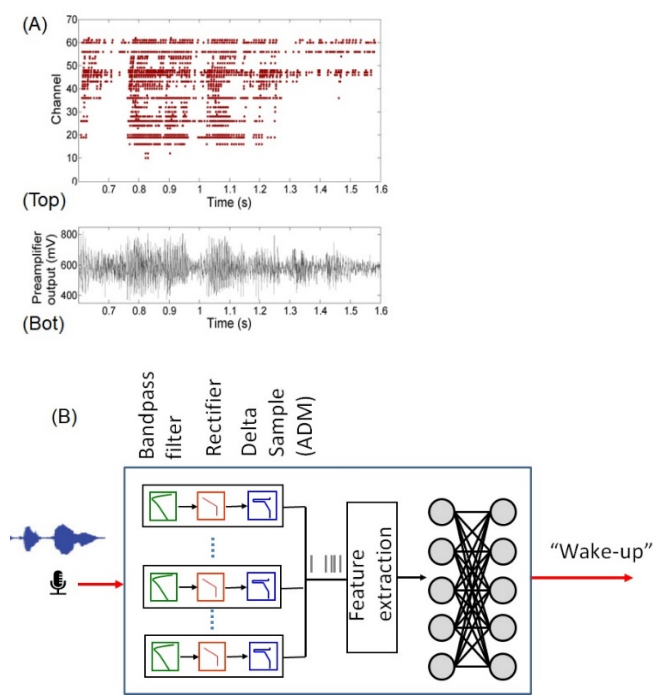

**Figure 1.** (A) Top shows the 64-channel cochlea output spike rasters for a speech sample in (Bot). (B) Architecture for an example audio keyword spotting task. Figure shows an ASIC block that combines the Dynamic Audio Sensor (DAS) front-end with a continuous-valued DNN [12][14] for an example "wake-up" keyword spotting task. Spike outputs of the local filter channels are generated using asynchronous delta modulation (ADM) [6]. The spike events can be used to drive a SNN directly [13].

With the advent of deep learning, deep neural networks (DNNs) have been used successfully for many audio benchmarks. Even though training of these networks requires global supervised methods and large datasets, they can be applied on features from spiking cochlea events for always-on low-level audio tasks such as voice activity detection (**VAD**) and key word spotting (**KWS**). These modules can then activate the more energy-expensive audio tasks such as speech recognition which require running larger networks on the cloud. The energy savings of these neuromorphic solutions come from the event-driven way of processing, i.e., the processing is triggered by asynchronous information carried by the frequency-selective events rather than on regular time-stepped spectrogram frames; compute savings come naturally with silent pauses and non-changing inputs. Hardware versions of this combination include a complete system using an FPGA recurrent network processing cochlea spikes for a continuous digit recognition task [12] and TrueNorth spiking network platform processing audio features [13]. ASIC implementations that include both sensor and network show extremely low-power consumption, e.g., the < 1 uW ASIC VAD chip [14].

**Mapping** spiking neural network (SNN) algorithms trained for a task to SNN hardware platforms require additional considerations such as the variability of the network parameters from transistor mismatch and the timing jitter noise of the transmitted spikes from the transmission of asynchronous spikes in real-time on a specified SNN hardware platform. While more work is needed on local spike-based learning rules [15] to configure SNNs to reach similar accuracy compared to continuous-valued artificial neural network (ANN) solutions on a specific task, conversion techniques that map trained

ANNs to SNNs [16] and global supervised training methods for SNNs have been effective for this goal [17].

**Advances in Science and Technology to Meet Challenges**

For edge applications, besides the high energy-efficiency needed of the hardware platform, maintaining high audio task accuracy is also important. More work is needed to show that neuromorphic audition systems can be competitive for these domains, e.g., through benefits of spike events for efficient models or local learning.

Audio spectrogram features are dominant features used in conventional audio applications. Studies such as [11][18] show that the accuracy on a simple speech recognition task using spiking cochlea features while showing lower numbers than spectrogram features for clean conditions, can maintain lower accuracy loss over decreasing signal-to-noise conditions. These ideas need further investigation for larger datasets and tasks. Features extracted from the spikes, e.g., exponential features [19], could help to determine the utility of dynamic information. The real-time extracted azimuth information allows source separation from a mixture of two speakers, e.g., by using the streamed source spikes for speaker identification and keyword spotting using deep network algorithms [19]. Future research is needed for models that operate robustly in wide dynamic range (> 60 dB) auditory scenes.

More extensive work is needed to develop training or spike-based learning methods to configure for high-accuracy networks on specified audio tasks. Training directly on spiking networks instead of using conversion ANN to SNN methods would capitalize better on implicit SNN encoding capabilities. The approximation of rate codes to analog activations of ANNs is not ideal. Temporal coding methods would be more energy efficient but need extension to more complex datasets. Combination of the auditory systems together with other modalities for a robotic platform would also be interesting for future developments.

**Technology**: Technology advances include the continual support of analog integrated circuit designs as technology nodes scale down for digital transistors. Of importance for low-power devices are transistors with low leakage current (for low standby power) and fabrication processes that still allow low transistor mismatch without requiring large transistor dimensions. The standby power consumption of the ASIC is an important metric for always-on audio devices. Other challenges include algorithms that are hardware aware, e.g., to the variability of the network parameters after the ASIC fabrication, and approaches to reduce memory access or to create predictable memory access patterns to reduce energy loss from the unpredictable memory accesses of SNNs. The emerging large-scale availability of high-density local memory is also an interesting component of future research for the ASIC development.

**Concluding Remarks**

Progress in neuromorphic audition critically depends on advances in both silicon technology and algorithmic development. Recent ASIC prototypes that combine binary networks and spectral features report competitive low-power numbers ( < 1 uW) for KWS [20]. Evaluation of these solutions with ASICs that use a spiking cochlea front-end will help determine the latency-accuracy-power tradeoff between these approaches for audio edge applications. Other considerations are the system-level power of a complete audio system including microphones (current commercial audio assistants are ~1W), that continue to benefit from the ultra-low power numbers of the audio ASICs. Also needed are event-driven algorithms to address more complex tasks beyond KWS and VAD while maintaining the energy efficiency and low latency benefits, and how we can capitalize on the new emerging high-density memory technologies to reduce the power dissipation from unpredictable memory accesses of events/spikes. The combination of the Dynamic Audio Sensor with other spiking sensors, e.g., the Dynamic Vision Sensor, will be interesting for robotic platforms. Hybrid solutions that bring knowledge from auditory science, signal processing, and machine learning will enrich future artificial neuromorphic technology solutions for audio application domains.

### Acknowledgements
We acknowledge the Sensors Group members and colleagues who have worked on the Dynamic Audio Sensor design and audio systems. Partial funding provided by the Swiss National Science Foundation, HEAR-EAR, 200021172553.

## 4.6 – Biohybrid Systems for Brain Repair

Gabriella Panuccio, Enhanced Regenerative Medicine, Istituto Italiano di Tecnologia, Italy

Mufti Mahmud, Department of Computer Science and Medical Technologies Innovation Facility, Nottingham Trent University, UK

**Status**

Biohybrid systems are established by biological and artificial components interacting in a unidirectional or bidirectional fashion. In this section, we specifically refer to neurons or brain tissue as the biological component of biohybrid systems for brain repair.

The first demonstration of a biohybrid dialogue was achieved in vitro at the beginning of the 1990's by Renaud-LeMasson and colleagues, who established a communication between a biological neuronal network and a computational model neuron [1]. Soon after, Chapin and colleagues brought the biohybrid paradigm to the in vivo setting by providing the first proof of concept of interfacing the brain with a robotic end-effector [2], a paradigm that has recently become a reality in clinical research [3].

Biohybrid systems are now a widespread approach to address brain dysfunction and devise novel treatments for it [4]. Representative examples are electronic devices coupled to biological neurons in vitro [5] or to the brain in vivo [6] and establishing a bidirectional communication through a closed-loop architecture. A key feature of such systems is the real-time processing and decoding of neural signals to drive an actuator for brain function modulation or replacement. To this end, enhancement of biohybrid systems with artificial intelligence is the emerging strategy to achieve an adaptive interaction between the biological and artificial counterparts. Neuromorphic engineering represents the latest frontier for enhancing biohybrid systems with hardware intelligence [7] and distributed computing [8], offering unprecedented brain-inspired computational capability, dynamic learning of and adaptation to ongoing brain activity, power-efficiency, and miniaturization to the micro-scale. In particular, the intrinsic learning and adaptive properties of neuromorphic devices present the key to bypass the typical trial-and-error programming along with the stiff pre-programmed behaviour of current brain implantable devices, such as those used for deep-brain stimulation. In turn, such a unique potential enables surpassing the drawbacks of current mechanistic approaches with a phenomenological (evidence-based) operating mode. Overall, these features serve as an asset to attain a physiologically-plausible interaction between the biological and artificial counterparts.

The latest avenue for biomedical applications is neuromorphic-based functional biohybrids for brain regeneration. These are hybridized brain tissue grafts (Fig. 1), wherein the neuromorphic counterpart(s) emulate and integrate brain function, aiming at guiding the integration of the biological graft into the host brain. This crucial aspect cannot be attained by a purely biological regenerative approach. Further advances in neuromorphic biohybrids are thus expected to bring unparalleled strategies in regenerative medicine for the brain: by providing symbiotic artificial counterparts capable of autonomous and safe operation for controlled brain regeneration, they herald a paradigm shift in biomedical interventions for brain repair, from interaction to integration.

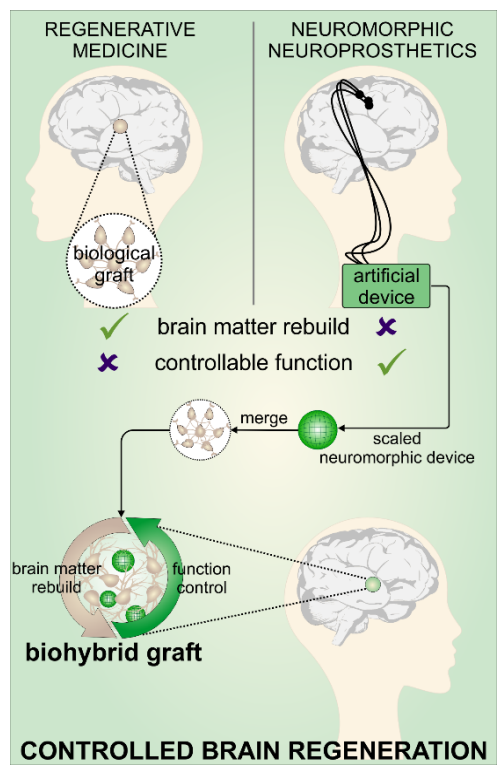

**Figure 1. Concept of functional biohybrids for brain regeneration.**
Functional biohybrids merge concepts from regenerative medicine (rebuild of brain matter) and neuromorphic neuroprosthetics (adaptive control of brain function). The symbiotic interaction between the biological and artificial counterparts in the biohybrid graft is expected to achieve a controlled brain regeneration process.

**Current and Future Challenges**

As biohybrid grafts for brain repair are fully integrated autonomous constructs, their achievement entails the full physical integration of the neuromorphic counterpart(s) within the bioengineered brain tissue to be grafted. However, while neuromorphic neuroprostheses coupled to bioengineered brain tissue may constitute in themselves a biohybrid system, they do not make up a biohybrid graft in the strict sense, being neuroprostheses conceived as an exogenous body.

In functional biohybrids, the integrated neuromorphic counterpart(s) must not perturb the mechanical equilibrium of the biological neurons; hence, depending on the intended application (i.e., extracellular or intracellular biosensing from neurons and their networks), these must be scaled down to meet the size of a small neuronal ensemble (<150 $\mu$m [9]) or even to achieve intracellular residency (<3 $\mu$m [10, 11]). In both cases, the primary challenge, and the pre-requisite, is the aggressive miniaturization of the integrated neuromorphic devices(s) without compromising their computational capability. Thus, functional biohybrids entail the realization of neuromorphic dust. Further challenges arise due to the physical confinement of the neuromorphic counterparts within the biohybrid graft, stemming in the requirement of:

(i) Power-autonomy: device powering cannot rely on a wired power supply unit, such as commonly used subcutaneous batteries: while continuous device operation without the need of battery replacement is of utmost importance in brain regeneration, wiring neural dust is unfeasible (the devices are ultrasmall and physically inaccessible). Further, the operation of an autonomous system, by definition, should not depend on external components.

(ii) Wireless operation: this is required to follow the graft's evolving function during the regeneration process, to enable wireless device re-programming and hardware failure monitoring.

(iii) On-chip learning, supported by application-specific integrated circuits for advanced signal processing, to follow the evolving temporal dynamics of the graft during its integration within the host brain, without the aid of an external controller.

(iv) Bioresorbable property: In aiming at healing brain damage, the neuromorphic counterparts should be regarded as a temporary aid in the process. Thus, they should be removable upon completion of brain repair. While non-invasive micro-surgery techniques, such as high-intensity focused ultrasound, may permit removal of mm-sized devices, this is not technically feasible in the case of ultrasmall (and, even more so, intracellular) devices. Thus, particularly relevant to functional biohybrids is that the neuromorphic counterparts should be bioresorbable.

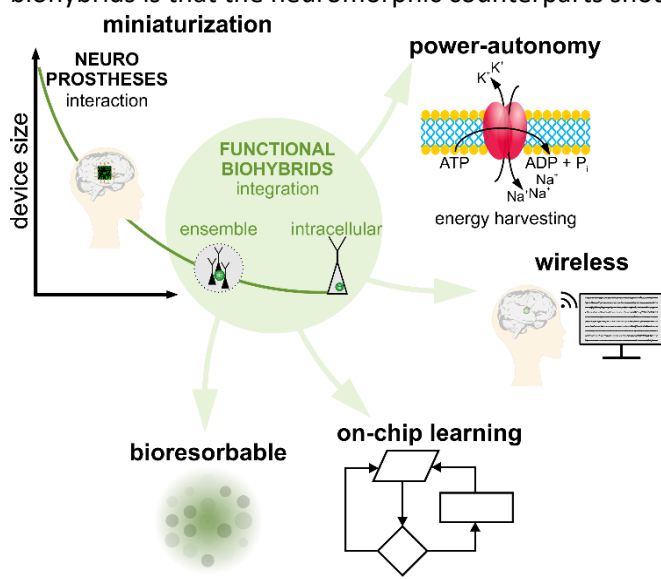

Figure 2. Challenges to achieve functional biohybrids for brain regeneration.

**Advances in Science and Technology to Meet Challenges**

**Miniaturization –** The device size will reflect the complexity of the required architecture. Such a complex system requires integrated interfacing circuits, for which the roadmap toward 1 nm-gate length of CMOS technology [12], along with avant-garde fabrication and integration technologies hold promise to meet this outstanding challenge. Further, novel materials for improved performance are expected to emerge in the near future to enable higher computational capability within smaller sizes. Organic materials hold great promise to merge advanced computation with extreme miniaturization [13].

**Power autonomy** – While radiofrequency [14] and ultrasonic waves [15] are successfully deployed for powering deep biomedical implants, they require an external transmitter. Energy harvesting from the human body may represent the strategy of the future [16], wherein ATP seems to be the most promising solution to a long-standing issue [17]. However, caution must be made to avoid negative physical effects on the patient, as the amount of harvested energy must be much less than what available. Thus, the device components must be ultra-low-power and energy efficient.

**Wireless operation –** While autonomous operation is a key feature of the neuromorphic dust, the need of patient monitoring, device fine-tuning, and hardware failure checks should not be underestimated for guaranteeing the patient's safety. To this end, wireless access to the device is fundamental. Thus, dedicated integrated circuits are required, which must be ultrasmall so not to introduce bottlenecks in device miniaturization. As stated above, advanced CMOS technology holds promise to enable these wireless features in neuromorphic dust. Further, protocols tailored to energy efficient wireless communication are needed.

**On-chip learning –** Understanding spatiotemporal patterns is a key feature to address the evolving dynamics of neuronal networks and reverse-engineer brain dysfunction. So far, these features have been achieved by pre-programming and the use of a microcontroller [18]. Further advances must be made in order to achieve the same level of performance through on-chip learning. This would enable to address the inter-individual variability of the human brain while overcoming the drawbacks of trial-and-error (re)programming and of the need of a wired controller.

**Bioresorbable materials –** The device materials must be fully biocompatible so not to release cytotoxic compounds in the patient's brain. While outstanding advances have been made in the biosensors field [19], a major efforts must be put in the field of neuromorphic engineering, where the performance of the device strongly depends on materials. In this regard, organic materials may present the key to beat this challenge.

**Concluding Remarks**

Much progress has been made to incorporate artificial devices within neural spheroids [20], or inside single living cells [10, 11]. However, these are biosensors of simple architecture, far from the complexity of a neuromorphic system. The achievement of functional biohybrids thus requires the parallel effort from diverse disciplines to address the numerous challenges linked to them. The exponential progress in fabrication and miniaturization strategies, energy harvesting, learning algorithms, wireless technology, and bioresorbable bioelectronics heralds the feasibility of cutting-edge biohybrid neurotechnology for controlled and safe brain regeneration.

**Acknowledgements**

*This work was funded by the European Union under the Horizon 2020 framework programme through the FET-PROACTIVE project HERMES – Hybrid Enhanced Regenerative Medicine Systems, Grant Agreement n. 824164.*

Roadmap on Neuromorphic Computing and Engineering[18] W. Wang et al., "Learning of spatiotemporal patterns in a spiking neural network with resistive switching synapses," *Science Advances,* vol. 4, no. 9, p. eaat4752, 2018, doi:https://doi.org/10.1126/sciadv.aat4752.

[19] X. Huang, "Materials and applications of bioresorbable electronics," *Journal of Semiconductors,* vol. 39, no. 1, p. 011003, 2018/01 2018, doi:https://doi.org/10.1088/1674-4926/39/1/011003.

[20] A. Lecomte, L. Giantomasi, S. Rancati, F. Boi, G. N. Angotzi, and L. Berdondini, "Surface-Functionalized Self-Standing Microdevices Exhibit Predictive Localization and Seamless Integration in 3D Neural Spheroids," *Advanced Biosystems,* vol. 4, no. 11, p. 2000114, 2020, doi:https://doi.org/https://doi.org/10.1002/adbi.202000114.

## 4.7 – Embedded Devices for Neuromorphic Time-Series Assessment

Arnab Neelim Mazumder, University of Maryland, Baltimore County
Morteza Hosseini, University of Maryland, Baltimore County
Tinoosh Mohsenin, University of Maryland, Baltimore County

**Status**

Neuromorphic computing aims to mimic the brain to create energy-efficient devices capable of handling complicated tasks. In this regard, analysis of multivariate time-series signals has led to advancements in different application areas ranging from speech recognition and human activity classification to electronic health evaluation. Exploration of this domain has led to unique bio-inspired commercial off-the-shelf device implementations in the form of fitness monitoring devices, sleep tracking gadgets, and EEG-based brain trauma marker identifying devices. Even with this deluge of work over the years, the necessity of evolving the research direction with day-to-day needs relating to this sphere is still pivotal. The key idea behind the wealth of research in these domains comes from the fact that it is very difficult to generalize human abilities and activities, and it is even more difficult to create devices that can operate at a level as accurate as human-level perception. This is where contemporary machine learning and the more modern deep learning frameworks shine. The current scenario of using automated devices for a variety of health-related applications requires that these devices become more sensitive, specific, user-friendly, and lastly accurate for their intended tasks. This relates to further advancements in the region of algorithm construction and constraint-based design of implementable hardware architectures. The current crop of research in this area investigates deep neural network (DNNs) architectures for the purpose of feature extraction, object detection, classification, etc. DNN models utilize the capacity of convolutional neural networks (CNNs), recurrent neural networks (RNNs), and even to some extent fully connected layers to extract spatial features for time-series assessment which was previously exhaustively calculated via different hand-engineered feature extraction techniques coupled with simple classification algorithms. Along with this, RNNs and their advanced equivalents in the form of long short term memory networks (LSTMs) and gated recurrent unit (GRU) has also been integrated into the deep learning architectures to handle timeseries signals. The idea behind this integration stems from the fact that RNNs and LSTMs are modeled in such a way that they can keep track of previous instances of the input data in order to make a prediction, which makes these architectures very effective for pattern and dependency detection within the time-series data. The other aspect of developing these diverse DNN models is to make them readily implementable in terms of hardware accelerators and therein lies the issue of hardware constrained efficient designs. As a consequence, the computation and model size specifications of different hardware-oriented approaches will result in the advancement of application-oriented software designs which will, in turn, increase the reliability and efficiency of these embedded devices.

**Current and Future Challenges**

There are several challenges associated with managing time-series signals for classification or recognition tasks. One of the foremost issues of time-series classification is to make these signals interpretable by the DNNs as these signals contain multiple variables relaying information about concurrent actions and it is difficult to process these signals in their raw form. Authors in [1] proposed a solution to this problem by transforming these multimodal signals into windowed images based on their sampling frequencies. Another obstacle that is related to time-series analysis pertains to skewed or imbalanced information belonging to multimodal variables as the data collection procedure with different sensors might not always be the same. As a way around, a common practice is to use weighted sampling of the input features during the training of the DNN models so as to balance the impact of all features. Pruning outliers in the dataset by eliminating unnecessary sensor data can alleviate this problem as demonstrated by the authors in [2], however, it is not always feasible to delete multimodal information as the sensor data for multiple variables might be correlated. In addition to this, many of the software frameworks dedicated to time-series classification do not consider the large computation overhead of the DNNs. This has a significant impact when these frameworks are replicated on to resource-limited and low-power embedded platforms where the use of off-the-chip-memories becomes essential. As a result, the performance remains limited by the memory bandwidth while the power consumption stays high due to the rapid accessing of off-thechip memories. The extent of these complications has introduced shallow networks [3], approaches to quantizing model parameters [4] along with ternary [5] and binary [6] models that focus on reducing the memory overhead for efficient resource-constrained hardware accelerator implementation. Authors in [7] provide an example of a fixed-point CNN classifier involving 4-bit fixed point arithmetic that suggests negligible accuracy degradation and authors in [8] present fast BNN inference accelerators to meet the FPGA on-chip memory requirements. Reducing memory footprints in hardware accelerators is also tied up to the cost-effective designing of memory units. On top of this, managing and limiting frequent accesses of these memory units also contribute to latency, power, and energy efficiency as a whole. Thus, a critical challenge in terms of hardware design is to maintain high frequency and energy efficiency with low energy consumption.

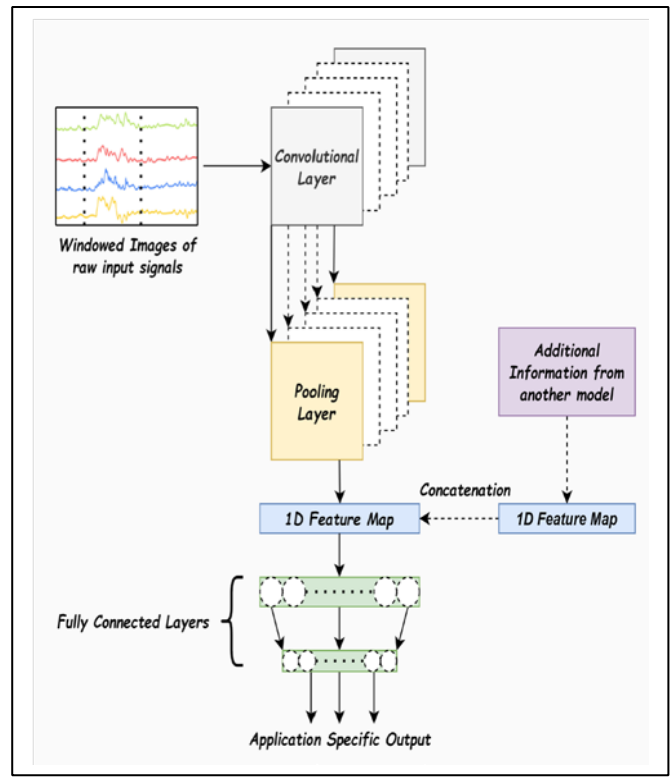

**Figure 1.** The deep learning framework takes in windowed images of the raw multimodal time-series signals as input to the convolutional layers. Correspondingly, feature extraction is achieved in convolutional layers which results in a two-dimensional feature map. The pooling layers contribute to reducing the feature map size while keeping the spatial features intact. This two-dimensional pooled feature map is reshaped to have one-dimensional form so that it can be forwarded to the next fully connected layers. Finally, the last fully connected layer will have neurons equal to the number of outputs as desired by the application. Furthermore, with regards to multi-input model, supplementary information coming from a separate model can be concatenated with the one-dimensional feature map to bolster the inference accuracy.

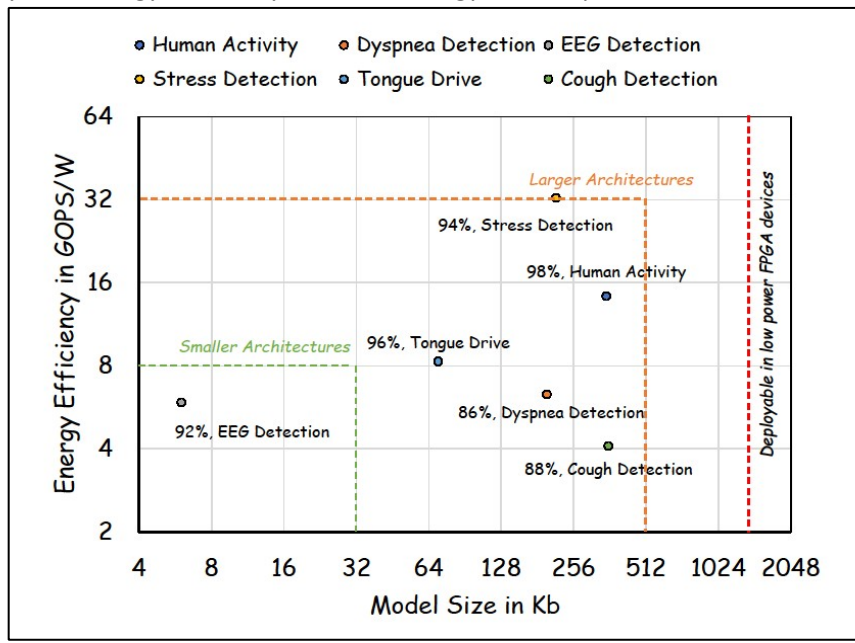

**Figure 2.** This figure illustrates the trend of energy efficiency against model size of different deep learning architectures deployed on the low power Artix–7 100t FPGA platform which has a memory of 1.65 Mb. The applications focused here are EEG detection [10], human activity recognition [1], stress detection [1], tongue drive systems [1] along with cough and dyspnea detection as part of respiratory symptoms recognition [9]. Depending on the model size, the frameworks can be tiny or large whereas the energy efficiency is dictated by the performance of the design. In the same vein, the plot also shows the device inference accuracy for the different models ranging from 86% up to 98% which further justifies that these architectures are specific enough for low power embedded deployment.

**Advances in Science and Technology to Meet Challenges**

Deep learning frameworks have been widely successful for classifying time-series signals. However, the challenges mentioned in the previous section make this task ever more difficult. To further boost the performance of deep learning methods for time-series data, some form of digital signal processing is commonly required. To this extent, a common practice is to convert these raw waveforms into windowed overlapping time-series frames. A sliding window of some specific size along with a stepping size is passed through all variables, creating a set of images of shape as desired by the user. Since most time-series signals contain label information at precise time intervals, it is fairly easy to determine the label of the images. Another facet of dealing with time-series signals requires feature extraction relevant to the application that is being targeted. With classical machine learning algorithms, this was achieved using several mathematical and analytical processes to determine the correlation between variables. In contrast, one of the strengths of using CNNs or RNNs in deep learning ensures that the relevant features are being extracted in image or time-space. A general practice in time-series classification is to deploy CNN or RNN layers in conjunction with pooling layers as illustrated in Fig.1. The pooling layers reduce the feature map size so that the cost of computation for the following fully connected layers is minimized. Additionally, the feasibility of hardware deployment of these deep learning algorithms depends on the computational complexity and size of these architectures. It is imperative that these frameworks are reduced in size via quantization, pruning, or by making the networks shallow in the first place so that they fit on embedded devices with small memories. Hence, there comes a point where the designer has to find the sweet spot between the accuracy of the model and the practicality of its size being suitable for low power embedded platforms while also ensuring that the energy efficiency of the target device is also satisfactory. Fig. 2 shows a comparison among different models with a variety of applications for their model size, classification/detection accuracy, and energy efficiency which establishes that depending on the application, deep learning models can fit on low-power embedded devices with standard performance. Also, a modification to these frameworks can take in additional information in the form of vectors from a separate model to enhance the overall accuracy of the model as demonstrated in
Fig.1.

**Concluding Remarks**

Human-related time-series data analysis encompasses a wide range of tasks including speech recognition, keyword spotting, health monitoring, and human activity recognition to name a few. This also allows the dedicated development of embedded devices suited for accelerating such tasks. Challenges in processing such time-variant data for device implementation range from pre-processing the raw signals and removing noise and outliers to interpreting long and short dependencies that exist within the nature of the data. Windowing the continuous stream of data into overlapping frames to be processed using a simple DNN or CNN is a common practice for real-world applications in which the long dependencies in data are negligible. On the other hand, novel approaches such as RNNs and LSTMs can improve the overall confidence of analysis for time-series data with long dependencies. When implementing all these methods on resource-bound hardware in which power, energy, memory footprint, and application latency are all limited, it is of utmost importance to design deep learning algorithms with small model sizes and low computation that meet all the application requirements and hardware limitations. In conclusion, there must be a trade-off between performance and implementation feasibility to justify the use of low-power embedded devices to replicate deep learning applications of time-series assessment.

**Acknowledgements**

*[Please include any acknowledgements and funding information as appropriate.]*

Roadmap on Neuromorphic Computing and Engineering[5]  H. Alemdar, . V. Leroy, . A. B. Prost and F. Pétrot, "Ternary neural networks for resource-efficient AI applications," in *International Joint Conference on Neural Networks (IJCNN)*, Anchorage, AK, USA, 2017.

[6]  M. Courbariaux, Y. Bengio and J. P. David, "Binaryconnect: Training deep neural networks with binary weights during propagations," *Advances in neural information processing systems,* vol. 28, pp. 3123-3131, 2015.

[7]  C. Y. Lo, F. C. M. Lau and C. W. Sham, "Fixed-Point Implementation of Convolutional Neural Networks for Image Classification," in *International Conference on Advanced Technologies for Communications (ATC)*, Ho Chi Minh City, Vietnam, 2018.

[8]  Y. Umuroglu, . N. J. Fraser, G. Gambardella, M. Blott, P. Leong, M. Jahre and K. Vissers, "Finn: A framework for fast, scalable binarized neural network inference," in *Proceedings of the 2017 ACM/SIGDA International Symposium on Field-Programmable Gate Arrays*, ACM, 2017, pp. 65-74.

[9]  H. Ren , A. N. Mazumder , H.-A. Rashid, V. Chandrareddy, A. Shiri and T. Mohsenin, "End-to-end Scalable and Low Power Multi-modal {CNN} for Respiratory-related Symptoms Detection," in *IEEE 33rd International System-on-Chip Conference (SOCC)*, 2020.

[10] M. Khatwani, H.-A. Rashid, H. Paneliya, M. Horton, N. Waytowich, W. D. Hairston and T. Mohsenin, "A Flexible Multichannel EEG Artifact Identification Processor using DepthwiseSeparable Convolutional Neural Networks," *ACM Journal on Emerging Technologies in Computing Systems (JETC),* 2020.

## 4.8 Electromyography processing using Wearable Neuromorphic Technologies

Elisa Donati Institute of Neuroinformatics, University of Zurich and ETH Zurich, Switzerland

**Status**

Electromyography (EMG) is a neurophysiological technique for recording muscle movements. It is based on the principle that whenever a muscle contracts, a burst of electric activity is propagated through the close tissue. The source of the electrical signal in EMG is the summation of action potentials of motor units (MUs) [1]. A MU is composed of muscle fibers innervated by axonal branches of a motorneuron, that is intermingled with fibers of other MUs. The recorded electric activity is linearly correlated to the strength of the contraction and the number of recruited MUs. EMG signals can be acquired both invasively, using needle electrodes, and superficially, by placing electrodes on the skin - called surface EMG (sEMG).

EMG signals have been and are relevant in several clinical and biomedical applications. In particular, they are extensively employed in myoelectric prosthetics control for classifying muscle movements. Wearable solutions for this application already exist, but they have a large margin for improvement, from increasing the granularity of movement classification to reducing computational resources needed and consequently power consumption.

Like any other signal, EMG is susceptible to various types of noises and interferences, such as signal acquisition noise, and electrode displacement. Hence, a pre-processing phase is the first step to perform proper signal analysis, which involves filtering, amplification, compression, and feature extraction both in time and frequency domains [2]. The mainstream approach for movement classification is machine learning (ML), which delivers algorithms with very high accuracy [3], although the high variability in test conditions and their high computational load limit their deployment to controlled environments. These drawbacks can be partially solved by using deep learning techniques that allow for better generalization to unseen conditions but remain computationally expensive, requiring bulky power-hungry hardware, that hinder wearable solutions [4].

Neuromorphic technologies offer a solution to this problem by processing data with low latency and lowpower consumption mimicking the key computational principles of the brain [5]. Compared to state-of-the-art ML approaches, neuromorphic EMG processing shows a reduction of up to three orders of magnitude in terms of power consumption and latency [6–8], with limited loss in accuracy(5–7%) [9, 10].

New approaches have been proposed that directly extract the motorneurons activity from EMG signals as spike trains [12]. They represent a more natural and intuitive interface with muscles but currently limit themselves by processing spikes with traditional ML techniques and do not consider the possibility of using more appropriate frameworks such as spiking neural networks (SNNs).

**Current and Future Challenges**

Although the performance of myoelectric prosthetics increased conspicuously in the last decade [13], they still can not be used in daily life. The fine-grained control is in fact limited by the number of electrodes. This issue can be overcome by using High-Density EMG (HD-EMG), which typically uses hundreds of electrodes, allowing to monitor larger areas and effectively increasing the precision of the measurements [14]. However, HD-EMG uses more computational resources, in terms of power and time required to classify movements and to generate motor commands. Current technologies are not able to process such an amount of data in-situ and with low latency simultaneously. For this reason, the EMG signals are transmitted, for example via Bluetooth, to a remote system that is quite bulky and heavy, making a wearable solution impractical.

Neuromorphic technologies represent a solution to all the described limitations by processing data in parallel, with low latency, and taking advantage of the low-power nature of analog computing and spiking communication, as the biological system they are inspired from. Although recent results show promising advances, the current challenge of neuromorphic technology is to fill the gap with state-of-the-art ML approaches, in terms of accuracy. One of the main reasons behind this gap is the different amount of resources invested in the respective research fields. In addition, current research that focuses on adopting ML methods and implementing them in neuromorphic hardware faces challenges governed by the unsuitability of such substrates which are primarily targeted for SNNs [11].

To get the most from neuromorphic computing we need a change of paradigm, where the neuromorphic technology can directly interface with motorneurons' spiking activity, instead of continuous sEMG signals. This represents a matching condition between inputs and outputs that optimize the information transfer between the muscle activity and the processing and control unit. The spike trains of motorneurons can be extracted from sEMG signals by means of decomposition algorithms. In particular, the spatial distribution of MUs action potentials can be assessed with activation maps obtained from HD-EMG signals [12]. Nevertheless, current implementations are still computationally expensive, and only recently it was possible for their deployment in real-time. After the decomposition, the spike trains are translated and processed using ML methods instead of better-suited SNNs [15].

Designing neuromorphic systems able to extract and process motorneurons activity from EMG signals will pave the way to a new class of wearable devices that can be miniaturized and directly interface with the electrodes.

**Advances in Science and Technology to Meet Challenges**

A concrete roadmap towards wearable neuromorphic EMG processing, see Figure 1, could be constructed with short and long-term objectives. In the short term, we should advance neuromorphic computation to bridge the gap with ML methods for EMG classification, and optimize decomposition algorithms to make them run real-time on embedded systems. In the long-term, the decomposition algorithm should be ported into a neuromorphic chip to implement a fully spiking pipeline while the technological breakthroughs in surface smart electrodes could potentially be able to record directly motorneurons' spike trains.

Bridge-the-Gap. The first step is to understand the requirements to improve the accuracy of EMG movements classification. The front-end, which includes pre-processing and spike conversion, has the largest margin for improvement. Signal-to-spike conversion produces spike trains required by neuromorphic devices. The most common signal-to-spike converter is the delta-sigma [7] which is widely applied in biomedical applications, thanks to its lower circuit complexity compared to multi-bit ADCs. However, the delta-modulator generates a high sampling rate and larger data size that can easily push the neurons' firing rate into saturation, making them insensitive to further input variations. Furthermore, SNNs for EMG classification should be optimized and learning algorithms could make them adaptable to different patients.

Embedded Decomposition sEMG decomposition into spike trains is generally based on shape-based algorithms, also called template matching [16] or blind source separation algorithms [17]. The decomposition of the complex sEMG is a computationally expensive procedure in a multidimensional constraint space. To run these algorithms on embedded platforms and in real-time it is imperative to i. reduce the complexity and ii. optimize it for the selected digital embedded architecture (e.g. PULP platform [18]) and exploit its hardware capabilities. The extracted spike trains are then sent to a neuromorphic chip, creating a hybrid digital-analog framework for spike encoding low-power computation.

Spike-based EMG decomposition To build a fully spiking pipeline that can be integrated into a single neuromorphic chip, the MUs identification algorithm needs to be translated into a spiking version. Embedding the entire process into a single chip that can be miniaturized and connected directly to the electrodes will allow online processing, which is optimal for real-time closed-loop applications and less vulnerable to interferences either caused by humans or the environment.

Smart electrodes Another long-term game-changer would be the technological breakthroughs that will allow the single electrode to be able to record directly the activity of a single MU, removing the need for decomposition algorithms.

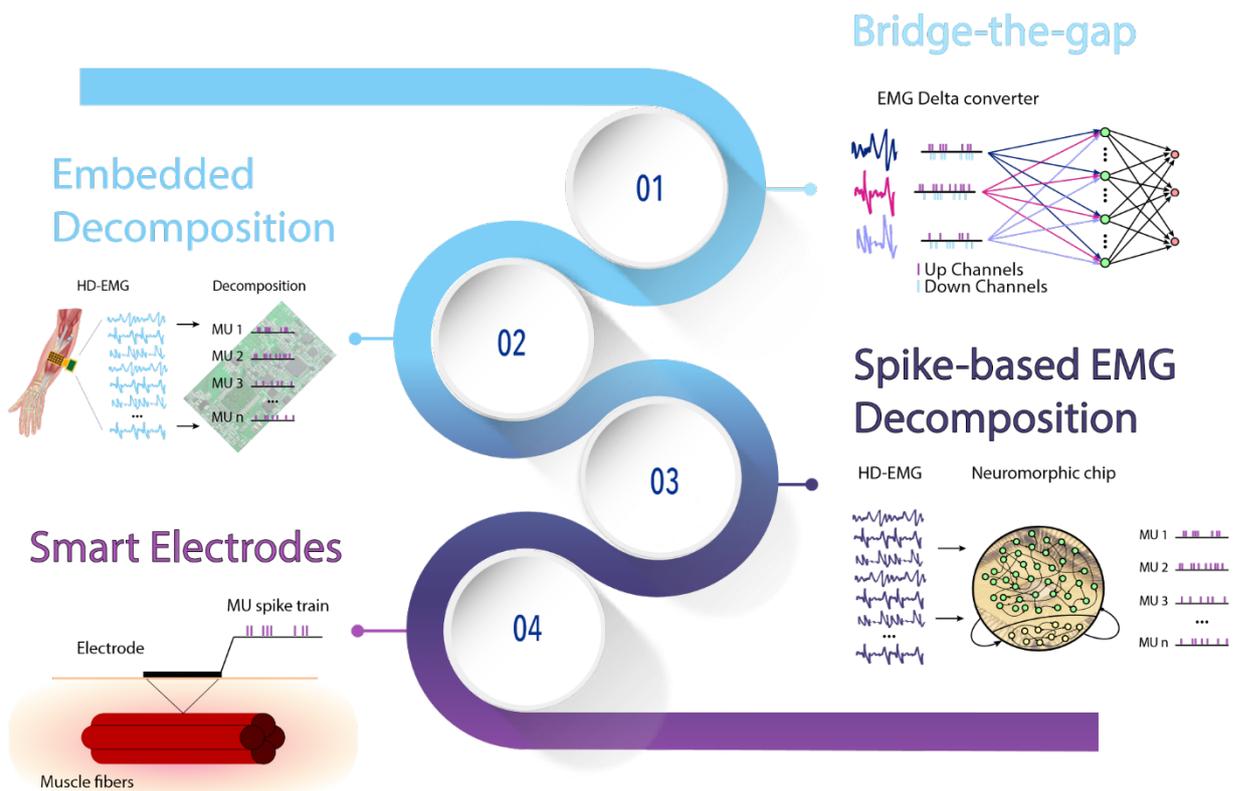

Figure 1: A concrete Roadmap towards neuromorphic Wearable Devices

**Concluding Remarks**

The need of improving myoelectric prosthetic control to increase the life quality of the patient poses new challenges for implementing real-time, compact, and low-power EMG processing systems. A wearable device based on neuromorphic technology can enable in-situ EMG signal processing and decomposition, without information transfer and external computation. In particular, mixed-signal SNNs implemented on neuromorphic processors can be integrated directly with the sensors to extract temporal data streams in real-time with lowpower consumption.

This roadmap presents the specific case of prosthetic control, nevertheless, the development of this technology could reveal useful to more applications where continuous monitoring is required. In clinical settings, continuous monitoring of EMG signals can be utilized to detect degenerative diseases of motorneurons [19] even for very large time spans such as weeks or months. In rehabilitation, EMG can be used as feedback to adapt the patient training accordingly to its muscular status, after a stroke or neurological impairments [20].

With the current rate of technological and computational improvements the proposed objectives could be realistically achieved within a decade. If successfully executed, this roadmap will bring technology that will improve the quality of life for amputees and patients with motorneuron diseases.

## 4.9 – Collaborative Autonomous Systems
Silvia Tolu, Roberto Galeazzi, Technical University of Denmark (DTU)

**Status**

Collaborative autonomous systems (CAS) (see Figure 1) are entities that can cooperate among themselves and with humans, with variable level of human intervention (depending on the level of autonomy) in performing complex tasks in unknown environments. Their behaviour is driven by the availability of perception, communication, cognitive and motor skills and improved computational capabilities (on/off-board systems). The high level of autonomy enables the execution of dependable actions under changing internal or external conditions. Therefore, CAS are expected to be able to: 1. perceive and understand their own condition and the environment they operate in; 2. dependably interact with the physical world despite of sudden changes; 3. intelligently evolve through learning and adaptation to unforeseen operational conditions; 4. self-decide their actions based on their understanding of the environment.

Currently, CAS (e.g., collaborative robots - cobots) show limited performances when accomplishing physical interaction tasks in complex scenarios [1]. Recent studies have demonstrated that autonomous robots can outperform the task they are programmed for, but they are limited in the ability to adapt to unexpected situations [2] and to different levels of human-robot cooperation [1]. These limitations are mainly due to the lack of generalization capabilities, i.e., cobots cannot transfer knowledge across multiple situations (environments, tasks, and interactions). One of the most viable pathways to solve this issue is to build intelligent autonomous cobots by incorporating Artificial Intelligence (AI)-based methods into the control systems [3]. These bio-inspired controllers [4] allow taking a different perspective from the classical control approaches, which require a deeper understanding of the mechanics of the interactions and of the intrinsic limitations of the systems beforehand. Main current research directions [5] are focused on the understanding of the biological working principles of the central nervous system (CNS) in order to build innovative neuromorphic computing algorithms and hardware that will bring significant advances in this field; In particular, they will provide computational efficiency and powerful control strategies for robust and adaptive behaviours.

In the next decades, there will be significant developments in CAS related to self-capabilities such as self-inspection, -configuration, -adaptation, -healing, -optimization, -protection, and -assembly. This will be a great enabler of systems acting in real-world unstructured scenarios, such as in remote applications (deep sea or space), in hazard situations (disasters), in healthcare interventions (assistive, rehabilitation, or diagnosis), and in proximity to people.

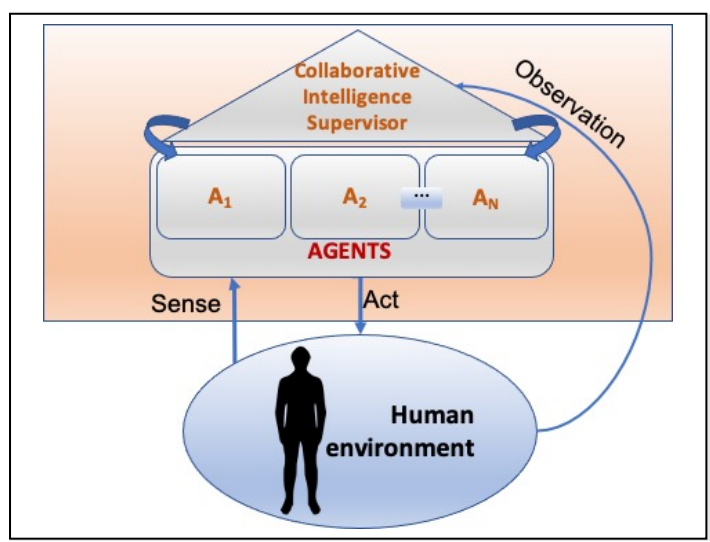

**Figure 1.** Overall idea of a Collaborative Autonomous Control System. The supervisor manages the entire system, observes and analyses the whole situation and provides information to each agent to improve their autonomous actions and optimize the operations.

**Current and Future Challenges**

Several fundamental challenges demand to be addressed to enable the deployment of heterogenous autonomous systems able to collaborate towards the achievement of common mission objectives.

These challenges span across different research topics including **online mission planning** and execution for multi-agent systems under uncertainty. Future mission planners [6] should integrate several factors to determine the optimal allocation of agents to the fulfilment of the mission tasks. These factors, among many, include energy availability and depletion rates, physical capabilities of the agents, probability of failures, and amount of collaboration needed. The mission execution demands the development of a revolutionary control paradigm that enables true collaboration among CAS with different functionalities. **Cooperative control** [7] has been so far limited to consensus and synchronization to enable the coordinated dynamic evolution of mostly homogeneous multi-agent systems to perform the same type of actions. The execution of tasks in uncertain environments calls for robust learning/adaptation methods to enable baseline control systems to ensure robust cooperation and coordination of heterogeneous multi-agent robots in various real applications [7]. Another big question is how to endow CAS with a high-level of **fault tolerance** capabilities in order to ensure dependability under a wide variety of operational conditions [8]. Despite the large research effort pursued by the community over the past four decades, condition monitoring and fault tolerant control are lacking efficiency due to the ever-increasing complexity of the systems.

Future work will aim to provide insights about how a CAS will show **robust, compliant, and intelligent physical interactions** with the environment, human beings, or other systems. In this regard, real-time, energy-efficient computing is required to advance the type of primitive collaborations that are achievable so far. With this aim, systems should be equipped with small processors able to ensure low energy consumption, and, at the same time, increase the memory bandwidth. Current alternatives (e.g., multicore central processing units - CPUs [9], new graphical processors - GPUs [10], parallel processing core - SpiNNaker [11]) still suffer from an extremely high energy demand that is not sustainable and they cannot be easily scaled. Additionally, a limited number of processes can run simultaneously, and the speed of the response is still low. Consequently, new neuromorphic architectures are the most promising alternatives to address the increasing demand to create CAS able of a seamless interaction with human beings.

**Advances in Science and Technology to Meet Challenges**

In this section, we discuss the foremost advances in science and technology that will address the main aforementioned challenges.

**Mission Planning** - Novel AI-based heuristic methods will be developed to equip mission planners with key functionalities that will increase the value for the human operators. These include: the close-loop decomposition of missions to achieve an adaptive task allocation by leveraging information gathered at mission execution; automated survivability prediction to assess the likelihood of vehicle loss based on faults and failures occurred in past missions; automated reliability assessment to forecast the probability of mission failure based on past missions' information; automated learning from previous missions' performance to tune the future missions' parameters; inclusion of services to extend the mission endurance [12, 13].

**Fault-tolerant and Cooperative Control** - Paradigms based on cooperation will be created to fulfil the advances in multi-agent systems [8]. Cooperation among agents offers the possibility of achieving fault-tolerance towards sensors and actuators faults through the design of diagnostic solutions that leverage shared proprioceptive and exteroceptive information. Prescribe-time fault tolerant cooperative control solutions for safety critical cyber-physical systems will be achieved; these will provide the basis for efficient fault-tolerant algorithms able to trade-off between fast convergence and acceptable fault-tolerance performance.

**Robust, Compliant and Intelligent Physical Interactions** - New physical mechanisms will be designed to provide passive properties to the system, to increase the physical interaction performances, and include advanced control aspects for achieving simultaneous robustness and compliance. The advances in Neuro-robotics and Neuromorphic Computing will influence the development of the next generation of intelligent agents [14]. Current neuromorphic computing systems already exploit learning and adaptive skills in systems compared to conventional von-Neumann machines thanks to non-volatile memories and power efficiency performance [15]. However, new types of sensors and actuators will be introduced to enhance the cognitive and learning functionalities of the systems and deal with safety and robustness concerns. Advanced bio-inspired platforms, e.g., brain-on-the-chip devices, will be designed for processing complex brain-inspired computing techniques that will support autonomy, more connectivity, increased decentralization, and high-performance computing. Indeed, neuromorphic technologies will be able to process complex unstructured data and learn to self-respond to external unknown stimuli enabling their use in critical edge applications, for example in autonomous navigation, human-machine interactions and smart healthcare markets.

Finally, innovative applications could be generated through the development of self-reconfigurable modular CAS, systems able to adapt their morphology and functionality to varied environments including unforeseen conditions [16]. This will require self-learning capabilities to develop new knowledge and to decide upon the previous accumulated experience.

**Concluding Remarks**

This paper has presented the future perspectives of collaborative autonomous systems and the main challenges and research issues that need to be addressed toward their realization. Further to these scientific and technological challenges, there are ethical, social, and legal issues when realising CAS, though these are beyond the scope of this article.

CAS working alongside humans have already been deployed and they support humans' work ensuring high productivity, speed, and accuracy [17]; they also relieve us of many heavy and time-consuming tasks and reduce the overall risk of collisions. CAS provide an economically viable entry-point to automation of processes, i.e., accelerated testing scenarios on products, environmental impacts.

Fusion of fundamental and applied research in both technical and natural sciences will facilitate the development of new theoretical frameworks for the design of intelligent CAS. Multiple disciplines will be merged to pursue a systematic innovation within cyber-physical systems with variable level of autonomy and cooperation; the use of AI and Internet of Everything technologies future proofs the system to address changing market demands and expectations in several technological areas. Applications will be many and varied including, and not limited to, manufacturing, health care, inspection and maintenance, precision farming, autonomous marine operations, and education.

## 5. 1 The ethics of developing neuromorphic technology

Martin Ejsing Christensen
Research Officer
The Danish Council on Ethics
mech@dketik.dk

Sune Holm
Associate Professor
Department of Food and Resource Economics
University of Copenhagen
suneh@ifro.ku.dk

Like the development of other forms of artificial intelligence, the development of neuromorphic technology may raise a number of ethical questions. [1], [2].

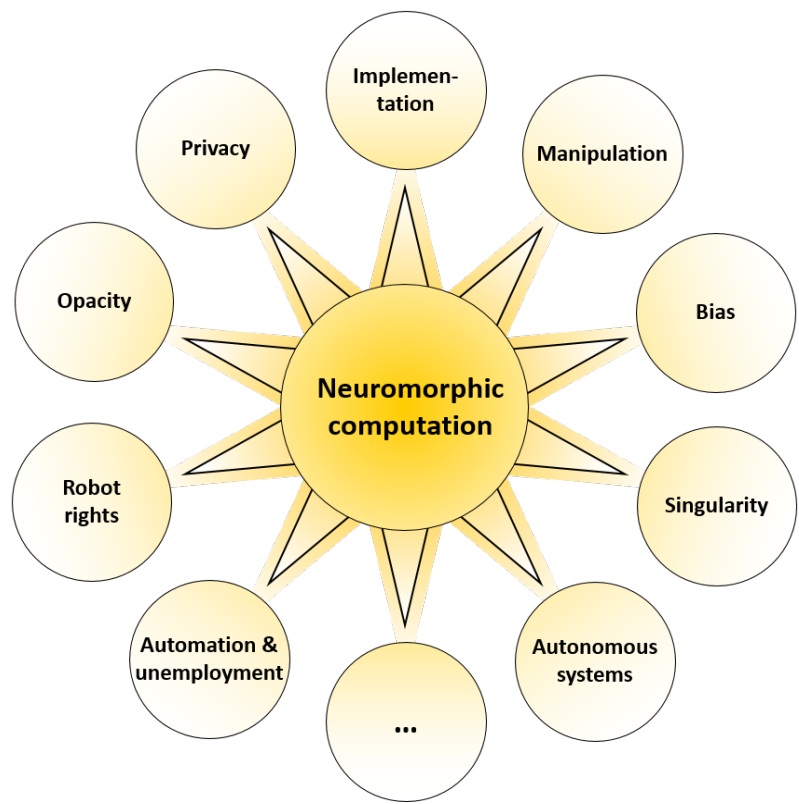

*Figur 1 Some of the most salient ethical issues raised by the development of neuromorphic technology*

One issue concerns privacy and surveillance. The development of most forms of artificial intelligence depends upon access to data, and as far as these data can be seen as private or personally identifiable, it raises a question about when it is (ethically) defensible to use such data. On the one hand, some argue that persons have a right to be let alone and exercise full control over information about themselves, so that any use of such data presupposes fully informed consent. On the other hand, others recognize the importance of privacy but argue that it may sometimes be outweighed by the fact that reliable applications for the good of everyone presuppose access to high quality representative data [3].

Another issue concerns opacity. Many forms of artificial intelligence support decision making based on complex patterns extracted from huge data sets. Often, however, it will be impossible not only for the person who makes the final decision but also for the developer to know what the system's recommendations are based on and it is in this sense that it is said to be opaque. For some such opacity does not matter as long as there are independent ways of verifying that the system delivers an accurate result, but others argue that it is important that the system is explainable [4]. In this way, a tension is often created between accuracy and transparency, and what the right trade-off is may often depend upon the concrete context.

Opacity is closely connected with the question of bias since opacity may hide certain biases. There are different forms of bias but in general, bias arises when automated AI decision support systems are based on data that is not representative of all the individuals that the system supports decisions in relation to [5]. There are different opinions as to when the existence of bias in automated decision support systems poses a serious problem. Some argue that 'traditional' unsupported human decision-making is biased, too, and that the existence of bias in automated AI decision support systems only pose a serious problem if the bias is more significant than the pre-existing human bias. Others argue that features such as opacity or the lack of suitable institutional checks and balances may tend to make the existence of bias in automated decision support systems more problematic than 'ordinary' human bias [6]. A separate problem is created by the fact that it sometimes will be easier to identify and quantify bias in AI systems than in humans, making a direct comparison more difficult.

The development of forms of artificial intelligence based on neuromorphic technology also raises questions about manipulation of human behavior, online as well as offline. One context in which such questions arise is advertising and political campaigning, where AI generated deep knowledge about individuals' preferences and beliefs, which may be used to influence them in a way that escapes the individuals' own awareness. Similar issues may also arise in connection with other forms of artificial intelligence such as chatbots and care or sex robots that simulate certain forms of human behavior without being 'the real deal'. Even if persons develop some form of emotional attachment to such systems, some argue that there is something deeply problematic and deceptive about such systems [7], while others point out that there is nothing intrinsically wrong with such systems as long as they help satisfy human desires [8]. If, as described in section 4.1, neuromorphic technologies will make it possible for robots to move from extremely controlled environments to spaces where they collaborate with humans and exhibit continuous learning and adaptation, it may make such questions more pressing.

A distinct set of issues are raised by the possibility of developing AI systems that do not just support human decision making but operate in a more or less autonomous way such as 'self-driving' cars and autonomous weapons. One question that such systems raise concerns the way in which they should be programmed in order to make sure that they make ethically justifiable decisions (in most foreseeable situations). Another question concerns how responsibility and risk should be distributed in the complex social system they are a part of. If, as described in section 4.2, neuromorphic engineering offers the kind of technological leaps required for achieving truly autonomous vehicles, the development of neuromorphic technologies may make such questions more pressing than at present.

A distinct issue relates to sustainability. As pointed out in the introduction, 5-15% of the world's energy is spent in some form of data manipulation (transmission or processing), and as long as a substantial amount of that energy comes from sources that contribute to climate change through the emission of greenhouse gases, it raises a question as to whether all that data manipulation is really necessary or could be done in a more energy efficient way. And in so far as neuromorphic technologies, as e.g. pointed out in section 4.7., shows a reduction of up to three orders of magnitude in terms of power consumption compared to state-of-the-art ML approaches, it seems to provide robust ethical support for the development of neuromorphic technologies.

As mentioned in the beginning of this section, the ethical questions raised by the development of neuromorphic technology is not unique to this technology but related to the development of artificial intelligence as such. The successful development of neuromorphic technology may make some of the issues more pressing, and a central task for future work on the ethics of neuromorphic technology will, accordingly, be to inquire into the exact way in which the issues are raised by the development of neuromorphic technology. But the existing forms of artificial intelligence already raise many of the questions described so far. Besides these questions, however, the development of neuromorphic technology (as well as other forms of artificial intelligence) may also raise a number of questions that are more speculative either because it is unclear whether the development will take place, when it will happen or what the precise consequences will be.

One such issue has to do with automation and unemployment. Artificial intelligence systems have already replaced humans in certain job functions (e.g., customer service), but it has been suggested that most job functions will be affected by the development of artificial intelligence at one point [9]. Because such a development has the potential to disrupt the social order (e.g., through mass unemployment) it raises an important ethical (and political) question as to how artificial intelligence systems should be introduced into society [10].

Another more speculative issue relates to artificial moral agents and so-called robot rights. If the development of neuromorphic (and other) forms of artificial intelligence leads to the creation of systems that possess some or all the traits that make us ascribe rights and responsibilities to humans, it may thus raise a question about whether such rights and responsibilities should be ascribed to artificially intelligent systems [11], [12].

Thirdly, some have also pointed out that the development of neuromorphic (and other) forms of artificial intelligence may create issues related to the so-called singularity. The idea is that the technological development may lead to the creation of general forms of artificial intelligence that surpass the human level of intelligence and then begin to control the further development of artificial intelligence in ways that may not be in the interests of the human species and perhaps even threaten its very existence. Whether such a scenario is likely has been questioned [13], but some argue that even a slight risk should be taken serious given the potentially devastating consequences [14].

No matter what one thinks is the right answer to the ethical questions raised by the development of neuromorphic technology, it is, finally, worth noticing that it still leaves an important practical question: how best to make sure that the actual development and implementation of neuromorphic technology will take place in an ethically defensible way. For some questions, governmental regulation may be the best means. For others, the best solution may be to trust the community of developers to make the right, value-based decisions when designing systems, while some questions, perhaps, should be left to the enlightened citizenry. In the end, however, it will probably be up to an inquiry into the concrete situation to decide when one or the other approach – or combination of approaches – provides the best means of securing an ethically defensible development of neuromorphic technology.